\documentclass[reprint,aps,prb,longbibliography]{revtex4-2}
\usepackage{amsmath,amssymb,color,comment}
\usepackage[makeroom]{cancel}
\usepackage[caption=false]{subfig}
\usepackage{mathrsfs}
\usepackage{ esint }
\usepackage{graphicx}
\usepackage{subfig}
\usepackage[countmax]{subfloat}
\usepackage[english]{babel}
\usepackage{ stmaryrd }
\usepackage{relsize}
\usepackage{ bbold }
\usepackage[bookmarks=true,colorlinks,linkcolor=blue,urlcolor=blue,citecolor=blue]{hyperref}
\usepackage{cleveref}
\usepackage{ textcomp }
\usepackage[section]{placeins}
\crefname{equation}{}{}

\usepackage{bm}

\newcommand{\eF}{\varepsilon_F}
\newcommand{\ek}{\varepsilon_\mathbf{k}}

\newcommand{\myim}{\operatorname{Im}}
\newcommand{\myre}{\operatorname{Re}}
\newcommand{\sgn}{\operatorname{sgn}}

\newcommand{\win}{\omega_\text{IN}}
\newcommand{\wmax}{\omega_{\max}}
\newcommand{\ep}{\varepsilon_p}
\newcommand{\dpq}{d^2_\mathbf{p}}
\newcommand{\dpv}{d^4_\mathbf{p}}
\newcommand{\dkq}{d^2_\mathbf{k}}
\newcommand{\dkFq}{d^2_{\mathbf{k}_F}}
\newcommand{\dkFv}{d^4_{\mathbf{k}_F}}
\newcommand{\dkF}{d_{\mathbf{k}_F}}

\begin{document}
\title{$\omega/T$ scaling and IR/UV-mixing in Ising-nematic quantum critical metals}
\author{Bernhard Frank}
\author{Francesco Piazza}
\affiliation{Max-Planck-Institut f\"ur Physik komplexer Systeme, 01187
 Dresden, Germany}
\begin{abstract}
The instability of a Fermi surface against Ising nematic order destroys the quasiparticle character of the low-energy degrees of freedom. Therefore, observables exhibit deviations from Fermi liquid behavior which gives rise to the term Ising nematic quantum critical metal. To obtain a theoretical description we use a finite-temperature version of Eliashberg theory which allows to treat the strong coupling between quantum and thermal fluctuations in the absence of well-defined quasiparticles. Here, we use this self-consistent, diagrammatic approach to compute, in particular, the nematic susceptibility and the non-Fermi liquid correlations. Upon decreasing the temperature, the susceptibility crosses over from a $(T \log T)^{-1}$ to $T^{-2/3}$ behavior, which is induced by the absence of quasiparticles and restores the Ising nematic critical scaling. Correspondingly, the fermions obey a simple $\omega/T$ scaling law at low enough temperatures. However, this regime is characterized by strong IR-UV mixing since the proportionality factors exhibit a dependence on the spectral width of the non-quasiparticle excitations and on the underlying lattice. Tuning the parameters of the model, therefore, gives rise to several scenarios for the breakdown of the scaling theory. We discuss them within the Eliashberg approach and estimate the related crossover scales. We also show that the leading order vertex corrections do not change the scaling with temperature or coupling constants.

\end{abstract}
\maketitle
\section{Introduction}\label{sec:intro}
Nematic correlations have been observed in a variety of strongly correlated electron systems like underdoped cuprates~\cite{ando2002, kohsaka2007,hinkov2008,daou2010,cyr2015,keimer2015} or iron-based compounds~\cite{nandi2010,chuang2010,chu2010,song2011,baek2015,watson2015,boehmer2017,coldea2018}. 
These nematic signatures~\cite{fradkin2010rev,fernandes2014,wang2015} are associated with a quantum critical point~\cite{sachdevBook} (QCP). 
In particular, the coupling between the low-energy Fermions to the critical modes close to a QCP is a standard route to the creation of a non-Fermi liquid state~\cite{loehneisen07} (non-FL) which is characterized by the absence of  well-defined quasiparticles (qp).    

In this work we consider the Ising nematic model~\cite{metzner2003} (INM) in $d=2$ that describes the coupling of the Fermi surface (FS) to an Ising order parameter which gives rise to a QCP accompanied by the reduction of the rotational symmetry of the electronic system from $C_4$ to $C_2$.  
While first studied within the Hertz-Millis-Moriya framework~\cite{hertz76,millis1993,moriya73} for Bose-Fermi-models it turned out that the underlying effective bosonic action contains an infinite set of highly singular, nonlocal terms~\cite{rech2006,metl10,thier11}. Therefore, various methods and expansion schemes have been devised to treat the bosonic and fermionic sectors of the INM and the very related problem of a FS coupled to an U(1) gauge field on equal footing, in order to access the low-energy physics down to  the ground state in a controlled way~\cite{lee1989,altshuler1994,kim1994,nayak1994,oganesyan2001,rech2006,dell2006,zacharias2009,
lee2009,metl10,
maslov2010,mross2010,thier11,yamase2012,dalidovich2013,fitzpatrick2014,hartnoll2014,sur2014,holder2015,
holder2015rapid,punk2016,eberlein2016,eberlein2017,lee2018rev,klein2020}. 
An essential ingredient are Landau overdamped Boson
that gives rise to the dynamical critical exponent $z=3$ in Hertz-Millis theory.
The Fermions, on the other hand, acquire in the ground state a self-energy that is non-analytic in the low energy regime with branch cuts rather than poles, which manifests the non-qp character of the excitations. For instance, at one loop one obtains the self-energies $\Sigma_B(\mathbf p, \omega_p)\sim \omega_p/p $ and $\Sigma_F(\mathbf{k}_F, \omega_k) \sim \win^{1/3}|\omega_k|^{2/3}$ where $\mathbf k_F $ denotes a momentum on the FS. However, it turns out that these expressions also solve the corresponding self-consistent theory~\cite{polchinski1994}, which is called Eliashberg theory (ET) for its formal similarity to the electron-phonon case. Corrections to the fermionic self-energy are found at three loops for the momentum dependence in form of a small anomalous dimension~\cite{metl10}, while the inclusion of four loops~\cite{holder2015,holder2015rapid} indicates a deviation from $z=3$. 

\begin{figure}[t]
\begin{center}
\includegraphics[width=1\columnwidth]{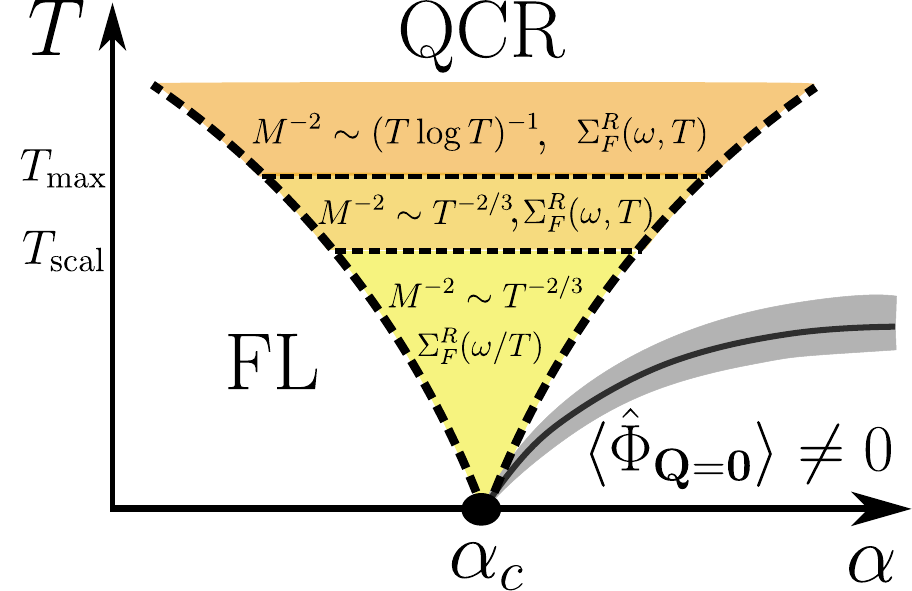}
\caption{
Phase diagram of the Ising nematic order parameter $ \langle \hat \Phi_{\mathbf Q=0}\rangle $; $\alpha$ denotes the effective coupling constant between the fermionic and bosonic sector. The nematic susceptibility $M^{-2}(T)$ scales like $T^{-2/3}$ below $T_{\max}$. The fermionic self-energy obeys a non-universal $\omega/T$ scaling form below $T_{\text{scal}}$.}
\label{fig:PD_INM}
\end{center}
\end{figure} 

At finite temperature the situation becomes more involved since quantum and thermal fluctuations become intertwined. This is of particular importance for the quantum critical regime (QCR) depicted in Fig.~\ref{fig:PD_INM}, which features two large correlations lengths: Spatial correlations extend over a distance $\xi \sim |\alpha-\alpha_c|^{-\nu}$ much larger than the intrinsic length scales of the system, while the correlations in the imaginary time domain exceed the standard interval set by the inverse temperature: $\xi_T \sim |\alpha -\alpha_c |^{-z \nu} \gg \hbar /(k_B T)$. Here, $\nu$ denotes the correlation length exponent and $\alpha$ refers to the effective coupling between the bosonic and fermionic modes. 

Regarding the INM, the static nematic susceptibility in the limit of long-wave lengths $-D(\mathbf p = \mathbf{0},\omega_p =0) = 1/M^2(T)$, which diverges at the QCP, can be parametrized at $T>0$ by a thermal mass gap $M(T)$ of the bosonic modes. Imposing scale invariance on $D(0,0)$ implies 
\begin{align}\label{eq:MTQCR}
M(T) \sim T^{1/z} \xrightarrow{z= 3} T^{1/3} \, . 
\end{align} 
Based on the assumption of scale invariance one can furthermore postulate a quantum critical scaling form for the single electron correlation functions that obeys $\omega/T$ scaling. The resulting temperature dependence of thermodynamic and transport quantities depends on $z$ and differs from the standard FL predictions~\cite{senthil08}. However, the low-energy physics of the INM in the ground state can be described in terms of a two-patch model which features an emergent U(1) gauge symmetry~\cite{metl10,mross2010,dalidovich2013}. This symmetry restricts $M(T) \equiv 0$ even at finite temperatures and a finite value of $M(T)$ is only generated when effects from higher energies like the underlying band dispersion are taken into account~\cite{hartnoll2014}. Indeed, Hertz-Millis theory and an equivalent expansion in the number of Fermion flavours yields~\cite{millis1993,hartnoll2014}
\begin{align}\label{eq:MTqp}
M(T) \sim (T \log T)^{1/2} \, .
\end{align} 
Combining this form with Eliashberg theory at finite temperatures furthermore leads to the break-down of $\omega/T$ scaling~\cite{dell2006} in the electronic correlations. This is caused by  the strong impact of static bosonic fluctuations with $\omega_p \to 0$ due to their large thermal occupation $\sim T/\omega_p$.
Using ET to adress the INM at finite temperatures is quite appealing since
a lot of results can be obtained in closed form. Moreover, such theories also give reasonable quantitative descriptions, sometimes even beyond their expected range of validity~\cite{chowdhury20,chubukov2020ET}. Quite remarkably in this regard, Klein et al.~\cite{klein2020} have shown recently for the INM that ET agrees very well with Quantum-Monte-Carlo (QMC) simulations~\cite{schattner16QMC,lederer17QMC,berg19QMCrev}, provided that it is properly modified to include thermal effects.
These unbiased, sign-problem-free QMC results provide the most reliable quantitative analysis available and confirm the temperature dependence of~\eqref{eq:MTqp}.

While these observations certainly settle the behavior at sufficiently high temperatures, the question how $M(T)$ vanishes at the lowest temperatures is not ultimately answered: The QMC simulations always work at finite temperatures and extracting ground state correlations poses an additional task~\cite{xu2020}. Furthermore, the non-FL is unstable against the transition to a superconducting phase such that nematic and pairing fluctuations become entangled which was both observed in QMC and in ET~\cite{raghu2015SC,lederer2015SC,wang2016SC,wang2017SC}. On the other hand, the existing analytic results of the form~\eqref{eq:MTqp} rely on methods that treat the electrons basically as quasiparticles. 

In this spirit, the major goal of the present work is to compute $M(T)$ in the QCR as function of temperature in a fully self-consistent manner in order to include the non-FL excitations. In order to focus on the nematic correlations we suppress the superconducting instability. Nevertheless, the resulting features of the spectral functions are expected to survive close and below $T_c$ in a large range of frequencies. Regarding $M(T)$, we indeed find a solution of the form~\eqref{eq:MTQCR} within ET. The proportionality factor depends on the spectral width $\wmax$ of the non-FL excitations and the corresponding domain $|\mathbf k - \mathbf k_F| \lesssim \Lambda$ in momentum space. ET allows to estimate these scales by determining the range-of-validity of the underlying Ising nematic scaling relations in the \emph{high}-energy limit. Thus, $M(T)$ exhibits IR/UV mixing which cannot be described within an effective low-energy theory focused on the IR only. In particular, extending the latter to all scales by taking the limit $\Lambda \to \infty, \wmax \to \infty$ leads to $M(T) = 0$ in agreement with the symmetry constraints. At sufficiently high temperatures, 
ET reproduces the form~\eqref{eq:MTqp}. The corresponding crossover temperature $T_{\max} \sim \wmax$ coincides with maximal spectral width of the non-FL excitations. For temperatures larger than this scale the thermal broadening of the excitations exceeds the inverse lifetime which restores a scenario rather typcial for quasiparticles. Irrespective of the temperature, we argue that the fermionic compressibility $(\partial n/\partial \mu)_T$ scales in the same way with $T$ as $M^2(T)$, when evaluated in ET. 
  
Regarding the electrons, we find that $\omega/T$ scaling is restored for temperatures below $T_{\text{scal}}\lesssim T_{\max}$. For instance, the imaginary of the retarded self-energy can be written as $\myim \Sigma^R_F(\omega,T) \sim T^{2/3} \tilde{\Sigma}^R_F(\omega/T)$ but the scaling function $\tilde{\Sigma}^R_F$ inherits a dependence on $\wmax$ and $\Lambda$ from the bosonic mass and thus exhibits IR/UV-mixing, too.  As a consequence, the breakdown of the scaling solution of ET can have different physical origins due to the appearance of various ratios of nonuniversal parameters. ET not only allows to identify them but also to estimate the corresponding $T_{\text{scal}}$.
In the range $T_{\text{scal}} \lesssim T \lesssim T_{\max}$ the inverse susceptibility is still of the form~\eqref{eq:MTQCR} but the fermionic self-energy cannot be described in terms of a specific scaling function. 
Fig.~\ref{fig:PD_INM} illustrates the different scaling regimes with respect to temperature.  

In principle, ET provides only an uncontrolled approximation to a strongly coupled problem. In particular, the neglected higher-order diagrams potentially give
rise to IR singularities that can change scaling forms obtained from ET. This can be seen by the anomalous fermionic dimension~\cite{metl10} or the anomalous correction to $z=3$~\cite{holder2015rapid,holder2015}. In addition, even the lowest order vertex diagram in the ground state shows a rather rich behavior in the low-energy limit, which depends delicately on how the latter is approached~\cite{chubukov2005Ward,rech2006,metl10}. At finite temperatures, the highly-occupied static bosonic fluctuations give rise to an additional contribution which is not perturbatively suppressed.  Nevertheless, we argue that at $T>0$ at least the total first order vertex correction merely changes the $O(1)$-prefactors of the self-energy, whereas the temperature dependence and the crossover scales are left invariant. 

The manuscript is organized as follows: In Sec.~\ref{sec:MaF} we introduce the INM and the Eliashberg equations. Furthermore, we discuss known results and the physical expectations for the correlations functions at finite temperature. 
In Sec.~\ref{sec:keyRes} we summarize the key results while the details of the underlying computations can be found in Sec.~\ref{sec:ET}. Before turning to $M(T)$, we make contact to previously established results of ET at finite temperatures and extend the calculations where necessary in Secs.~\ref{sec:bosDamp} and~\ref{sec:SF}. Sec.~\ref{sec:bosMass} is dedicated to the computation of the bosonic mass and the resulting scaling solution of the fermionic self-energy. Sec.~\ref{sec:vertexCorr} considers the vertex corrections  before we conclude in Sec.~\ref{sec:conclude}.  
\section{General Framework}\label{sec:MaF}
In the following, we will first briefly introduce the model and state the Eliahsberg equations in Sec.~\ref{sec:Model}. Afterwards, we summarize the known behavior of the relevant Green's functions in the ground state and discuss the expected changes at finite temperatures in Sec.~\ref{sec:ansatz}. 
\subsection{Model}\label{sec:Model} 
The INM consists of a two-dimensional Fermi surface (FS) of spin-1/2 electrons and a real bosonic field, whose expectations value is the Ising nematic order parameter. The two sectors are coupled by a Yukawa interaction. The corresponding  
euclidean action reads
\begin{align}\label{eq:defAction}
S =& \sum_{\mathbf k, \omega_n,\sigma} \bar{c}_{\mathbf k, \omega_n, \sigma} (- i\omega_n + \xi_{\mathbf k}) c_{\mathbf{k},\omega_n,\sigma} \notag \\
& + \frac{1}{2}\sum_{\mathbf{p},\Omega_n} \phi_{\mathbf p, \Omega_n}D_0^{-1}(\mathbf p, \Omega_n)\phi_{-\mathbf p, -\Omega_n} \\
&+ \frac{g}{\sqrt{\beta V}} \sum_{\substack{\mathbf k,\omega_n,\sigma\\ \mathbf p, \Omega_n}} d_{\mathbf k} \phi_{\mathbf p,\Omega_n} \bar{c}_{\mathbf k+\mathbf p/2,\omega_n + \Omega_n,\sigma} c_{\mathbf k-\mathbf p/2, \omega_n,\sigma} \, . \notag
\end{align}
Here, $c_{\mathbf k, \omega_n,\sigma}$ denote fermionic fields with momentum $\mathbf k$, Matsubara frequency $\omega_n$ and spin index $\sigma$. We will use units with $\hbar = 1 =k_B$ throughout. The dispersion $\xi_\mathbf{k} = \ek-\mu$ is measured with respect to the chemical potential and 
we will assume an isotropic FS with $\ek = k^2/(2m)$ rather than one generated by filling electronic bands of a $d=2$ square lattice. If necessary, one can always restore the non-trivial directional dependence of the density of states in a genuine band by introducing an angle-dependent Fermi velocity $v_F(\varphi_{\mathbf k})$. However, this will merely affect the angular weights entering the numerical prefactors of order one.
  
The bare nematic susceptibility corresponds to the propagator of a noninteracting real bosonic field: $D_0(\mathbf p, \Omega_n)=[-(i \Omega_n)^2 +c_{B,0}^2 p^2 + M_0^2)]^{-1}$. Here, we keep both the bare frequency-dependence and a bare mass term which can be assumed to be of Curie-Weiss form $M_0^2 \sim T-T_0$. Both of them turn out to be irrelevant once the contributions from the interactions are considered. The variable $c_{B,0}$ refers to the bare speed of the bosonic modes.   

The strength of the interactions is set by $g$, $V$ denotes the volume and $\beta=1/T$. Furthermore, the coupling term also carries the $d$-wave nematic form factor which on a square lattice of constant $a$ reads as
\begin{align}
d_\mathbf{k} = \cos(k_x a) - \cos(k_y a)  \, .
\end{align}
In the vicintiy of the spherical symmetric FS we use instead the approximation 
\begin{align}
d_\mathbf{k} \simeq \cos (2 \varphi_\mathbf k) \, ,
\end{align}
where $\varphi_{\mathbf k}$ denotes the angle between $\mathbf{k}$ and the x-axis.
The action~\eqref{eq:defAction} is invariant under rotations of $\pi/2$ provided that the Bose field transforms as $ \phi \to -  \phi$ which means that the Ising $\mathbb{Z}_2$-symmetry is conserved. For finite expectation values of $\langle \Phi_{\mathbf Q= \mathbf 0} \rangle \neq 0$, however, this symmetry is spontaneoulsy broken.
In the presence of a finite, conserved density of Fermions, a symmetry-broken phase with a homogoneous order parameter can always be realized at sufficiently large coupling $g$. The resulting phase diagram is most conveniently described in the $\alpha-T$ plane, where $\alpha \sim g^2$ turns out to be the relevant effective coupling constant. It is schematically shown in Fig.~\ref{fig:PD_INM}.
 In the following, we will always focus on the critical value $\alpha=\alpha_c$ such that the QCP is approached from the QCR in the limit $T \to 0$.
  
\begin{figure}[t]
\begin{center}
\includegraphics[width=0.7\columnwidth]{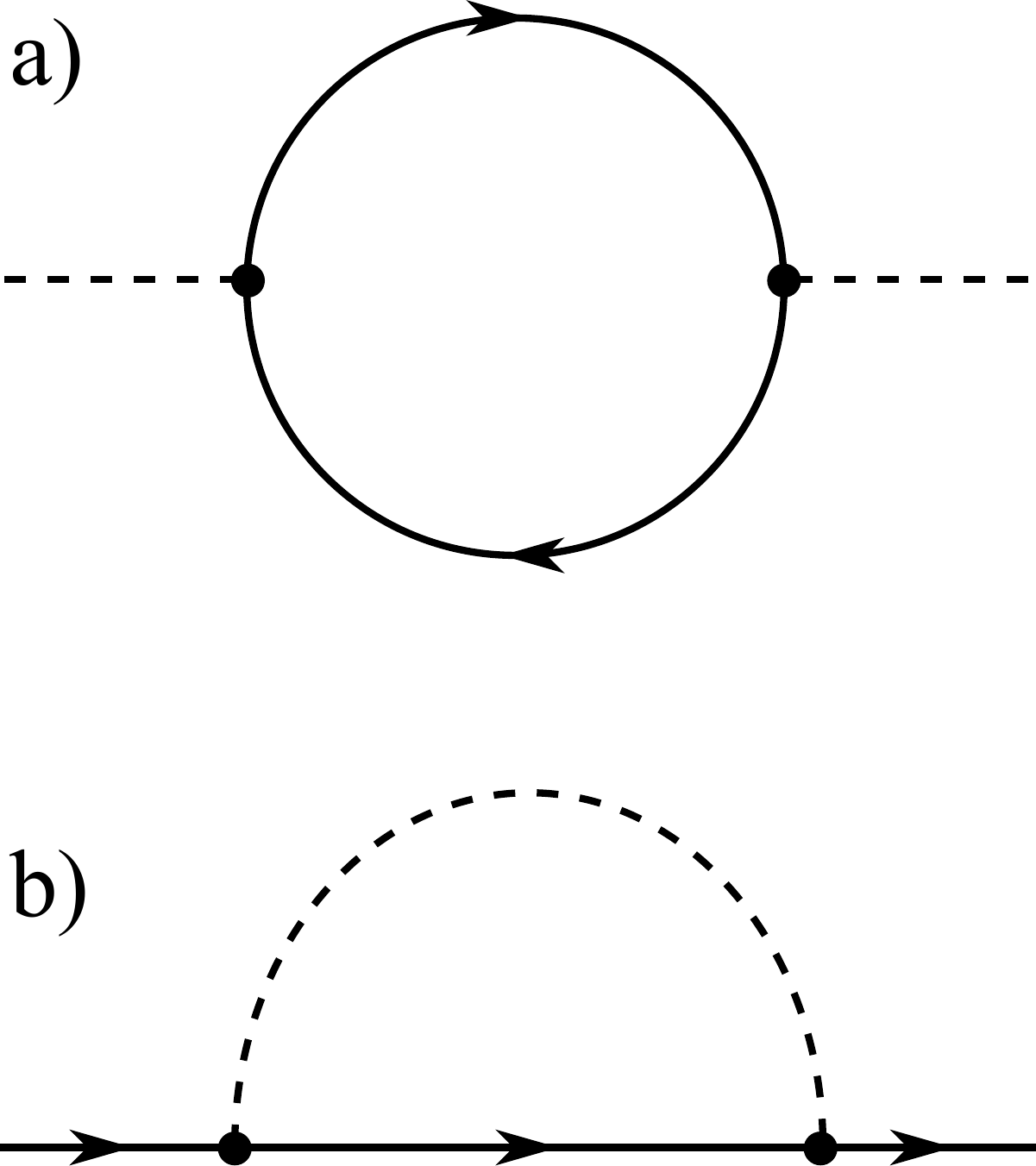}
\caption{
Bosonic self-energy $\Sigma_B$ (a) and fermionic self-energy $\Sigma_F$ (b);
Solid lines represent $G$ and dashed ones $D$.}
\label{fig:SE}
\end{center}
\end{figure}

Within ET, the effects of interactions are included via the two lowest-order diagrams depicted in Fig.~\ref{fig:SE}. 
In Matsubara frequencies the corresponding bosonic and fermionic self-energies read (spin indices are supressed): 
\begin{align}
\begin{split}
\Sigma_F(k) & = \frac{g^2}{\beta V} \sum_{\mathbf p , \Omega_n} d^2_{\mathbf k - \mathbf p/2} D(p) G(k-p)\\
\Sigma_B(p) & = -2 \frac{g^2}{\beta V} \sum_{ \mathbf k,\omega_n} d^2_{\mathbf k + \mathbf{p}/2} G(k+p) G(k)\, , 
\end{split}
\end{align}
where $k=(\mathbf k, \omega_n)$ and $p=(\mathbf p, \Omega_n)$ and the factor of two results from the sum over spins. 
In addition to the Fock diagram, also the Hartree diagram contributes to $\Sigma_F$ at one loop. However, it merely induces a shift of the chemical potential which we absorb into the physical value $\mu = \mu (n)$ at density $n$. 
 
To obtain information about the dynamics and in particular the electronic spectral functions we analytically continue the Matsubara self-energies to their retarded counterparts by the prescription~\cite{fetter2003Book} $\Sigma_F^R(\mathbf k, \omega_k)=\Sigma(\mathbf k, i \omega_n \to \omega_k + i 0^+$) and equivalently $\Sigma^R_B(\mathbf p,\omega_p) = - \Sigma_B(\mathbf p,i \Omega_n \to \omega_p + i 0^+)$, which applies in the same way for the continuation $G,D \to G^R,D^R$. Moreover, the retarded Green's functions allow to define the spectral functions~  
$A_F(\mathbf k, \omega_k) = -2 \myim G^R(\mathbf k, \omega_k)$ and $A_B(\mathbf p, \omega_p) = -2 \myim G^R(\mathbf p, \omega_p)$ that connect the Matsubara and the retarded Green's function via the Hilbert transformation
\begin{align}\label{eq:defHilbert}
\begin{split}
G(\mathbf{k},\omega_n) & = \int \frac{d\omega}{2\pi} \frac{A_F(\mathbf{k},\omega)}{i \omega_n - \omega} \\
D(\mathbf{p},\Omega_n) & = \int \frac{d\omega}{2\pi} \frac{A_B(\mathbf{p},\omega)}{i \Omega_n - \omega} \,.
\end{split}
\end{align}
Introducing the new variables $k=(\mathbf k, \omega_k)$ and $p=(\mathbf p,\omega_p)$, this yields for the imaginary parts  
\begin{subequations}\label{eq:defSgen}
\begin{align}
\myim\Sigma_F^R(k) & = -2 g^2 \int \frac{d^2 p\, d \omega_p}{(2 \pi)^3} d^2_{\mathbf k-\mathbf p/2}\myim G^R(k-p) \notag\\
& \qquad \quad\times \myim D^R(p)\left[n_B(\omega_p) + n_F(\omega_p - \omega_k)\right] \label{eq:defSF} \\ 
\myim \Sigma^R_B(p) & = 4 g^2 \int \frac{d^2k \,d\omega_k}{(2\pi)^3} d^2_{\mathbf k + \mathbf p/2} \myim G^R(k+p) \notag \\& \qquad \quad\times\myim G^R(k) \left[n_F(\omega_k) - n_F(\omega_k + \omega_p)\right] \label{eq:defSB} \, ,
\end{align}
\end{subequations}    
which contain information about the inverse lifetimes of the excitations. 
The functions $n_{F,B}$ denote the Fermi and Bose functions, respectively. The real parts $\myre \Sigma^R_{F,B}$ are obtained from $\myim \Sigma^R_{F,B}$ via the Kramers-Kronig relations
\begin{align}\label{eq:defKK}
\myre \Sigma^R_{F,B}(\mathbf q ,\omega) = \fint \frac{d \omega'}{\pi} \frac{\myim \Sigma^R_{F,B}(\mathbf q , \omega')}{\omega'-\omega} \, ,
\end{align}
where necessary.
Inserting $\Sigma^R_{F,B}$ into the Greens's functions $G^R$ and $D^R$ via the Dyson equation leads to
\begin{subequations}\label{eq:Dyson}
\begin{align}
G^R(k) & = \frac{1}{\omega_k-\ek + \mu - \Sigma^R_F(k)} 
\label{eq:GRex} \\
D^R(p)& = \frac{1}{\omega_p^2 -c_{B,0}^2 p^2 -M_0^2 - \Sigma_B^R(p)} \, . \label{eq:DRex}
\end{align} 
\end{subequations}
Eliashberg theory considers the one-loop expressions \cref{eq:defSgen,eq:defKK} as functionals $\Sigma^R_{F,B}[G^R,D^R]$ and the Dyson equations turn into a coupled, self-consistent problem.
Formally, the one-loop approximation can be seen as the leading term of perturbative expansion in the effective coupling constant with units $(\text{energy})^2$
\begin{align}\label{eq:defAlpha}
\alpha = \frac{m g^2}{\pi}\, ,
\end{align}
which is assumed to be small in the sense $\alpha/\eF^2 \ll 1$. 
However, naive power counting generically breaks down due to strong fluctuations in the vicinity of the QCP and the problem cannot be treated in a perturbative manner. In the absence of another expansion parameter, one cannot establish a strict hierarchy of diagrams in this situation and higher order diagrams like vertex corrections typically become important, too. 
In the ground state this not the case, because the correlation functions obey a Ward identity which allows to neglect vertex corrections provided that the typcial scaling relations of the INM are not violated~\cite{metl10}. However, depending on how the limit of small external energies and momenta is approached, the vertex function may vary between a perturbatively small value or even a divergence~\cite{chubukov2005Ward,rech2006}. Since the self-energy at finite temperatures is not guaranteed to be exclusively described within the $T=0$ scaling limit of the INM~\cite{klein2020}, the stability of ET against vertex corrections requires further justification. Furthermore, it has been shown that the thermally occupied static fluctuations of the order parameter give rise to large vertex corrections.
In the following, we will evaluate the Eliashberg equations first without vertex corrections and return to them in Sec.~\ref{sec:vertexCorr}, where we show that they in total affect only the $\mathcal{O}(1)$ prefactors but do not affect the scaling with $T$ and $\alpha$ at the lowest temperatures. 

\subsection{Approach and validity}\label{sec:ansatz}
Before solving the Eliashberg equations~\cref{eq:defSgen,eq:defKK,eq:Dyson}, we recapitulate the $T=0$ results and state the expected changes at finite temperatures. Furthermore, we estimate the region in frequency and momentum space where ET applies. 

The dressed bosonic Green's function~\eqref{eq:DRex} at small energies will acquire the form
\begin{align}
D^R(p) = \frac{1}{\omega_p^2 -c_B^2 p^2 -M^2(T) +  i \Gamma_B(p)} \, ,  \label{eq:ansDR}
\end{align} 
which is amenable to an analytic treatment within ET. Here, we do not consider larger momenta where nesting effects across the FS imprint a nontrivial momentum structure on the nematic susceptibility~\cite{altshuler1995,holder2012,punk2015}.
We introduce for the imaginary part $\myim \Sigma_R^B(p \to 0)= - \Gamma_B(p)$, which encodes the decay of the modes. $\Gamma_B(p)$ must be odd in frequency since it inherits its symmetry properties from the retarded expectation value $i D^R(t)=\theta(t)\langle[\hat\Phi (t), \hat \Phi(0)]\rangle$ such that $\omega_p\, \Gamma_B(\mathbf p, \omega_p) > 0$. In the ground state the bosonic modes relevant for the generation of the non-FL correlations acquire a Landau damping form 
\begin{align}\label{eq:defLandau}
\Gamma_B(\mathbf p, \omega_p)|_{T=0} = \alpha \dpq\frac{\omega_p}{v_F p}\, ,
\end{align}
which corresponds to strongly overdamped excitations. Furthermore, if Landau damping is present, the dynamical critical exponent attains the value $z=3$. As we will discuss in Sec.~\ref{sec:bosDamp}, Landau damping is cut off either by thermal fluctuations or by sufficiently large frequencies. The first case leads to $\Gamma_B(\mathbf p, \omega_p) = \alpha \dpq\omega_p/\Gamma_F(T)$, which implies the reduced dynamical critical exponent $z=2$ as pointed out by Punk~\cite{punk2016}. In addition, Landau damping does not apply in the vicinity of the angles $\varphi_{\mathbf p} \in \{1,3,5,7\}\cdot\pi/4$ due to the nematic form factor. In the computations these directions will be suppressed by additional powers of $d_{\mathbf{p}}$ anyway and we will not discuss them further.
The real part of $[D^R]^{-1}$ encodes the dispersion of the modes which resembles the standard relativstic form $\omega_p=\pm \sqrt{c_B^2 p^2 + M^2(T)} $. We parametrize the renormalized, physical velocity  as $c_B =A v_F$ and assume $A = \mathcal O(1)$ for simplicity. The main focus of this work is the physical energy gap $M^2(T)$.
In order to describe the quantum critical regime, we have to impose a renormalization scheme that guarantees to recover the quantum critical point which means $M(T) \to 0$ in the limit $T \to 0$. In mathematical terms, we absorb the bare parameter $M_0^2$ from Eq.~\eqref{eq:DRex} together with the ground-state self-energy in the condition
\begin{align}\label{eq:renorm}
M^2_0+\myre \Sigma^R_B(0)|_{T=0} =0 \, .
\end{align}
At $T>0$ we thus obtain the bosonic mass as
\begin{align}\label{eq:m2def}
\begin{split}
M^2(T) = -(D^R)^{-1}(0)& = M^2_0 + \myre \Sigma^R_B(0)|_T   \\ 
& = \myre \Sigma^R_B(0)|_{T}-\myre \Sigma^R_B(0)|_{T=0}\, . 
\end{split}
\end{align}
Note that $M^2(T)$ must be positive since negative values correspond to the symmetry-broken state. 
The requirement of scale invariance for the boson propagator $D$ would imply $M(T) \sim T^{1/z}$. However, the effective gauge symmetries of the two-patch model~\cite{metl10,mross2010,dalidovich2013} prohibit an anomalous dimension of the bosonic propagator and enforce $M(T) \equiv 0$. A nonzero $M(T)$ can then only be introduced by considering non-universal effects~\cite{hartnoll2014} like the band structure, which goes beyond the scope of the infrared theory. In order to compute $M(T)$, we,  therefore, need to know the range of validity of ET, which we will estimate at the end of this section . 

Physically, $M(T)$ can be observed via the nematic susceptibility $D^R(0) = -1/(2 M^2(T))$. Moreover, $M^{2}(T)$ is quite closely related to the compressibility~\footnote{We omit the additional factors of density as compared to the standard definition of the compressibility $\kappa_T = n^{-2} (\partial n/\partial\mu)_T$.} $(\partial n/\partial \mu)_T$ of the Fermions: In any ergodic system the static limit of the density-density response function (i.e., the Lindhard function) $\chi_{nn} (\omega, \mathbf q)$ must coincide with the thermodynamic susceptibility for densitiy fluctuations $\chi_{nn}^T = (\partial n/\partial \mu)_T$ which means
$\lim_{\omega \to 0}\chi_{nn} (\omega, \mathbf{q}=\mathbf 0) = (\partial n/\partial \mu)_T $. Because ET neglects the vertex corrections, the Lindhard function is diagrammatically represented by the particle-hole bubble for the bosonic self-energy in Fig.~\ref{fig:SE}, except for the replacement of the bare coupling $g\, d_\mathbf{p} \to 1$. 
Since the nematic form factor only affects the numerical prefactor of the diagrams, both $M^2(T)$ and $(\partial n/\partial \mu)_T-(\partial n/\partial \mu)_{T=0}$ scale identically~\footnote{The presence of additional coupling constants in $M^2(T)$ could give rise to a different $T$ dependence when vertex corrections are taken into account. However, since the latter do not change the leading terms only the subleading terms differ(cf. Sec.~\ref{sec:vertexCorr}).} with $T$. We will verify this connection explicitly in Sec.~\ref{sec:bosMass}.

Let us turn to the electrons now.
In the ground state, the Eliashberg equations~\cref{eq:defSgen,eq:defKK,eq:Dyson} are usually solved in imaginary frequencies $\Omega$. The electronic self-energy acquires a noninteger power-law typical for the non-FL  behavior in the vicinity of the FS $$\Sigma_F(\mathbf{k},\Omega)=\Sigma_F(\Omega,\varphi_{\mathbf k}) = -  2 i \win^{1/3} |d_\mathbf{k}|^{4/3}\sgn(\Omega)|\Omega|^{2/3},$$ for angles not too close to directions where $d_{\mathbf k}$ vanishes. 
The scale
\begin{align}\label{eq:defwin}
\win = \frac{1}{8^3 \cdot 3^{3/2} A^4} \frac{\alpha^2}{\eF^3}\, ,
\end{align} 
is set by the interactions.  The nonanalytic scaling $\Sigma_F^R \sim \alpha^{2/3}$ indicates the break-down of naive perturbation theory. Furthermore, it is indepedent of the magnitude $|\mathbf k|$ of the momentum.
By analytic continuation to real frequencies $i \Omega \to \omega_k+ i 0^+$, we find the corresponding retarded self-energy
\begin{align}\label{eq:SFgs}
\Sigma^R_F(k)|_{T=0}= -|d_{\mathbf{k}}|^{4/3} (\sqrt{3} \sgn(\omega_k) + i ) \win^{1/3} |\omega_k|^{2/3}\, , 
\end{align}
consistent with the Kramers-Kronig relation~\eqref{eq:defKK}, see also App.~\ref{sec:KK}. 
Regarding the momentum dependence, Ref.~\cite{dell2006} shows that deviations from the frequency-dependence $|\omega_k|^{2/3}$ by the finite momentum difference $\delta k = |\mathbf k -\mathbf k_F|$ appear only below a very small scale suppressed by $\delta k^3$. Moreover, corrections to the bare dispersion $\ek$ appear in form of a small anomalous dimension only when three loop diagrams are included~\cite{metl10}, while corrections to the exponent $2/3$ at $T=0$ appear only at four loops~\cite{holder2015rapid}. As a consequence, using Eq.~\eqref{eq:SFgs}, which only depends on frequency and the angular variable, as a starting point for the ET-calculations at finite temperatures seems well justified. 
More precisely, we expect the following asymptotics of the self-energy for a low-energy Fermion
\begin{align}\label{eq:SFans}
\Sigma^R_F(k)\simeq \Sigma^R_F(\omega_k,\varphi_{\mathbf k}) \simeq 
\begin{cases}
\Sigma^R_F(k)|_{T=0}\, ,  & \text{if } |\omega_k| \gtrsim \omega_> \\
- i \Gamma_{F,\mathbf k} (T) \, , &\text{if } |\omega_k| \lesssim \omega_<
\end{cases} .
\end{align}
The upper line is based on standard quantum critical scaling arguments~\cite{sachdevBook,loehneisen07}: Within the QCR the scaling forms of the underlying QCP are recovered at energies above the crossover scale $\omega_>$. On the other hand, below $\omega_< $ the Fermions acquire a thermal decay rate $\Gamma_{F,\mathbf{k}}(T) = \Gamma_F(T) |d_{\mathbf{k}}|^x$ that inherits an angular dependence from the interaction term with an exponent $x>0$. A nonzero damping rate is expected on quite general grounds: Finite temperatures broaden the thermal distribution functions $n_{B,F}$ in Eq.~\eqref{eq:defSF} as compared to their $T=0$ forms $n_{B,F}(\omega)|_{T=0} = \mp \theta(-\omega)$. As a consequence, the low-energy single-particle states are no longer protected against energy dissipation, irrespective whether a quasiparticle picture applies or not. However, one expects that $\Gamma_F(T)$ scales more slowly to zero than $T$ if quasiparticles are absent. On the other hand, one finds in a FL $\Gamma_F(T) \sim T^2$ such that the thermal decay rate is less important than the the thermal broadening of the distribution. Regarding the real part we do not introduce a modification of $\myre \Sigma^R_F$ for frequencies below $\omega_<$ since the latter is forced to vanish by symmetry for small frequencies and momenta in the vicinity of the FS. Moreover, we show in App.~\ref{sec:KK} that solving the Eliashberg equations with the approximation $\myre \Sigma^R_F(k) = \sqrt{3} \win^{1/3} |d_{\mathbf k}|^{4/3} \sgn (\omega_k) |\omega_k|^{2/3}$ at all relevant frequency scales is consistent with the Kramers-Kronig relations.

Let us discuss the implications of $\omega/T$ scaling, introduced in Sec.~\ref{sec:intro}, for the low-temperature regime in further detail. If it applies, the self-energy can be written in terms of a \emph{positive} scaling function $\tilde{\Sigma}_F$ 
\begin{align}
\myim \Sigma^R_F (k) \sim  -T^{2/3} \tilde{\Sigma}^R_F(\omega/T ,\varphi_{\mathbf k}) \, ,
\end{align}    
whose limiting behavior for $T \to 0$ obeys $\tilde{\Sigma}^R_F \sim (\omega/T)^{2/3} $, while the total prefactor becomes $-|\dkF|^{4/3} \win^{1/3}$ to match the ground state result.   Correspondingly, the damping rate must scale like $\Gamma_F(T) \sim T^{2/3}$ since $\tilde\Sigma^R_F(0,\varphi_k)$ is independent of temperature. In the analysis of quantum critical systems the scaling functions are typically universal in the sense that they contain only dimensionless ratios of the relevant scaling variables and universal numbers that can be fully determined within an effective low-energy theory.
As we will see below, however, here the proportionality factor of dimension $(\text{Energy})^{1/3}$ consists in the weak-coupling limit of a combination of $\alpha$ and $\eF$ which is fixed by UV-physics. At larger couplings an admixture of parameters from the band structure is possible.

Quite generically, one expects that the non-FL correlations exist only in a certain low-energy regime close to the FS. 
This allows to restrict $\delta k = k-k_F \leq \Lambda $ by a momentum cut-off $\Lambda$ which we parametrize as
\begin{align}\label{eq:defcut}
v_F \Lambda := h_\Lambda \eF \ll \eF\, ,
\end{align}
or equivalently $h_\Lambda \ll 1$. One constraint on this number follows from the range in momentum space within which the dispersion can be linearized around the Fermi surface:
\begin{align}\label{eq:deflin}
\varepsilon_{\mathbf k_F+ \mathbf{p}} - \mu \simeq v_F p \cos(\varphi_{\mathbf{k}_F} -\varphi_{\mathbf p})  \, ,
\end{align}
where $\mathbf p$ denotes a typical momentum transfer while the angles refer to the orientation of $\mathbf{k}_F$ and $\mathbf{p}$, respectively. Thus, the curvature of the Bloch band gives rise to one bound on $h_\Lambda$.
Furthermore, we obtain one constraint that arises intrinsically from the scaling structure of the INM. 
In the ground state, the non-FL correlations are generated by the exchange of energy and momentum between the Fermions via low-energy Bosons which can be far off-shell due to Landau damping~\eqref{eq:defLandau}. At finite $T$ we expect from quantum critical scaling that this processes still dominate at energies beyond $\omega_>$. The physical picture translates into a criterion for the applicability of ET: Nematic fluctuations transfer typical momenta $p$ and energies $\omega_p$ between Fermions with $|\mathbf k| \simeq k_F$ and $\omega_k \to 0$, that obey~\cite{altshuler1994, metzner2003, chubukov2005Ward, rech2006, dell2006}
\begin{align}\label{eq:ETcond1}
v_F p \gg \left| \omega_p -\Sigma^R_F(\omega_p, \varphi_{\mathbf k_F})|_{T=0}\right|\, .
\end{align}
Then, the electronic $|\omega_k|^{2/3}$ correlations arise from the momentum transfers that scales like
\begin{align}\label{eq:ETcond2}
v_F p \sim (\alpha \omega_p)^{1/3}\, .
\end{align} 
The connection to the computations is detailed in Sec.~\ref{sec:SFQ}, in particular below Eq.~\eqref{eq:defMbar}. 
We combine the last two relations by equating the left-hand-sides to find an upper frequency bound for the validity of Eliashberg theory and consequently for the non-FL regime, too. First of all, for frequencies on the scale $\omega_p \sim \win$ this implies $v_F p \sim \alpha/\eF \gg \win$ because $\win \sim \alpha^2/\eF^3$ and $\alpha/\eF^2 \ll 1$. This sets the minimal possible value $h_{\min} = \alpha /\eF^2 \ll 1$ since for smaller $h_\Lambda$ the typical scale $\win$ for the non-FL behavior is discarded. On the other hand, we find that the maximally allowed frequency scales like $\omega_{\max} \sim \alpha^{1/2} \gg \win$, where now the linear term on the left-hand side of Eq.~\eqref{eq:ETcond1} dominates. A similar estimate has also been given in Ref.~\cite{rech2006}. To actually satisfy the inequality~\eqref{eq:ETcond1}, we define $\wmax = h_\omega \alpha^{1/2}$ with $h_\omega \ll 1$. Translated to momentum this corresponds to the maximal cut-off $h_{\max} = h^{1/3}_\omega\alpha^{1/2}/ \eF \ll 1$. 
Note that these bounds result exclusively from the structure of the interactions and apply in the weak-coupling limit $\alpha$ of interest here. Therefore, we use in the following  
\begin{align}\label{eq:deffullcut}
\begin{split}
h_{\max} = h^{1/3}_\omega\frac{\alpha^{1/2}}{\eF} \qquad \qquad  \wmax =h_{\omega} \alpha^{1/2} \, .
\end{split}
\end{align}
However, for bigger values of $\alpha$ one has to check whether the last definition violates the linearization criterion~\eqref{eq:deflin}. In this case one has to take the constraints from the nonlinearity of the band dispersion into account by decreasing the cut-offs appropriately. We emphasize that any value $h_{\min} \leq h_\Lambda \leq h_{\max}$ gives rise to the same temperature scaling for all quantities. However, the range where $\omega/T$ scaling emerges, as well as the $T$-independent prefactors of $\Gamma_F(T)$ and $M^2(T)$ are strongly influenced. This is exemplified on the basis of the minimal cut-off scheme using $h_{\min}$ and $\win$ in App.~\ref{sec:minCut}.  

Finally, we expect the non-FL metal correlations to decay quickly as function of $\omega_k$ or $k$ when at least one of these variables violates the given bounds for the non-FL regime. Then the $\Sigma^R_F$ approaches the standard FL form due to residual Fermi-Fermi interactions not considered in the model.

\section{Key results}\label{sec:keyRes}
The major goal of the present work is to obtain the temperature-dependence of $M^2(T)$ at the onset of finite temperatures in a fully self-consistent approach, thereby finding also $\Gamma_{F, \mathbf k_F} (T)$ and the crossover scales $\omega_{\gtrless}$.
In addition, one answers also the question whether an $\omega/T$ scaling form $\tilde{\Sigma}^R_F$ exists. 
We point out that the behavior of $\Sigma^R_F(\omega)$ in the regime $\omega_< \leq |\omega| \leq \omega_>$ is in general not guaranteed to follow a simple functional dependence that can be extracted in an analytic way. However, within the QCR it turns out possible to determine the dominant behavior of the solution of the Eliashberg equations by analytic means. 
In the following, we summarize our main results for the Eliashberg equations and give their solution. The details of the underlying calculations can be found in the next section. 

The fermionic self-energy can be decomposed~\cite{dell2006,punk2016,klein2020} as a sum of a quantum $\Sigma^R_{F,Q}$ and a thermal component $\Sigma^R_{F,T}$, discussed in Secs.~\ref{sec:SFQ} and~\ref{sec:SFT}, respectively.  
The first function contains the non-FL correlations in the ground state and obeys $\omega/T$ scaling at finite temperatures, as has been pointed out already in Ref.~\cite{dell2006}, provided that $M(T)/(\alpha T)\ll 1$ to avoid the FL regime. In particular, we find the asymptotic behavior for large frequencies (cf. Eqs.~\eqref{eq:resSFQ1} and~\eqref{eq:resSFQdelta}):
\begin{widetext}
\begin{align}
\begin{split}
\myim \Sigma^R_{F,Q}(\mathbf k_F, |\omega_k| \gg T)& = \myim \Sigma^R_{F}|_{T=0} (\mathbf k_F, |\omega_k| \gg T) + \myim \delta\Sigma^R_{F}| (\mathbf k_F, |\omega_k| \gg T) \\
& \quad - \win^{1/3} |\dkF|^{4/3} |\omega_k|^{2/3} + 0.318919...\dfrac{|\dkF|^{4/3} \alpha^{2/3}}{A^{4/3}\eF} T^{2/3} \, .
\end{split}
\end{align}
\end{widetext}
The frequency-independent correction $ \myim \delta\Sigma^R_{F,Q} \sim T^{2/3}$ has not been discussed so far. 
Its scaling is is intrinsically related to the non-FL quantum fluctuations. Moreover,
it turns out to be responsible for the temperature dependence of the bosonic mass gap $M^2(T)$. 

On the other hand, the thermal part $\Sigma^R_{F,T}$ dominates the behavior of the total self-energy for small frequencies $|\omega_k| \ll T$. Its most general form  reads as
\begin{align}
\begin{split}\label{eq:keyResSFTfin}
&\myim \Sigma^R_{F,T}(\mathbf k_F,\omega_k) = \\ 
&  \myim \dfrac{\frac{ \alpha \dkFq T}{4 \eF A^2} \operatorname{arcosh}\left(-i \dfrac{(\omega_k-\Sigma^R_F(\omega_k,\varphi_{\mathbf k_F}))}{M/A}\right)}{(\omega_k-\Sigma^R_F(\omega_k,\varphi_{\mathbf k_F})) \sqrt{1 + \dfrac{M^2/A^2}{(\omega_k-\Sigma^R_F(\omega_k,\varphi_{\mathbf k_F}))^2}}}\, ,
\end{split}
\end{align}
which is derived in Eq.~\eqref{eq:SFTfin}. 
This poses a self-consistent problem since the total self-energy $\Sigma^R_{F,Q} + \Sigma^R_{F,T}$ appears on the right-hand side.
This form is closely related to the results by Klein et. al~\cite{klein2020} obtained in Matsubara frequencies. The expansion~\eqref{eq:arcsecExp} admits to find simple analytic expressions in different regimes. By taking the limit $\omega_k \to 0$, we can extract the thermal damping rate.
If $M \ll \Gamma_F(T)$, which includes the vicinity of the thermal phase transition but also the QCR at sufficiently large temperatures, it becomes (up to log(log...) terms)
\begin{align}\label{eq:keyresGammaFTc}
\Gamma_{F, \mathbf{k}_F} = \frac{|\dkF|}{2A}\sqrt{\frac{ \alpha T}{\eF} \log \left(\frac{ \sqrt{ \alpha T} |\dkF|}{\sqrt{\eF} M/A^2} \right)} \, ,
\end{align}
which agrees with the analytic continuation of the results in Ref.~\cite{klein2020,damia2020}.
On the other hand, if $M(T) \gg \Gamma_F(T)$, which applies to the QCR at the smallest temperatures, we find
\begin{align}\label{eq:keyresGammaFQCR}
\Gamma_{F, \mathbf{k}_F}(T) = \frac{\pi}{8} \frac{\alpha \dkFq}{\eF A} \frac{T}{M(T)} \, ,
\end{align} 
in agreement with the results by Dell'Anna and Metzner~\cite{dell2006}, as well as by Punk~\cite{punk2016}.
Taking the latter two equations together, as well as the condition  $M(T \to 0) \to 0$ algebraically in a critical system, we conclude that $\Gamma_F(T)$ vanishes at least like $\sqrt{T}$ (possibly enhanced by logarithmic terms) but not faster than $T$. This reveals the non-quasiparticle behavior since the thermal decay rate approaches zero more slowly than the temperature. However, to satisfy the necessary condition $\Gamma_F(T)\sim T^{2/3}$ for $\omega/T$ scaling, one needs $M(T) \sim T^{1/3}$ according to Eq.~\eqref{eq:keyresGammaFQCR}, which is the marginal case for the existence of Fermi liquid correlations, as discussed in Sec.~\ref{sec:SFQ}. 

As becomes obvious from Eqs.~\cref{eq:keyResSFTfin,eq:keyresGammaFTc,eq:keyresGammaFQCR}, the bosonic mass determines the overall behavior of $\Sigma^R_{F,T}$ and, therefore, $M^2(T)$ is an important ingredient for a complete solution of Eliashberg theory. In Sec.~\ref{sec:bosMass} we show that at the smallest temperatures in the QCR, the main contribution to $M^2(T)$ arises from $\Sigma^R_{F,Q}$.
In particular, the bosonic mass inherits the temperature dependence $T^{2/3}$ of the frequency-independent $\myim \delta \Sigma^R_{F,Q}$ but aqcuires a cut-off dependent, nonuniversal prefactor since the underlying integration is sensitive to the full spectral width of the non-FL excitations in the ground state. In agreement with the expectations~\cite{hartnoll2014} from the gauge-symmetry constraints of the low-energy two-patch model~\cite{metl10,mross2010,dalidovich2013}, $M^2(T)$ vanishes when one sends the cut-off $\Lambda$ to infinity, for details see Eq.~\eqref{eq:MQdom} below. For finite cut-off values the result is thus formally equivalent to Eq.~\eqref{eq:MTQCR} and reads explicitly for the maximal cut-off scheme from Eq.~\eqref{eq:deffullcut}:
\begin{align}\label{eq:keyResM2-}
M^2(T) \simeq 0.08241... h_\omega \frac{\alpha^{7/6} T^{2/3}}{\eF} \, .
\end{align}
Here, the parameter $h_\omega$ incorporates the influence of the UV-physics.
The corresponding calculation for the minimal cut-off is presented in App.~\ref{sec:minCut}, where the same $T^{2/3}$ scaling is obtained.
Notice that this term is more important at small temperatures than the estimate from Eq.~\eqref{eq:MTqp} $M^2(T) \sim T \log T$ obtained from the RG approach of Millis~\cite{millis1993} and the equivalent diagrammatic calculation of Hartnoll et al.~\cite{hartnoll2014}. In the first reference $M^2(T)$ is obtained within a purely bosonic theory that takes Landau damping  and a standard $\Phi^4$ interaction with coupling constant $u$ into account. In the second reference, $M^2(T)$ is determined from a self-consistent equation derived from diagrams up to three loop order. These arise from adding further bosonic lines to the polarization bubble in Fig.~\ref{fig:SE}. However, the Fermions are still treated as noninteracting with infinite lifetimes. The anomalous scaling $T \log T$ is then caused by the strongly damped, almost critical bosonic fluctuations that interact with particle-hole excitations of the Fermi gas. Our result is instead governed by the thermal corrections to the anomalous non-FL correlations, which incorporate the non-quasiparticle character of the low-energy degrees of freedom. As a consequence, taking them properly into account, e.g. by dressing $G^R$ self-consistently, is necessary to determine the leading scaling behavior. However, once $T$ becomes comparable to the extent of the non-FL regime in frequency space ($\wmax$ for the maximal cut-off),
 the temperature dependence of $M^2(T)$ is dominated by the asymptotic tails of $\myim \Sigma^R_{F,T}$, which are insensitive to the non-FL correlations. Furthermore,
the width of the thermal distribution starts to exceed the broad spectral features characteristic of the non-quasiparticles, such that a qp-picture is restored. 
As a result, ET reproduces $M^2(T) \sim T$ times logarithmic terms for temperatures $T \gtrsim T_{\max} = \wmax$, when the cut-off scheme~\eqref{eq:deffullcut} is used (see the discussion below Eq.~\eqref{eq:Mm3T}). If the minimal cut-off $\win$ is employed instead, $T_{\max}$ is substituted by $\win$, but the connection between the spectral width of the non-FL regime and the temperature remains unchanged, as is shown in App.~\ref{sec:minCut}. In fact, we show that the connection between the crossover temperature above which $M^2(T) \sim T $ with logarithmic corrections and the extent of the non-FL correlations in the ground state is generic.

\begin{figure}[ht]
\begin{center}
\includegraphics[width=\columnwidth]{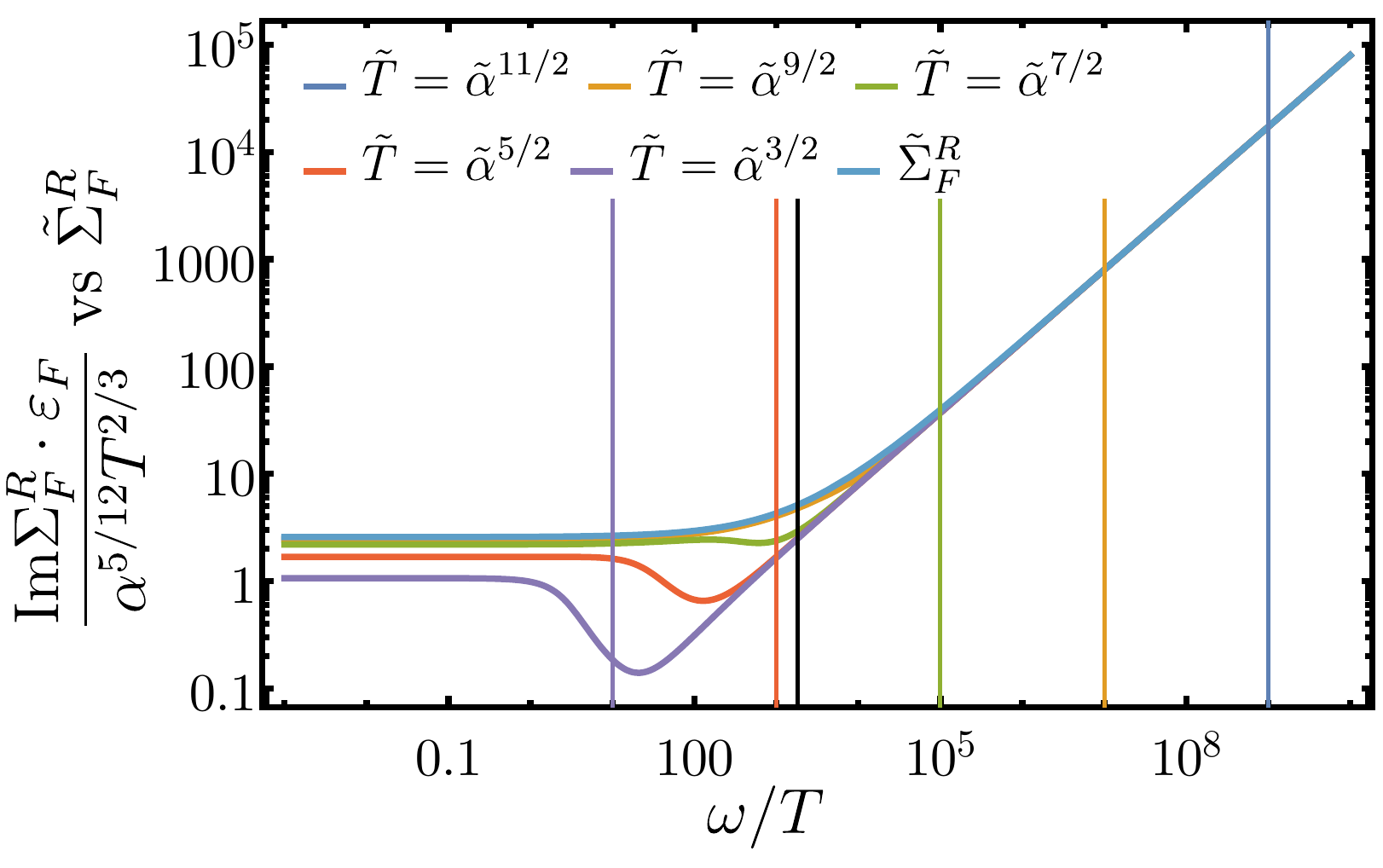}
\caption{Comparison of the total, self-consistent self-energy $ \Sigma^R_{F,Q}\cdot \eF/(\alpha^{5/12} T^{2/3}) $ with the scaling form $\tilde{\Sigma}^R_F$ given in Eq.~\eqref{eq:keyResGammaFscal} at different temperatures $\tilde T = T /\varepsilon_F$ for the parameters
 $\tilde \alpha = \alpha/\eF^2 = 10^{-2}$, $h_\omega=10^{-1},A=1$. 
 For low enough temperatures the scaling form $\tilde{\Sigma}^R_F$ agrees perfectly with the total self-energy, i.e the curve for $\tilde{T} = \tilde{\alpha}^{11/2}$ complete overlaps with $\tilde{\Sigma}^R_F$. At larger temperatures one observes deviations. In particular, Eq.~\eqref{eq:keyResTscal} predicts $T_{\text{scal}} \simeq 0.002\, \win$ from the second relations with the very large value $B=100$ to obtain an error of at most one percent by deviations of the scaling $\Gamma_F(T)\sim T^{2/3}$. Indeed, the self-energy at the second lowest temperature shows this, yet hardly visible. Upon increasing the temperature further, this effect becomes more pronounced while the nonmonotonic behavior emerges, too. The colored vertical lines indicate the position of $\wmax/T$ whereas the black line refers to the crossover of the scaling function at $\omega_{\lessgtr}$ according to Eq.~\eqref{eq:keyRescross} (with fully restored numerical prefactor). For temperatures below $T_{\text{scal}}$ we have $\omega_{\lessgtr}/T \ll \wmax/T$ such that non-FL correlations of the ground state indeed dominate in the regime $\omega_{\lessgtr} \lesssim |\omega| \lesssim \wmax$. On the other hand, this is not the case for the highest presented temperature.} 
\label{fig:scalCol}
\end{center}
\end{figure}

Next, we summarize (within the maximal cut-off scheme) the scaling relations, which are detailed in Sec.~\ref{sec:domScal}. 
According to the discussion below Eq.~\eqref{eq:m2def}, the fermionic compressibility scales like $(\partial n/\partial \mu)_T - (\partial n/\partial \mu)_{T=0} \sim M^2(T)\sim T^{2/3}$. Furthermore, the scaling $M(T) \sim h^{1/2}_\omega \alpha^{7/12} T^{1/3} \eF^{-1/2}$ implies $M(T) / (\alpha T)^{1/3} \ll 1$ such that Fermi liquid correlations are irrelevant. For the thermal damping rate, we find from Eq.~\eqref{eq:keyresGammaFQCR}
\begin{align}\label{eq:keyResGammaFscal}
\Gamma_F(T) \sim \frac{\alpha^{5/12}}{h^{1/2}_\omega\eF^{1/2}} T^{2/3}\, ,
\end{align}
which indeed satisfies $M(T) \gg \Gamma_F(T)$ for $T \to 0$ and is compatible with $\omega/T$ scaling. Moreover, the dominant temperature dependence of the total fermionic self-energy admits a representation in terms of a simple but nonuniversal scaling form (see Eq.~\eqref{eq:resSFtot2})
\begin{align}\label{eq:keyResSFtot2}
\begin{split}
&\myim \Sigma^R_{F} (\mathbf{k}_F, \omega_k,T,h_\omega)   =  -\frac{\alpha^{5/12 }}{\eF^{1/2}}T^{2/3} \tilde{\Sigma}^R_F\left(\frac{\omega_k}{T},\varphi_k,h_\omega\right) \\
& \tilde{\Sigma}^R_F\left(\frac{\omega_k}{T},\varphi_k,h_\omega\right)= \frac{b_1}{ h_\omega^{1/2}}\dkFq \! +\! \frac{ \alpha^{1/4}|\dkF|^{4/3}}{8 \cdot 3^{1/2} A^{4/3} \eF^{1/2}}\!\left[\frac{|\omega_k|}{T} \right]^{2/3}\!\! .
\end{split}
\end{align}
where $b_1 = 1.367...$. The first part, which is sensitive to the UV physics, originates from $\myim \Sigma^R_{F,T}$, whereas the second is associated with $\Sigma^R_{F,Q}$.
By comparing them we obtain the single crossover scale  
\begin{align}\label{eq:keyRescross}
\omega_\lessgtr \sim  h_{\omega}^{-3/4}\frac{\eF^{3/4} }{\alpha^{3/8}} T \, ,
\end{align}
which impies that both $\omega_<$ and $\omega_>$, defined in Eq.~\eqref{eq:SFans}, coincide with the latter scale.
The scaling function is compared to the full numerical evaluation in Fig.~\ref{fig:scalCol}, which indeed reveals a scaling collapse at sufficiently low temperatures. 
With increasing temperature, however, $\tilde \Sigma^R_F$ ceases to describe the total fermionic self-energy.
Within ET, we obtain thresholds for the range-of-validity of the given scaling function. In particular, it can be applied for temperatures below
\begin{align}\label{eq:keyResTscal}
T_{\text{scal}} = 
\begin{cases}
h_\omega^{15/8} \left(\dfrac{\alpha^{1/2}}{\eF}\right)^{23/8} \eF \, ,&\text{if } \dfrac{\alpha}{\eF^2} \ll \dfrac{h_\omega^{6/5}}{B^{16/5}} \\
B^{-3} h_\omega^3 \alpha^{1/2}\, ,&\text{if } \dfrac{\alpha}{\eF^2} \gg \dfrac{h_\omega^{6/5}}{B^{16/5}} 
\end{cases} \, ,
\end{align}
which is derived in Eq.~\eqref{eq:defTscal}: Due to the nonuniversal character of the scaling function we must distinguish several effects that destroy $\omega/T$ scaling in different parameter regimes: 
The first line applies to the extreme weak-coupling limit. In this case, the total self-energy becomes a nonmonotonic function of $\omega_k$ because an intermediate regime emerges, which is governed by the tails of $\myim \Sigma^{R}_{F,T}$.  For larger values of $\alpha$,  the thermal scattering rate is instead no longer given by Eq.~\eqref{eq:keyresGammaFQCR} but approaches the form of Eq.~\eqref{eq:keyresGammaFTc}. This results in a scaling mismatch of the low- and high-frequency regimes. In particular, the second form of $T_{\text{scal}}$ originates from the minimal ratio $M(T)/\Gamma_F(T) \gtrsim B \simeq 100$, needed for the evaluation of $\Gamma_F(T)$ via Eq.~\eqref{eq:keyresGammaFQCR}.

Finally, we argue in Sec.~\ref{sec:vertexCorr} that the vertex correction obtained from the self-consistent propagators change the numerical $O(1)$ prefactors but do not affect the found scaling with $T$ and $\alpha$.

\section{Computational details of Eliashberg theory at finite temperature}\label{sec:ET}
We present now the calculations necessary to derive the results of the previous section. We begin with the bosonic damping rate $\Gamma_B$ in Sec.~\ref{sec:bosDamp}. This allows to compute the fermionic self-energy in Sec.~\ref{sec:SF}. The calculation of the inverse order parameter susceptibility is presented in Sec.~\ref{sec:bosMass}, where we elucidate the connection with the fermionic compressibility, too. Furthermore, we will find that $M^2(T)$ can be related to the finite temperature corrections of the non-FL terms contained in $\Sigma^R_{F,Q}$ at the lowest temperatures while at temperatures above $T_{\max}$ the tails of  $\Sigma^R_{F,T}$ generate the $M^2(T) \sim T \log T$ scaling.
Eventually, we solve the Eliashberg equations and establish boundaries on the validity of $\omega/T$ scaling.
   
\subsection{The bosonic damping function}\label{sec:bosDamp}
\begin{figure}[ht]
\centering
\includegraphics[width=\columnwidth]{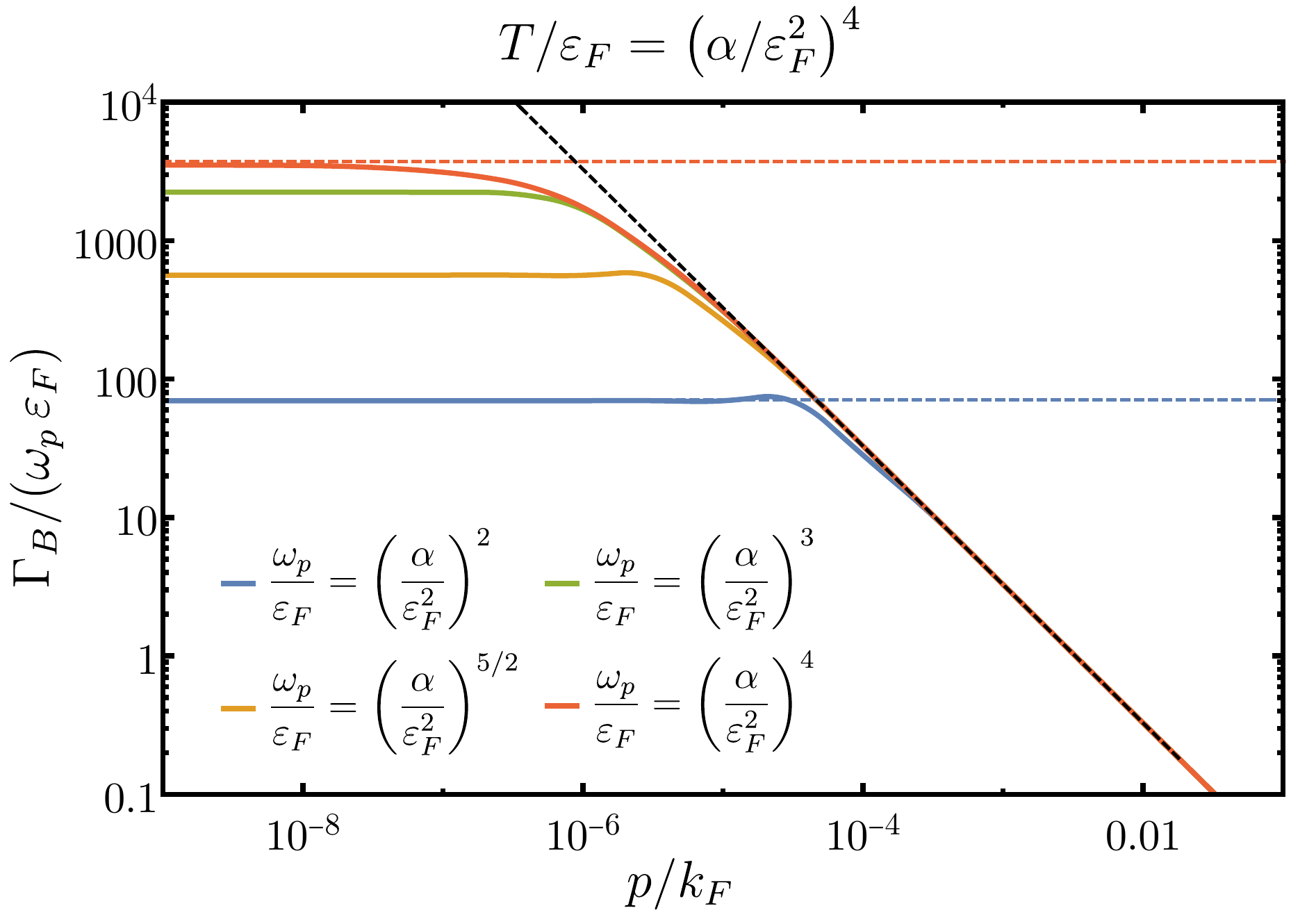}
\includegraphics[width=\columnwidth]{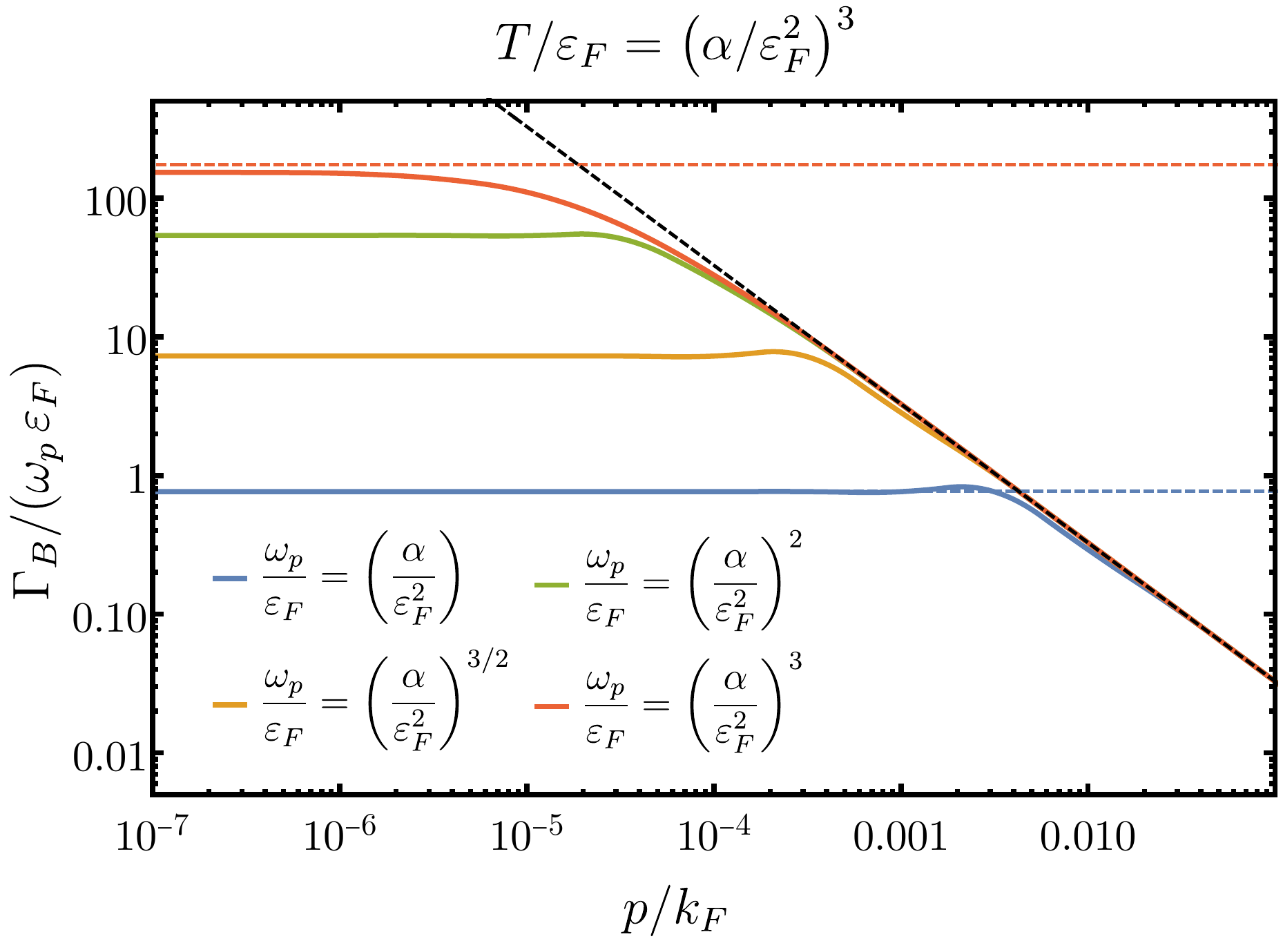}
\caption{Numerical evaluation of the bosonic damping from Eq.~\eqref{eq:resImSBgen} for two different temperatures with the self-consistent self-energy from Eq.~\eqref{eq:keyResSFtot2} and the parameters $\tilde\alpha = \alpha/\eF^2 = 1/100$, $A=1$ and $\varphi_{\mathbf k} = \pi/10 $. In fact, all frequencies $\omega_p/\eF$ smaller than the minimal one shown in the figure cannot be distinguished from the red curve. The dashed lines indicate the asymptotic expressions summarized in Eq.~\eqref{eq:resGammaBsum}. Black dashed: Landau damping. Blue dashed: Numerical evaluation of~\eqref{eq:numSBimT0}. Red dashed: asymptotic result Eq.~\eqref{eq:resSBimPunk}. The intermediate frequencies (yellow and green lines) cannot be described by the asymptotic expressions because both the low- and the high-frequency regime of $\myim\Sigma^R_F$ contribute. This is also the source of the remaining discrepancy between the solid and dashed red lines results.}
\label{fig:SBim}
\end{figure}

In this section we compute the damping of the bosonic modes $\Gamma_B(p) = -\myim \Sigma^R_B(p)$, defined below Eq.~\eqref{eq:ansDR}, at finite temperature. We keep the fermionic decay rate $\Gamma_F(T)$ and $\win$ as parameters in order to close the self-consistent loop.
In addition, we will see how and for which momenta the important Landau damping form~\eqref{eq:defLandau}, found in the ground state, is recovered. The results are presented for two different temperatures in Fig.~\ref{fig:SBim}.
The retarded bosonic self-energy is given in Eq.~\eqref{eq:defSB}. 
Obviously, it vanishes when $\omega_p \to 0$ in agreement with the general constraint that $\myim D^R(p)$ is odd in frequency (see the discussion below~\eqref{eq:ansDR}).
We insert the explicit forms of $G^R$ from Eq.~\eqref{eq:GRex} and introduce the new momentum variable $k' = k-k_F$. We keep only terms of first order in $k'$, thereby discarding processes far away from the FS:
\begin{align}
\begin{split}
&\myim\Sigma^R_{B}(p)  \simeq -4 g^2 \int \frac{d\omega_k}{2\pi} 
\int_{-\Lambda}^\Lambda \frac{dk' k_F}{2\pi}\int_{0}^{2\pi} \frac{d \varphi_{\mathbf k}}{2\pi}\, \dkq \\
 &\times \myim \frac{1}{\omega_p+\omega_k -\Sigma^R_F(k'+p) -v_F p \cos(\varphi_{\mathbf k}-\varphi_{\mathbf p})-v_F k'}\\
 & \times \myim \frac{1}{\omega_k -\Sigma^R_F(k') -v_F k'} \left[n_F(\omega_k) - n_F(\omega_k + \omega_p)\right] \, .
\end{split}
\end{align}
In addition, we have also linearized the expression in the small ratio $p /k_F\ll 1$, which allows to replace $\varphi_{\mathbf k + \mathbf p} \to \varphi_{\mathbf k}$ and to neglect $\ep$. Indeed, $\mathbf p$ corresponds to the momentum transfer onto a Fermion in the vicinity of the FS which is restricted by $p \leq \Lambda \ll k_F$ according to the statements above Eq.~\eqref{eq:defcut} for the deviations of momenta away from the FS. Therefore, we have introduced the cut-off explicitly in the integral. In case of $\Gamma_B$, the integrals are UV-finite and we take the limit $\Lambda \to \infty$ in the following. Introducing rescaled frequencies $\bar{\omega} =\omega/T$ and momenta $\bar{k}'= k'/k_F$ we arrive at
\begin{align}
\begin{split}
\myim &\,\Sigma^R_{B}(p)  = -\frac{2 \pi \alpha T}{ \eF} \int \frac{d\bar\omega_k}{2\pi} 
\int \frac{d\bar k'}{2\pi}\int_0^{2\pi} \frac{d \varphi_{\mathbf k}}{2\pi}\, \dkq \\
 &\times  \myim \frac{1}{\dfrac{E(T(\bar\omega_p+\bar\omega_k,\varphi_{\mathbf k}))}{2\eF}-\bar p \cos(\varphi_{\mathbf k}-\varphi_{\mathbf{p}})-\bar k'} \\
& \times \myim \frac{1}{\dfrac{E(T \bar \omega_k,\varphi_{\mathbf{k}})}{2 \eF} -\bar k'}\left[\bar n_F(\bar \omega_k) - \bar n_F( \bar\omega_k + \bar\omega_p)\right]   \, ,
\end{split}
\end{align}
where we have defined the functions
\begin{align}\label{eq:defP}
\begin{split}
E(\omega,\varphi) & = \omega -\Sigma^R_F(\omega,\varphi)\\
\bar n_{F,B}(\bar\omega) & = \frac{1}{\exp(\bar \omega) \pm 1}  \, .
\end{split}
\end{align}
Furthermore, we have replaced $g^2$ by the effective coupling $\alpha$ via Eq.~\eqref{eq:defAlpha}. For small $T \ll \eF $, the first factor in the second line can be approximated by $- \pi \delta (\bar k)$ like for noninteracting Fermions, which yields
\begin{align}\label{eq:resImSBgen}
\begin{split}
\myim \Sigma^R_{B}(p) & \simeq  \frac{ \pi \alpha T}{\eF} \int \frac{d\bar\omega_k}{2\pi} 
\int_0^{2\pi} \frac{d \varphi_{\mathbf k}}{2\pi}\, \dkq \\
&\quad\times\myim \frac{1}{\dfrac{E(T(\bar\omega_p+\bar\omega_k),\varphi_{\mathbf k})}{2\eF}-\bar p \cos(\varphi_{\mathbf{k}}-\varphi_{\mathbf{p}})} \\
& \quad \times\left[\bar n_F(\bar \omega_k) - \bar n_F( \bar\omega_k + \bar\omega_p)\right] \, .
\end{split}
\end{align}
This result establishes the most general form that can be obtained by analytic means. 
However, the function admits further simplifications in various limits which can be classified by the ratio of $\bar p $ and $|E|/\eF$. Note that this implies  a frequency-dependent threshold determined by the external frequency $\bar \omega_p$ because $\bar{\omega}_k$ varies on the same scale as $\bar \omega_p$ due to the thermal distribution functions.
In the case $\bar p \gg |E|/\eF$, the remaining imaginary part can be replaced by a Dirac delta, too. This allows to solve the integral over the angle by picking the two points on the FS whose tangential vectors are parallel to $\mathbf{p}$:
\begin{align}\label{eq:resLandau}
\begin{split}
&\myim \Sigma^R_{B}(p\gg |E(\omega_p)|/v_F, \varphi_{\mathbf p} ,\omega_p)  \simeq \\
& -\frac{\pi \alpha T}{\eF \bar{p}} \dpq\! \int \frac{d\bar\omega_k}{2\pi} \left[\bar n_F(\bar \omega_k) - \bar n_F( \bar\omega_k + \bar\omega_p)\right]= - \alpha \dpq \frac{\omega_p}{v_F p} ,
\end{split}
\end{align}
and the Landau damping term from Eq.~\eqref{eq:defLandau} is recovered. As a result, it exists also at finite temperatures provided that the bosonic damping is evaluated at sufficiently large momenta. Note that the dependence on $\Sigma^R_F$ has completely dropped out and the evaluation coincides with the perturbative one based on free Green's functions. This explains the fact that in the ground state the self-consistent solution is obtained after a single interation initialized with bare Green's functions.
 
Landau damping breaks down when $\mathbf p $ is tangential to the FS in any of the noninteracting directions along the diagonals of momentum space $\varphi_{\mathbf p} = \pi/4 \cdot \{1,3,5,7\}$. In agreement with Eq.~\eqref{eq:resLandau} the $p^{-1}$ tail vanishes, but it is replaced by a $p^{-2}$ tail from the angular integral in Eq.~\eqref{eq:resImSBgen} (see also Ref.~\cite{hartnoll2014}). In the vicinity of the noninteracting directions the two tails can even mix, yet in such a way that the resulting $|\myim \Sigma^R_B(k)|$ is always larger or equal to the Landau damping tail and no additional poles arise in the dressed $D^R$. Since the diagonals are always suppressed by the d-wave form factor it will not be necessary to consider them any further. 

Next, we examine the opposite regime $\bar p \ll |E|/\eF$, which is important to understand how Landau damping is regularized at small momenta. After setting $\bar p = 0$ in Eq.~\eqref{eq:resImSBgen}, we obtain:
\begin{align}\label{eq:SBimint1}
\begin{split}
\myim \Sigma^R_{B}(0,\omega_p)  \simeq & \, 2\pi \alpha T \int \frac{d\bar\omega_k}{2\pi} 
\int_0^{2\pi} \frac{d \varphi_{\mathbf k}}{2\pi}\, \dkq
\\
& \times \myim \frac{1}{T(\bar\omega_k+\bar\omega_p)-\Sigma^R_F(T(\bar \omega_k+ 
\bar \omega_p),\varphi_{\mathbf{k}})}
 \\ &\times \left[\bar n_F(\bar \omega_k) - \bar n_F( \bar\omega_k + \bar\omega_p)\right] \, .
\end{split}
\end{align}
Again, we can establish asymptotic forms for different limits.
First, we consider the case $|\omega_p| \gtrsim \omega_>$ which includes the behavior in the ground state when $T \to 0$ (see Eq.~\eqref{eq:SFans}). Here, we undo the $\bar \omega= \omega/T$ transformation and introduce instead $\hat \omega= \omega/\win$. In addition, we replace the thermal distribution functions by their ground state counterparts. This leads to
\begin{align}\label{eq:numSBimT0}
\begin{split}
&\myim \Sigma^R_B(\mathbf p =0, |\omega_p| \gtrsim \omega_>) \simeq 
 -2 \pi \alpha \int\frac{d \varphi_{\mathbf k}}{2\pi} \dkq \int_0^{\hat \omega_p} \frac{d \hat \omega_k}{2\pi} \\ & \frac{|\hat \omega_k|^{2/3} |d_\mathbf{k}|^{4/3}}{(\hat \omega_k +\sqrt{3} \sgn(\hat \omega_k)|\hat \omega_k|^{2/3} |d_\mathbf{k}|^{4/3})^2 +(|\hat \omega_k|^{2/3} |d_\mathbf{k}|^{4/3})^2} \\
& \qquad\qquad\qquad\qquad\qquad\quad=: -2 \pi \alpha \hat{\Gamma}_B(\hat \omega_p)\, ,
 \end{split}
\end{align}
where we have inserted the self-energy at the QCP from Eq.~\eqref{eq:SFgs} and substituted $\hat \omega_k \to \hat \omega_p- \hat \omega_k$.
Since the integrand varies like $ |\hat\omega_k|^{-2/3}$ at the origin, 
$\myim \Sigma^R_B(\mathbf p =0 ,|\omega_p| \gtrsim \omega_>) \sim \sgn(\hat \omega_p)|\hat \omega_p|^{1/3}$, which is odd but not linear in $\hat \omega_p$.

In the opposite limit $|\omega_p| \lesssim \omega_<$ we 
expand the integral~\eqref{eq:SBimint1} to first order around $\bar{\omega}_p=0$, which yields
\begin{align}\label{eq:numSBimw0}
\begin{split}
&\myim\Sigma^R_{B}(\mathbf p =0 , |\omega_p| \lesssim \omega_<)  \simeq 8 \pi \alpha \bar\omega_p \int \frac{d\bar\omega_k}{2\pi} 
\int_0^{2\pi} \frac{d \varphi_{\mathbf k}}{2\pi} \dkq\\
& \qquad\qquad\quad \times \dfrac{1}{\cosh^2\left(\dfrac{\bar\omega_k}{2}\right)}
\myim \frac{1}{T\bar\omega_k-\Sigma^R_F(T
\bar \omega_k, \varphi_{\mathbf k})} \, .
\end{split}
\end{align}
The first factor in the second line becomes a $\delta$ function in the limit of $T \to 0$ while the denominator of the last factor approaches $-\Sigma^R_F(0, \varphi_{\mathbf k}) = i\Gamma_{F, \mathbf k} = i \Gamma_F(T) \dkq$. Here, we have set the angular dependence of the damping rate according to Eq.~\eqref{eq:resGammaFQCR} for the QCR. As a result, we obtain 
\begin{align}\label{eq:resSBimPunk}
\myim \Sigma^R_{B}(\mathbf p = 0, |\omega_p| \lesssim \omega_<) = -\frac{ \alpha \omega_p }{\Gamma_F(T)} \, ,
\end{align} 
which agrees with the observation by Punk~\cite{punk2016}.
  
All in all, we can summarize the results for the bosonic damping rate $\Gamma_B(p)= \Sigma^R_B(p)$ that appears in $D^R(p)$ by two scaling subregimes that are distinguished by the frequency argument $\omega_p$:
\begin{align}\label{eq:resGammaBsum}
\begin{array}{l}
|\omega_p|\gtrsim \omega_>: \\[0.2cm]
\Gamma_B(p) \to \begin{cases}  \alpha \dpq\dfrac{\omega_p}{v_F p}\, , & \text{if } v_F p \gg |E(\omega_p)|_{T=0} \\[.5 cm]
2\pi \alpha  \hat{\Gamma}_B\left( \dfrac{\omega_p}{\win}\right)\, , & \text{if } v_F p \ll |E(\omega_p)|_{T=0}
\end{cases} 
\\[1.5cm]
|\omega_p| \lesssim \omega_<:\\
\Gamma_B(p) \to \begin{cases}  \alpha \dpq\dfrac{\omega_p}{v_F p}\, ,  &\text{if } v_F p \gg \Gamma_F(T)=|E(\omega_p \to 0)|\\[0.5 cm]
 \dfrac{\alpha \omega_p}{\Gamma_F}\, , & \text{if } v_F p \ll \Gamma_F(T)=|E(\omega_p \to 0)|
\end{cases} 
\end{array}\!\! .
\vspace{2cm}
\end{align}
The index $T=0$ indicates that the ground-state self-energy $\Sigma^R_{F}|_{T=0}$ is used for the evaluation of $P$. Fig.~\ref{fig:SBim} presents the numerical evaluation and compares to these analytic results. As function of the momentum one observes that $\Gamma_B$ crosses over quickly between the limiting regimes. 
 
\subsection{Calculation of the fermionic self-energy}\label{sec:SF}
The expression for $\myim \Sigma^R_F$ is given in Eq.~\eqref{eq:defSF}.
We are mostly interested in the behavior in the vicinity of the FS and focus therefore on momenta $\mathbf k_F$ on the FS with polar angle $\varphi_{\mathbf k_F}$ not too close to the noninteracting directions.
The major difference between the calculation at zero temperature and finite temperature is caused by the change of the analytic structure through the appearance of the $T/\omega_p$ pole in $n_B$ at small frequencies. To deal with this effect, we separate the self-energy in a quantum and a thermal part~\cite{dell2006,punk2016}
\begin{align}
\myim \Sigma^R_F(\mathbf k_F,\omega_k) = \myim \Sigma^R_{F,Q}(\mathbf k_F,\omega_k)+\myim \Sigma^R_{F,T}(\mathbf k_F,\omega_k) \, .
\end{align}
where the thermal part contains the contribution from the additional pole only.
In the vincity of the QCP this term is closely related the contribution from the  the static $\Omega_n = 0$ component of the order-parameter field~\cite{klein2020}.
With our definition the two terms read
\begin{align}\label{eq:defSFQ}
\begin{split}
\myim \Sigma^R_{F,Q}(\mathbf k_F,\omega_k)& = 
- 2 g^2\int \frac{d^2 p\, d\omega_p}{(2\pi)^3} d^2_{\mathbf k_F - \mathbf p/2}   \myim D^R(p) \\
 & \qquad \quad\times \myim G^R_F(\mathbf k_F - \mathbf p,\omega_k-\omega_p) \\
 & \qquad \quad \times \left[n_B(\omega_p)- \frac{T}{\omega_p} + n_F(\omega_p - \omega_k)\right]
\end{split}
\end{align}
and
\begin{align}
\begin{split}\label{eq:defSFT}
\myim \Sigma^R_{F,T}(\mathbf k_F,\omega_k) & =
-2 g^2 T\int \frac{d^2 p\, d\omega_p}{(2\pi)^3} d^2_{\mathbf k_F - \mathbf p/2} \frac{\myim D^R(p)}{\omega_p} \\
& \qquad \qquad \times\myim G^R_F(\mathbf k_F -\mathbf p,\omega_k-\omega_p) \, .
\end{split}
\end{align}
Apparently, 
$\myim \Sigma^R_{F,T}$ approaches zero when $T \to 0$, provided that the integral exists, but has a finite limit when $\omega_k \to 0$. On the other hand, the factor in the third line of Eq.~\eqref{eq:defSFQ} has no singularities but converges for $T \to 0$ to the combination $-\theta(-\omega_p)+  \theta(\omega_k-\omega_p)$. Therefore, we anticipate that the strange metal correlations $\sim |\omega_k|^{2/3}$ are encoded in $\myim \Sigma^R_{F,Q}$ while the dominant contribution to the damping rate $\Gamma_F(T)$ emerges from $\myim \Sigma^R_{F,Q}$. We will now compute each function first individually and then discuss how they mix within the self-consistent loop.
  
\subsubsection{Quantum component of the self-energy}\label{sec:SFQ}
\begin{figure}[ht]
\begin{center}
\includegraphics[width=1\columnwidth]{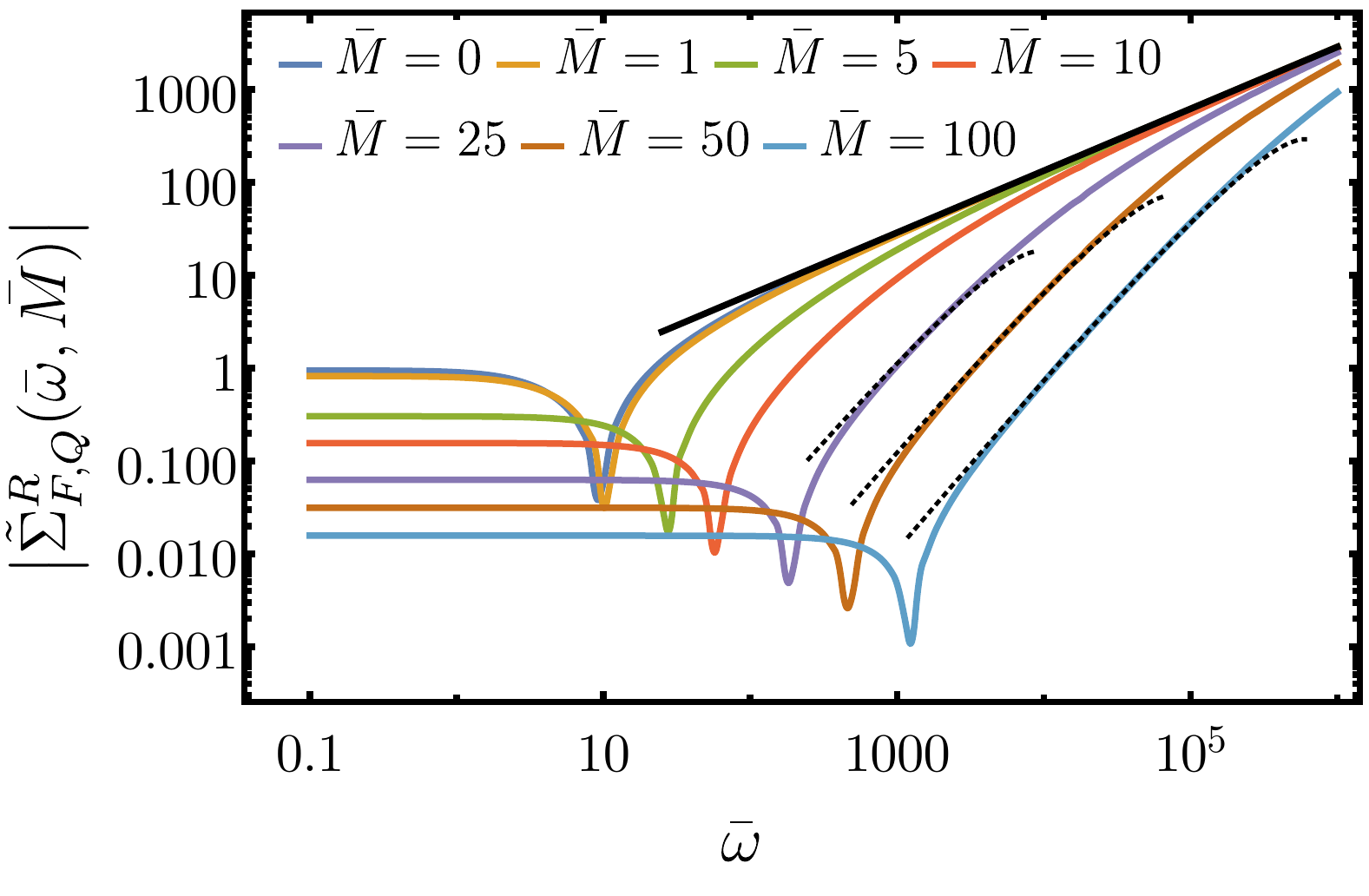}
\caption{Numerical result for $\tilde{\Sigma}^R_{F,Q}(\bar \omega, \bar M)$ for different values of the dimensionless mass.
The solid black line indicates the non-FL result~\eqref{eq:TSFQ}.
The black dashed lines show the intermediate FL regime observed for $\bar M \gtrsim 25$. For smaller masses it cannot be resolved since it becomes entangled with the crossover to the value  $\tilde{\Sigma}^R_{F,Q}(0, \bar M)$. The sign change of $\tilde{\Sigma}^R_{F,Q}$ corresponds to the minimum in this plot.}
\label{fig:TSFQ}
\end{center}
\end{figure}

\begin{figure}[ht]
\begin{center}
\includegraphics[width=1\columnwidth]{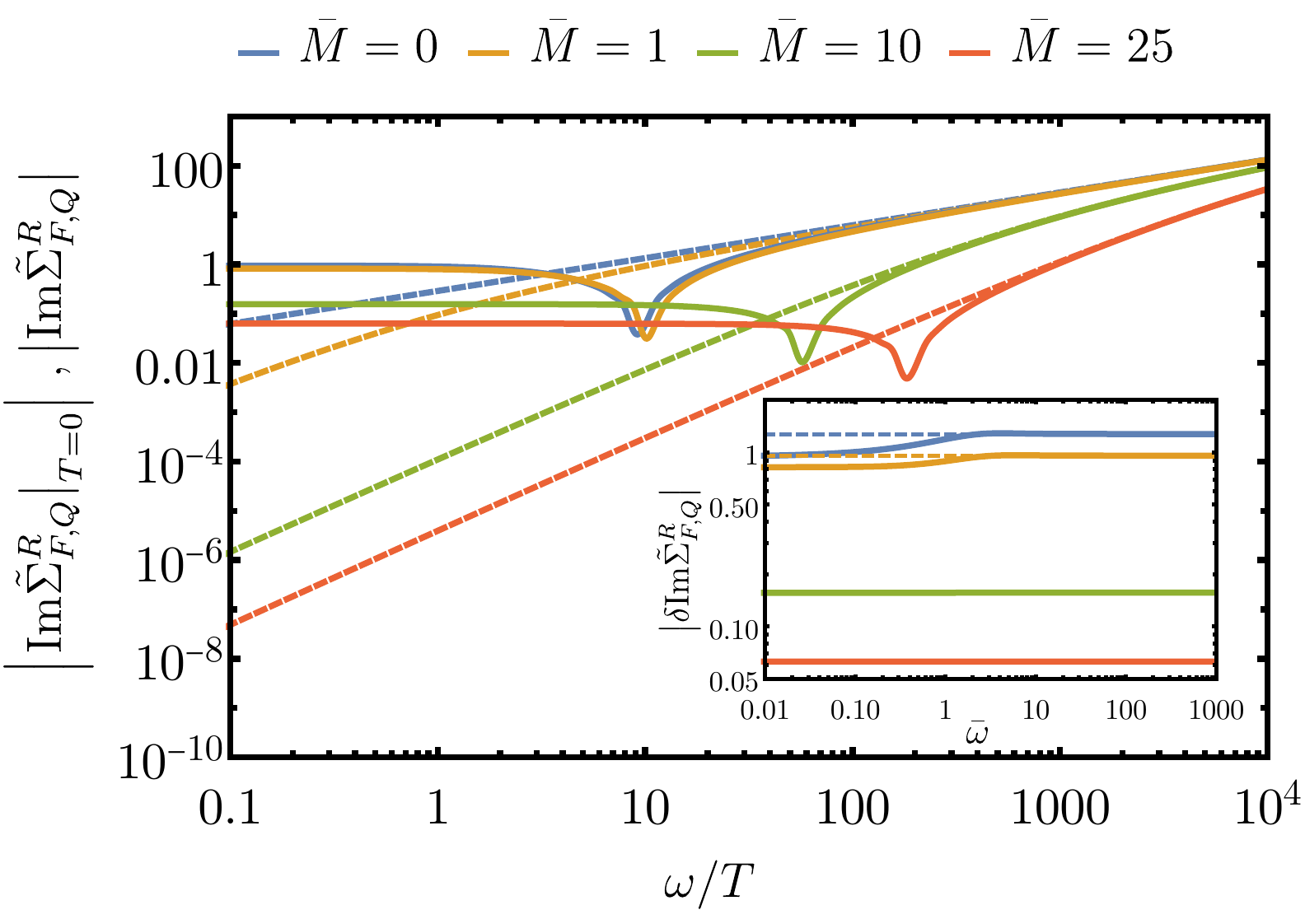}
\caption{Comparision of the quantum part of the self-energy $\Sigma^R_{F,Q}$ at finite temperatures (solid lines) vs. $\myim\left.\Sigma^R_{F,Q}\right|_{T=0}$ (dashed lines) obtained with ground state distribution functions for different masses. The latter are not cut-off by thermal fluctuations such that the FL or non-FL correlations persist in the limit $\bar \omega \to 0$. The inset shows the difference $\myim \delta \Sigma^R_{F,Q}$ defined in Eq.~\eqref{eq:defSFQdelta} (solid lines) with the same colour-coding. The dashed lines correspond to the high-frequency asymptotics~\eqref{eq:SFQdeltaAsy}. For the largest two masses they cannot be distinguished anymore.}
\label{fig:deltaSigma}
\end{center}
\end{figure}

$\myim \Sigma^R_{F,Q}$ has already been studied by Dell'Anna and Metzner~\cite{dell2006}. 
In particular, $\myim \Sigma^R_{F,Q}$ attains a universal scaling function $\myim \tilde \Sigma^R_{F,Q}$. Here, we extend the previous analysis in two ways: The scaling of $M(T)$ in our solution of the Eliashberg equations potentially gives rise to an instability of the non-FL regime, which has to be taken into account carefully. 
Moreover, we include thermal corrections to the non-FL asymptotics $\myim \Sigma^R_F(\mathbf{k}_F,\omega_k)_{T=0} \sim |\omega_k|^{2/3} $ for frequencies $|\omega_k| \gg T$, which is necessary for the computation of $M(T)$. 
The result for $\tilde{\Sigma}^R_{F,Q}$ as function of the dimensionless ratios $\omega_k/T$ and $M/(\alpha T)^{1/3}$ is presented in Fig.~\ref{fig:TSFQ}.
The difference between the result in the ground state and at finite temperatures, which illustrates the second issue, is shown in Fig.~\ref{fig:deltaSigma}.

Anticipating that the magnitude of the bosonic momentum $p$ is much less than $k_F$ we can approximate the angular variable
$\varphi_{\mathbf k_F + \mathbf p} \simeq \varphi_{\mathbf{k}_F}$ and $d^2_{\mathbf k_F-\mathbf p/2} \simeq \dkFq$.  
For the explicit calculation it is convenient to split the $p$ integral in Eq.~\eqref{eq:defSFT} into two parts separated by the threshold $v_F p=|E(\omega_k- \omega_p,\varphi_{\mathbf k_F})|$, where $P$ is defined above in Eq.~\eqref{eq:defP}.
We show in App.~\ref{sec:negSFQ} that the contribution from $v_F p \leq |E(\omega_k- \omega_p,\varphi_{\mathbf k_F})|$ is negligible and we consequently consider only $v_F p \geq |E(\omega_k- \omega_p,\varphi_{\mathbf k_F})|$ in the following. In this regime one encounters the Landau damping form $\Gamma_B(p) =\alpha \dpq \omega_p /(v_F p)$
according to Eq.~\eqref{eq:resLandau}.
In addition, we rescale frequencies via $\bar\omega = \omega/T $ and employ the transformation $\bar p = v_F A^{2/3}/(\dkFq \alpha T)^{1/3}\cdot p $ for momenta. Together with the definition of $\alpha$ in Eq.~\eqref{eq:defAlpha}, this brings the integral to the form:
\begin{align}\label{eq:SFQint1}
\begin{split}
&\myim \Sigma^R_{F,Q}(\mathbf k_F,\omega_k)=
 \frac{\pi}{\eF} \frac{ |\dkF|^{4/3} \alpha^{2/3} T^{2/3} }{A^{4/3}} \\
& \times \int_0^{2\pi} \frac{d \varphi_{\mathbf p}}{2\pi} \int\frac{d\bar\omega_p }{2\pi} \int_{\bar E }^\infty \frac{d\bar p}{2\pi} \, \bar p\,  \myim \frac{1}{\bar E + \bar p \cos(\varphi_{\mathbf p}- \varphi_{\mathbf{k}_F})}
 \\ & \frac{
\dfrac{\dpq}{\dkFq} \bar\omega_p \bar p \left[\bar n_B(\bar\omega_p)-\frac{1}{\bar \omega_p} +\bar n_F(\bar\omega_p-\bar\omega_k)\right]}{\bar p^2\!\left[\!\dfrac{T^{4/3}\bar\omega_p^2}{\alpha^{2/3} A^{2/3} |\dkF|^{4/3}} -\bar p^2 - \dfrac{M^2}{(A \dkFq \alpha T)^{2/3}}\!\right]^2 \!+ \dfrac{\dpv}{\dkFv}\bar\omega_p^2} ,
\end{split}
\end{align}
with the short-hand notation
$$\bar E=A^{2/3}(\dkFq \alpha T)^{-1/3} E(T(\bar\omega_k- \bar\omega_p),\varphi_{\mathbf{k}_F}). $$
Since the cut-off regularisation is irrelevant here, we have dropped it from the beginning.
The last line contains the bosonic propagator.
Its $T^{4/3} \bar \omega_p^2$ term, which originates from the dispersion of Eq.~\eqref{eq:ansDR}, is always irrelevant in the $T \to 0$ limit. However, the mass term can influence the results quite dramatically: If $M(T)$ vanishes faster than $T^{1/3}$, we can drop it. This scenario has been discussed in Ref.~\cite{dell2006}. On the other hand, if $M(T)$ decays more slowly than $T^{1/3}$, it is the most relevant parameter and we have to evaluate the function in the limit $M \to \infty$, which corresponds to a FL state. Finally, the marginal case $M(T) \sim T^{1/3}$ requires a careful analysis, since an additional $T$-independent variable emerges whose value has to be determined self-consistently. In fact, the solution presented in Eq.~\eqref{eq:keyResM2-} falls into this class. Therefore, we define the dimensionless parameter
\begin{align}\label{eq:defMbar}
\bar M =M/(A \dkFq \alpha T)^{1/3}
\end{align} 
and examine $\myim \Sigma^R_F(\mathbf k_F,\omega_k)$ as function of $\omega_k, \alpha, T$ and $M$ in the following.

Returning to the expression~\eqref{eq:SFQint1}, we recall that the frequencies of interest range between 0 and $\wmax$. The thermal distribution functions imply that $\omega_p$ varies predominantly on the same scale as the external frequency $\omega_k$. To simplify the integrals we consider different regimes of $\omega_k$:
For the smallest frequencies $|\omega_k| \lesssim \omega_<$, the function $E(T(\bar\omega_k  - \bar\omega_p),\varphi_{\mathbf k_F}) $ approaches the thermal damping rate $\Sigma^R_F (\mathbf k_F,0) =- i \Gamma_{F, \mathbf k_F} $. Provided that the latter vanishes faster than $T^{1/3}$ for small temperatures, which is in fact guaranteed by the Eliashberg equations (see Eq.~\eqref{eq:keyresGammaFQCR}), we can set $\bar E \to 0^+$. Thus, 
the imaginary part of the fermionic Green's function $\myim[\bar{E} + \bar p \cos(\varphi_\mathbf{p}- \varphi_{\mathbf k_F})]^{-1}$ turns into a Dirac delta $-\pi \delta(\bar p \cos(\varphi_{\mathbf p} -\varphi_{\mathbf{ k}_F}))$, similar to the case of Landau damping, and the integration over $\bar p$ includes all positive values. On the other hand, for the largest frequencies $|\omega_k| \gtrsim\omega_>$ we recover the non-FL correlations. In this case, the self-energy attains the non-FL form $\sim |\omega_k-\omega_p|^{2/3}$ such that $\bar E \sim T^{1/3}$. In the limit of small temperatures, the regime of the largest frequencies admits the same simplifications as the low-frequency regime discussed above. Furthermore, we expect that the intermediate regime $\omega_< \lesssim |\omega| \lesssim \omega_>$ can also be treated with these approximations since, otherwise, one finds a highly nonmonotonic $\left|\Sigma^R(\omega_< \lesssim|\omega| \lesssim \omega_>)\right|$. This assumption will be justified a posteriori by our self-consistent solution.
These considerations show that the quantum component satisfies the relation~\eqref{eq:ETcond1}, which states the condition for the applicability of Eliashberg theory at $T=0$, also at finite temperatures.

Next, we use the Dirac delta $-\pi \delta(\bar p \cos(\varphi_{\mathbf p} -\varphi_{\mathbf{ k}_F}))$ to integrate the over the angle. This picks out the pair of bosonic modes whose wave vectors are tangential to the FS a $\mathbf k_F$ and reduces effectively the two-dimensional integration measure $d\bar p\, \bar p$ to its one-dimensional counterpart $d\bar p$:
\begin{align}
\begin{split}
& \myim \Sigma^R_{F,Q}(\mathbf k_F,\omega_k) =
 -\frac{\pi}{\eF} \frac{ |\dkF|^{4/3} \alpha^{2/3} T^{2/3} }{A^{4/3}} \\
& \times \int\frac{d\bar\omega_p }{2\pi} \int_0^\infty \frac{d\bar p}{2\pi} \,  \frac{ \bar\omega_p \bar p \left[\bar n_B(\bar\omega_p)-\frac{1}{\bar \omega_p} +\bar n_F(\bar\omega_p-\bar\omega_k)\right]}{\bar p^2\left(\bar p^2 + \bar M^2\right)^2+\bar\omega_p^2} \, . \\
\end{split}
\end{align}
The largest contribution to the $\bar p$ integral arises from
the regime $\bar{p} \sim \max(\bar{\omega}_p^{1/3}, \bar M)$. In the limit $\bar M to 0$ this is equivalent the standard scaling argument of Eliashberg theory for the INM~\eqref{eq:ETcond2}, just formulated in the rescaled variables $\bar \omega, \bar p$. 
We can solve the integral over the magnitude $\bar p$ by the variable transformation $u =\bar p^2$. This leads to
\begin{align}\label{eq:SFQim1}
\begin{split}
&\myim \Sigma^R_{F,Q}(\mathbf k_F,\omega_k)\simeq
  \frac{ |\dkF|^{4/3} \alpha^{2/3} T^{2/3} }{4 A^{4/3} \eF}
 \int \frac{d\bar\omega_p }{2\pi} \, \bar \omega_p \\
 &\times\!\!\sum_{j=1}^3\! \dfrac{\log(u_j)}{\bar M^4 + 4 \bar M^2 u_j +3 u_j^2}\!\!  \left[\bar n_B(\bar\omega_p)-\frac{1}{\bar \omega_p} +\bar n_F(\bar\omega_p-\bar\omega_k)\right]\!,
\end{split}
\end{align}
where the $u_j$ denote the three solutions of
\begin{align}\label{eq:defuj}
u (u + \bar M^2)^2 + \bar\omega_p^2=0 \, .
\end{align} 
According to Eq.~\eqref{eq:SFQim1} the quantum component admits a representation in terms of a universal scaling form $\tilde{\Sigma}^R_{FQ}$
\begin{align}\label{eq:SQFdefScal}
\myim \Sigma^R_{F,Q}(\mathbf k_F,\omega_k)=
  \frac{ |\dkF|^{4/3} \alpha^{2/3} T^{2/3} }{4 A^{4/3} \eF} \tilde{\Sigma}^R_{F,Q}(\bar \omega_k ,\bar M) \, .
\end{align}
This representation is closely related to the results by Dell'Anna and Metzner~\cite{dell2006}. 
It reveals that $\myim \Sigma^R_{F,Q}(\mathbf k_F,\omega_k)$ indeed obeys $\omega/T$ scaling when $\bar M \to 0$. As long as the scaling function is not singular in this limit, we expect that $\omega/T$ scaling applies also in the regime $\bar M \ll 1$ up to small corrections from the finite mass.
The numerical evaluation of $\tilde{\Sigma}^R_{F,Q}$ for different values of the mass is presented in Fig.~\ref{fig:TSFQ}. First, of all, we observe that the scaling function is positive at small frequencies, which contradicts the physical constraint of a positive spectral function $A(\mathbf k, \omega_k) = -2 i \myim G^R(\mathbf k, \omega_k)$. This will be rectified as soon as the thermal component is taken into account, as can be seen by adding Eqs.~\cref{eq:defSFQ,eq:defSFT}. In fact, the sign change of $\tilde{\Sigma}^R_{F,Q}$ is expected due to a "sum rule for the first Matsubara frequency"~\cite{chubukov12sumrule1,chubukov12sumrule2,klein2020}. 

Regarding the behavior when $\bar M$ is varied, we indeed observe only small corrections to $\tilde \Sigma^R_{F,Q}(\bar \omega_k,0)$ for $\bar M \ll 1$. Furthermore, even values $\bar M \leq 1$ predominantly affect the low-frequency regime of positive sign, while the high-frequency regime is unchanged to logarithmic accuracy and described by the non-FL asymptotics~\eqref{eq:SFgs} (see also Eq.~\eqref{eq:TSFQ} below). At this stage $\omega/T$ scaling with non-FL exponents remains possible, even in the presence of a mass $\bar M \leq 1$ since the regime $\bar \omega_k \to 0$ is anyway dominated by the thermal component.  
On the other hand, if $\bar{M} \gg 1$, the scaling function differs strongly from $\tilde{\Sigma}^R_{F,Q}(\bar \omega_k ,0)$ on all frequency scales because the system is in the FL-regime (see also Eq.~\eqref{eq:TSFQ}) and $\omega/T$ scaling with non-FL exponents is ruled out.
As a result of these findings, we have to check the condition $\bar M \ll 1$ when we close the self-consistent loop.
Finally, the crossover between the low- and the high-frequency regime of $\tilde{\Sigma}^R_{F,Q}$ takes place at the rather large scale $\bar{\omega} \simeq 10$ when $\bar M \to 0$ and grows with $\bar M$.
To better capture the properties of the quantum component we can obtain closed forms of the scaling functions in different limits. To this end, we inspect the zeros $u_j$ from Eq.~\eqref{eq:defuj} in greater detail.
First, we distinguish two regimes that are smoothly connected: For $\bar M \to 0$, we have $u_j\sim |\bar  \omega_p|^{2/3}$ whereas in the opposite regime $\bar{M} \gg \bar\omega_p$, we find $u_1 \to 0$ and $u_{2,3} \sim \bar M^2$. We can estimate the crossover scale by the criterion $|\bar\omega_p|^{2/3} \sim \bar M^2$, which leads to $\bar 
\omega_p \sim \bar M^3$. By restoring dimensions this is equivalent to $\omega_p = M^3/\alpha$.  
More precisely, we obtain the following asymptotic results for the sum from Eq.~\eqref{eq:SFQim1}:
\begin{widetext}
\begin{align}\label{eq:SFQexp}
\begin{array}{c | c | c}
 & |\bar \omega_p| \gg \bar M^3  & |\bar \omega_p| \ll \bar M^3
 \\[.2 cm]
   \hline
 u_{1,2,3} &  |\bar \omega_p|^{2/3} \cdot \{-1, e^{\pm i \pi /3} \}  & -\dfrac{\bar \omega_p^2}{\bar M^2},\,  - \bar M^2 \pm \dfrac{\bar{\omega}_p}{\bar M} + \dfrac{\bar \omega_p^2}{\bar M^2} \\[.5 cm]
  \sum_{j=1}^3 \dfrac{\log(u_j)}{\bar M^4 + 4 \bar M^2 u_j +3 u_j^2} & \dfrac{-2\pi}{3 \sqrt{3} |\bar \omega_p|^{4/3}}
& 2\dfrac{\log\left(\sqrt{e}\dfrac{|\bar{\omega}_p|}{\bar M^3}\right)}{\bar M^4}
\end{array}\, .
\end{align}
\end{widetext}
Let us now consider the limit $T \to 0$ to check the convergence towards the established properties in the ground state. Approaching this limit eventually implies for any finite $\omega_k$ that $|\bar \omega_k| \gg 1$.
In this regime the integral is dominated by the internal frequencies $|\bar \omega_p| \gg 1$. As a consequence, the thermal distribution functions in Eq.~\eqref{eq:SFQim1} can be replaced by their ground state forms $\bar n_B(\bar\omega_p) \to -\theta(-\bar \omega_p)$ and $\bar n_F(\bar \omega_p - \bar \omega_k) \to \mp \theta (\bar\omega_k - \bar \omega_p)$, whereas the $\bar\omega_p^{-1}$ term that originates from subtracting the thermal pole of the Bosons is negligible. The resulting asymptotic function reads
\begin{align}\label{eq:TSFQ}
\begin{split}
\left.\tilde\Sigma^R_{F,Q}(\bar \omega_k, \bar M)\right|_{T= 0}  
\! \!=\int_0^{\bar \omega_k} \frac{d\bar\omega_p }{2\pi}  \bar \omega_p
 \sum_{j=1}^3 \dfrac{\log(u_j)}{\bar M^4 + 4 \bar M^2 u_j +3 u_j^2} \\
  \to 
 \begin{cases}
 -\dfrac{|\bar \omega_k|^{2/3}}{2\sqrt{3}} \, ,\!\! & \text{if } \bar\omega_k \gg \bar M^3 \\
 \dfrac{\bar \omega_k^2}{2\pi \bar M^4} \log\left(\dfrac{|\bar\omega_k|}{\bar M^3}\right)\, ,\!\! & \text{if } \bar{\omega}_k \ll \bar M^3
 \end{cases} \, ,
 \end{split}
\end{align}
which follows from inserting the expansions given in Eq.~\eqref{eq:SFQexp}. 
With Eq.~\eqref{eq:SQFdefScal} and the definition of $\bar M$ in Eq.~\eqref{eq:defMbar}, we find for the self-energy
\begin{align}\label{eq:resSFQ1}
\begin{split}
&\myim \Sigma^R_{F,Q}(\mathbf k_F,\omega_k)_{T = 0} \to \\ 
&
\begin{cases}
-\dfrac{1}{8\sqrt{3} \eF} \left(\dfrac{\dkFv \alpha^2}{ A^4}\right)^{1/3} |\omega_k|^{2/3}\, , &\text{if } |\omega_k| \gg M^3/\alpha  \\[.5 cm]
\dfrac{\dkFv \alpha^2}{8\pi \eF  M^4}  \omega_k^2 \log\left(\dfrac{A \dkFq \alpha |\omega_k|}{M^3} \right) , &\text{if } |\omega_k| \ll  M^3 /\alpha
\end{cases} 
 \, ,
\end{split}
\end{align}
where the temperature has dropped out as expected. In the high-frequency regime we indeed recover the non-FL scale 
$\win$ in agreement with Eq.~\eqref{eq:defwin} by identifying the first line with $-|\dkF|^{4/3} \win^{1/3} |\omega_k|^{2/3} $. In contrast, for frequencies below $M^3/\alpha$ we instead observe a two-dimensional Fermi liquid behavior~\cite{bloom1975} in agreement with Ref.~\cite{dell2006}. 
In Fig.~\ref{fig:TSFQ} we show how $\myim \tilde{\Sigma}^R_{F,Q}$ at finite temperatures approaches $\Sigma^{R,}_{F,Q}|_{T= 0}$ in the high-frequency limit. In addition, the inset of Fig.~\ref{fig:deltaSigma} presents the difference between $\myim \tilde{\Sigma}^R_{F,Q}$ and $\Sigma^{R,}_{F,Q}|_{T= 0}$, which will be discussed in much greater detail below.

\begin{figure}[t]
\begin{center}
\includegraphics[width=.85\columnwidth]{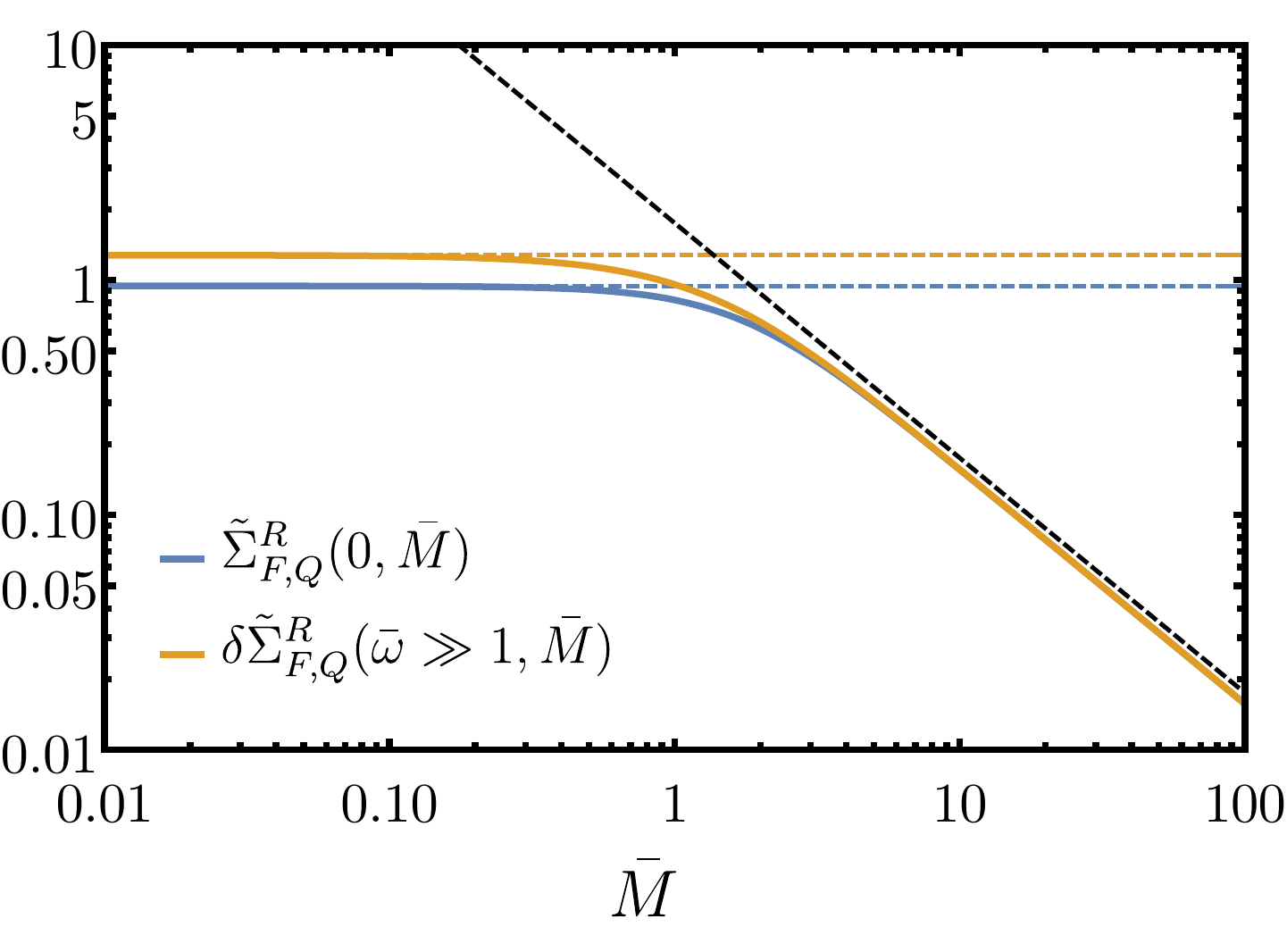}
\caption{$\bar M$-dependence of the scaling function $\tilde{\Sigma}^R_F(0,\bar M)$ (blue) and the thermal corrections $\delta \tilde{\Sigma}^R_{F,Q}(|\bar \omega_k| \gg 1, \bar M)$ in the high-frequency regime.
The dashed lines indicate the asymptotic expressions from Eqs.~\cref{eq:SFQT0Ml,eq:SFQdeltaAsy}, respectively. 
In particular, the black dashed line shows our estimate for the $\bar M^{-1}$ tail, which deviates from the numerical result by approximately 5\%.}
\label{fig:MdepComb}
\end{center}
\end{figure}

Next we turn briefly to the regime $|\bar \omega_k| \ll 1$ and evaluate $\tilde\Sigma^R_{F,Q}(0, \bar M)$,  shown in Fig.~\ref{fig:MdepComb}. 
The universal value $\tilde{\Sigma}^R_{F,Q}(0,0) = 0.944103...$ has to be determined by numerical integration while $\tilde\Sigma^R_{F,Q}(0, \bar M \to \infty)$ scales like $1/\bar M$. This is detailed in App.~\ref{sec:negSFQ}, where we also provide an estimate for the coefficient of the tail:
\begin{align}\label{eq:SFQT0Ml}
\tilde{\Sigma}^R_{F,Q}(0,\bar M) \to  
\begin{cases}
0.944103...\, , & \quad \text{if } \bar M \ll  1 \\
1.75024...\bar M^{-1}  \, , & \quad \text{if } \bar M \gg 1
\end{cases}\, .
\end{align} 

According to Eq.~\eqref{eq:SQFdefScal}, these findings entail the following low-frequency behavior of the quantum component
\begin{align}\label{eq:resSFQw0}
\begin{split}
&\myim \Sigma^R_{F,Q}( k_F,\bar \omega_k=0) \simeq\\
& \quad\frac{ 1}{\eF} 
\begin{cases}
0.236025... \cdot \dfrac{(\dkFq \alpha T)^{2/3}}{A^{4/3}}\, , &  \text{if } M \ll  \alpha^{1/3} T^{1/3} \\
0.437561... \cdot \dfrac{ \dkFq \alpha T }{A M}  \, , & \text{if } M \gg \alpha^{1/3} T^{1/3}
\end{cases}\, .
\end{split}
\end{align}
To close the self-consistent loop at the onset of finite temperatures,  it will turn out to be crucial to understand the thermal corrections to the non-FL asymptotics $\myim \Sigma^R_{F,Q} \sim |\omega_k|^{2/3}$ in the high-frequency regime $|\omega_k| \gg T$, too. To investigate them we define
\begin{align}\label{eq:defSFQdelta}
\begin{split}
\myim\delta\tilde\Sigma^R_{F,Q}&(\bar \omega_k ,\bar M ) := \\ &\myim \tilde\Sigma^R_{F,Q}(\bar \omega_k,\bar M ) - \left.\myim \tilde\Sigma^R_{F,Q} (\bar \omega_k, \bar M)\right|_{T=0} \, , 
\end{split}
\end{align}
and obtain the following expression from Eq.~\eqref{eq:SFQim1}
\begin{align}\label{eq:SFQdelta}
\begin{split}
&\myim\delta\tilde\Sigma^R_{F,Q}(\bar \omega_k ,\bar M ) = \int \frac{d\bar\omega_p }{2\pi}  \bar \omega_p \sum_{j=1}^3 \dfrac{\log(u_j)}{\bar M^4 + 4 \bar M^2 u_j +3 u_j^2} \\
&\times\! [n_B(\bar \omega_p)-\frac{1}{\bar \omega_p}+\theta(-\bar \omega_p)+n_F(\bar \omega_p -\bar \omega_k)-\theta(\bar \omega_k -\bar \omega_p)].
\end{split}
\end{align}
The last two terms in the second line contribute only substantially when $|\bar \omega_p - \bar \omega_k| \lesssim 1$. Since this combination is multiplied with an UV-integrable function, it gives rise to an algebraically decreasing contribution that is negligible for large arguments $|\bar \omega_k| \gg 1$. As a result, we obtain corrections that are to first approximation independent of the frequency:
\begin{align}\label{eq:resSFQHF}
\begin{split}
\myim \delta\tilde\Sigma^R_{F,Q}(|\bar \omega_k| \gg 1 ,\bar M )
 &\simeq  \!
\int \!\!\frac{d\bar\omega_p }{2\pi}    [\bar n_B(\bar \omega_p)-\frac{1}{\bar \omega_p}+\theta(-\bar \omega_p)]
  \\
 & \quad \times \omega_p\sum_{j=1}^3 \dfrac{\log(u_j)}{\bar M^4 + 4 \bar M^2 u_j +3 u_j^2} \, .
\end{split}
\end{align}
which agrees very well with the numerical evaluation, as confirmed in Fig.~\ref{fig:deltaSigma}.
Like above, we can find asymptotic results for this function in the limits $\bar{M} \ll 1$ and $\bar M \gg 1$. In the first one we set $\bar M =0$ which allows to replace the sum by $-2\pi/(3 \sqrt{3} |\bar \omega_p|^{4/3})$ (see Eq.~\eqref{eq:SFQexp}) and evaluate the integral numerically. In the opposite limit of large masses the analysis shown in App.~\ref{sec:negSFQ} reveals the same $\bar M^{-1}$ tail as in the $\omega_k \to 0$ case, see Eq.~\eqref{eq:SFQT0Ml}.  Altogether, we have the following positive corrections to the asymptotic behavior:
\begin{align}\label{eq:SFQdeltaAsy}
\delta \myim \tilde\Sigma^R_{F,Q}(|\bar \omega_k| \gg 1 ,\bar M ) \to 
\begin{cases}
1.27567... , & \!\text{if } \bar{M} \ll 1 \\
1.75024... \bar M^{-1}  , & \!\text{if } \bar{M} \gg 1 
\end{cases}  ,
\end{align}
which is plotted in Fig.~\ref{fig:MdepComb}. Restoring dimensions via Eqs.~\eqref{eq:SQFdefScal} and~\eqref{eq:defMbar} we obtain
\begin{align}\label{eq:resSFQdelta}
\begin{split}
&\delta \myim \Sigma^R_{F,Q}(\mathbf k_F,|\omega_k| \gg T) \to  \\
&\;\;\frac{1}{\eF}\begin{cases}
0.318919...\dfrac{|\dkF|^{4/3} \alpha^{2/3}}{A^{4/3}} T^{2/3} \, , &\text{if } M \ll (\alpha T)^{1/3}\\
0.437561... \dfrac{\alpha \dkFq}{A \eF} \dfrac{T}{M} \, , &\text{if } M \gg (\alpha T)^{1/3}
\end{cases} .
\end{split}
\end{align}
For later convenience we introduce $b=0.318919$.
Quite importantly, the leading correction scales itself like $T^{2/3}$ provided that the mass is small. This is intimately connected to the non-FL correlations in the ground state with the difference that the frequency integration in $\delta \myim \tilde{\Sigma}^R_{F,Q}$ is not cut-off by the external frequency but by the temperature.
This result turns out to have a strong impact on the temperature scaling of the bosonic mass $M(T)$, as discussed in Sec.~\ref{sec:bosMass}. 
Before turning to this function we analyze the classical component in the next section to obtain a complete understanding of the fermionic damping rate and the crossover scales. 

\subsubsection{Thermal component of the self-energy}\label{sec:SFT}
\begin{figure}[ht]
\begin{center}
\includegraphics[width=\columnwidth]{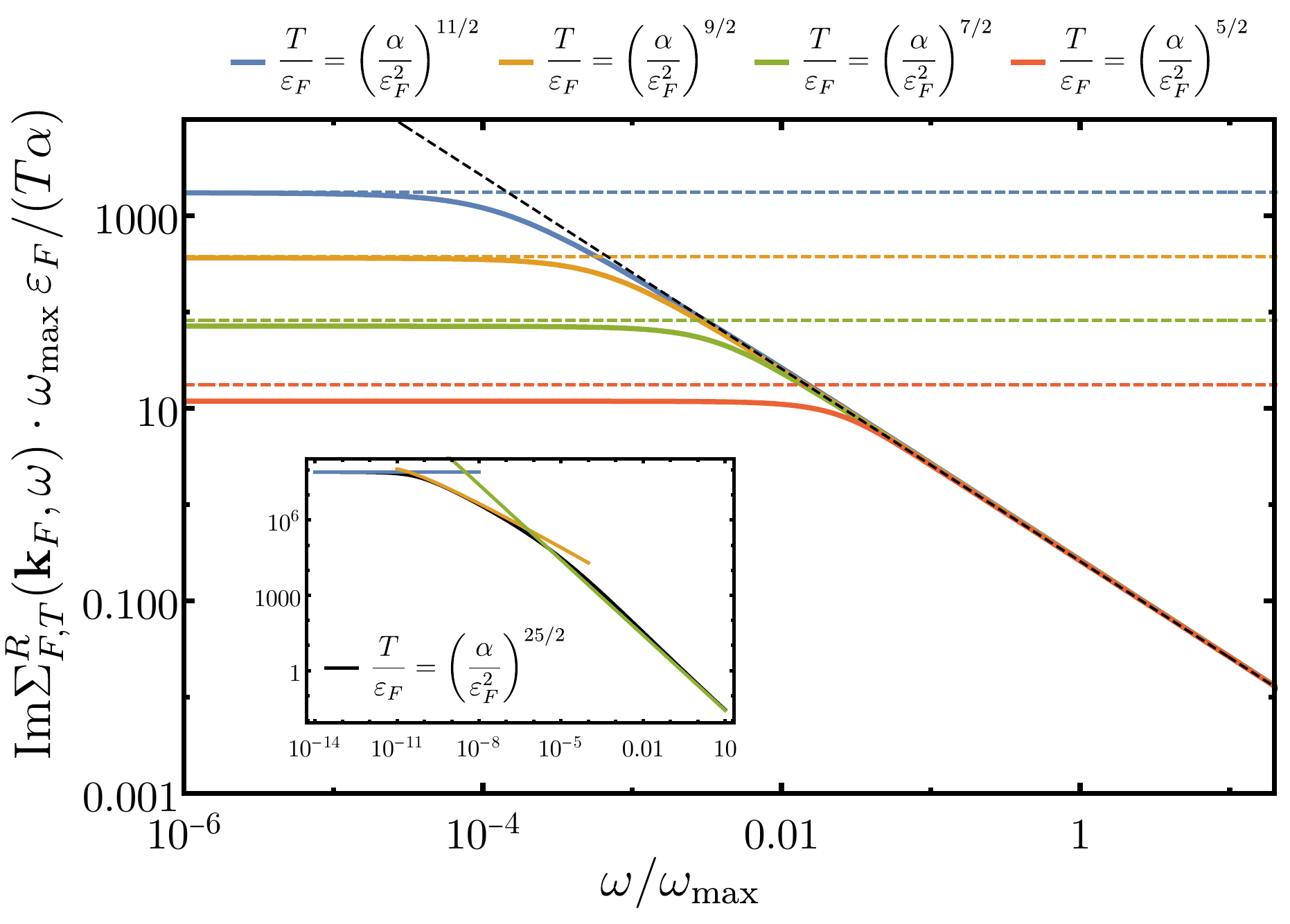}
\caption{Numerical, self-consistent solution of Eq.~\eqref{eq:SFTfin} at four different temperatures for $\alpha/\eF^2=10^{-2}$ and $M(T)$ from Eq.~\eqref{eq:keyresGammaFQCR} with $h_\omega =10^{-1}$ and $A=1$. The colored dashed lines represent the asymptotic result~\eqref{eq:resGammaFQCR}. However, at the two largest temperatures the latter deviates from the full result since the constraint $\Gamma_F(T) \ll M(T)$ ceases to be satisfied. The black dashed line indicates the $|\omega_k|^{-1}$ tail of Eq.~\eqref{eq:resSFTsum}.
The inset has the same axes as the main plot and presents $\myim \Sigma^R_{F,T}$ at an extremely small temperature. The blue line corresponds again to the thermal damping rate of~\eqref{eq:resGammaFQCR}. In addition to the green line for the $1/|\omega_k|$ power law, we observe also the intermediate regime governed by the non-FL self-energy shown in yellow (see Eq.~\eqref{eq:resSFTsum} for details).}
\label{fig:SFT}
\end{center}
\end{figure}

We turn now to the thermal part of the fermionic self-energy defined in Eq.~\eqref{eq:defSFT}.  A graphical representation of the self-consistent solution is given in Fig.~\ref{fig:SFT}. It reveals a plateau at small frequencies whose value determines the thermal scattering rate $\Gamma_F(T)$. For larger frequencies the plateau crosses over to algebraic tails of different nature depending on the temperature.

To perform the explicit calculation, we approximate the angular dependence in the limit $p \ll k_F$ and insert the physical bosonic propagator~\eqref{eq:ansDR}:  
\begin{align}
\begin{split}\label{eq:SFTint1}
\myim \Sigma^R_{F,T}(\mathbf k_F,\omega_k) & = 2 g^2 T\dkFq \int \frac{d^2 p\, d\omega_p}{(2\pi)^3} 
\frac{1}{\omega_p} \\
& \quad\times \myim G^R_F(\mathbf k_F - \mathbf p,\omega_k -\omega_p) 
\\ 
& \quad \times
 \frac{\Gamma_B(p)}{(\omega_p^2 -c_B^2 p^2 -M^2(T))^2 +  \Gamma^2_B(p)} \, .
\end{split}
\end{align}
To determine the leading behavior of this function, we 
first solve the frequency integral. To this end, we concentrate on the factors in the first and last line since the fermionic Green's function will turn out below to play only a subordinate role in this case.
In the regimes where the bosonic damping attains a form linear in the frequency, i.e $\Gamma_B(p) = \omega_p \tilde{\Gamma}_B(\mathbf p)$ (see Eq.~\eqref{eq:resGammaBsum}), we observe that the largest contribution arises from the scale
\begin{align}\label{eq:scaleDR}
\Omega_p =  \frac{c^2_B p^2+M^2(T)}{\tilde\Gamma_B(\mathbf p)}\,   ,
\end{align}
to lowest order in $\omega_p$.
The simple pole structure points towards solving the $\omega_p$ integration by the residue theorem. However, we recall from Eq.~\eqref{eq:resGammaBsum} that $\Gamma_B(p) \sim \alpha \sgn(\omega_p)|\omega_p/\win|^{1/3}$ 
in the regime $|\omega_p| \gg T$ with simultaneously $v_F p \ll \win^{1/3} |\omega_p|^{2/3}$. These branch cuts potentially impede the calculation of the $\omega_p$ integral but
we can argue that they give rise to irrelevant contributions only: We first note that the $1/\omega_p$ pole of the Bose distribution is not evaluated at the origin because of the constraint $|\omega_p| \gg T$. Next, we
determine the frequency scale where the denominator of $\myim D^R(p)$ becomes minimal for the given form of $\Gamma_B(p)$. This yields
$$ \Omega_p \sim  \win \left[\frac{c^2_B p^2+M^2(T)}{\alpha} \right]^3\, ,$$
to lowest order in $\omega_p$. 
If $ c_B p \sim v_F p \lesssim M(T)$, we have $\Omega_p \sim M^6(T)$ which scales at least with $T^2$ since we are only interested in solutions where $M(T) \sim T^{1/3}$ or faster to avoid entering a FL regime. As a result, this scenario contradicts the condition $\Omega_p \gg T$. In the opposite case $c_B p \gg M(T)$, we find that $v_F p \gg \eF$ by inserting $\Omega_p \sim \win (c_B p)^6/\alpha^3$ into the momentum constraint $v_F p \ll \win^{1/3} \Omega_p^{2/3}$. Such high-energy processes do not play any role at all for the  low-energy physics studied here and can be safely discarded. In other words, the main contributions to $\myim \Sigma^R_{F,T}$ arise from those regions in the $(\mathbf p,\omega_p)$ space where the bosonic damping is linear in frequency. 
Next, we note that the $\myim$ symbol in Eq.~\eqref{eq:SFTint1} can be taken in front of the integral and that $G^R(\omega_k - \omega_p)$ is analytic in the lower half of the complex $\omega_p$ plane. Hence, we can close the contour below to avoid the fermionic poles such that only the pole of $D^R$ at $\omega_p = - i \Omega_p$ from Eq.~\eqref{eq:scaleDR} is enclosed:
\begin{align}
\begin{split}\label{eq:SFTint2}
\myim \Sigma^R_{F,T}(\mathbf k_F,\omega_k) & =  g^2 T\dkFq \myim \int \frac{d^2 p}{(2\pi)^2} \frac{1}{c_B^2 p^2 +M^2(T)}
\\& \qquad\qquad\qquad\times G^R_F(\mathbf k_F - \mathbf p,\omega_k +i\Omega_p) \, .
\end{split}
\end{align}
Note that the factor $\tilde \Gamma_B(\mathbf p)$ that carries the momentum dependence of the bosonic damping has disappeared. Furthermore, we see that typical momenta scale like $v_F p \sim M(T)$. In case of Landau damping, that is $\tilde{\Gamma}_B(\mathbf p)\sim 1/(v_F|\mathbf p|)$, we have $\Omega_p \sim M^3(T) \sim T$ where the last power law applies for the result~\eqref{eq:keyResM2-}.  
On the other hand, if Landau damping is absent, we have $\tilde{\Gamma}_B \sim \Gamma_F^{-1}(T)$ and thus $\Omega_p \sim \Gamma_F(T) M^2(T)\sim T^{4/3}$, where the last relation holds for temperatures below $T_{\text{scal}}$. This justifies the omission of the $\omega_p^2$ term of the dispersive part of $D^R$ in Eq.~\eqref{eq:scaleDR}. In addition, $\Omega_p$ vanises at least like $T$ in both cases when the self-consistent results from Sec.~\ref{sec:keyRes} are inserted and the $T \to 0$ limit is taken. We also set $\Omega_p = 0$ in $\myim G^R$, 
which suffices to determine the leading behavior of the thermal component, since all crossovers scale in a more relevant way with $T$ or obtain a prefactor that is much less suppressed in $\alpha$ than $\Omega_p$.  
To proceed with the calculation, we linearize the dispersion around the Fermi surface as before
\begin{align}
\begin{split}\label{eq:SFTint3}
\myim \Sigma^R_{F,T}(\mathbf k_F,\omega_k) = g^2 \dkFq T \myim \int \frac{d^2 p}{(2\pi)^2}  \frac{1}{c_B^2 p^2 + M^2} \\
\frac{1}{\omega_k-\Sigma^R_F(\omega_k,\varphi_{\mathbf k_F})- v_F p \cos(\varphi_{\mathbf p}-\varphi_{\mathbf k_F})} \, ,
\end{split}
\end{align}
to solve the angular integral ($u=v_F p$) which leads to:
\begin{align}
\begin{split}
\myim \Sigma^R_{F,T}(\mathbf k_F,\omega_k)  = -\frac{ g^2 \dkFq T}{v_F^2 A^2} \myim \int_0^\infty \frac{du\,  u}{2\pi}  \frac{1}{u^2+ M^2/A^2} \\ \times \frac{i}{\sqrt{u^2 - (\omega_k-\Sigma^R_F(\omega_k,\varphi_{\mathbf k_F}))^2}} \, .
\end{split}
\end{align}
The final result reads then
\begin{align}\label{eq:SFTfin}
\begin{split}
&\myim \Sigma^R_{F,T}(\mathbf k_F,\omega_k) = \\ 
&  \myim \dfrac{\frac{ \alpha \dkFq T}{4 \eF A^2} \operatorname{arcosh}\left(-i \dfrac{(\omega_k-\Sigma^R_F(\omega_k,\varphi_{\mathbf k_F}))}{M/A}\right)}{(\omega_k-\Sigma^R_F(\omega_k,\varphi_{\mathbf k_F})) \sqrt{1 + \dfrac{M^2/A^2}{(\omega_k-\Sigma^R_F(\omega_k,\varphi_{\mathbf k_F}))^2}}}\, .
\end{split}
\end{align}
Here, arcosh denotes the inverse function of $\cosh(z)$ in the complex plane. 
This result can be obtained also from the corresponding Matsubara self-energy, which has been computed in Refs.~\cite{klein2020,damia2020}, by analytic continuation. 

In order to identify the dominant behavior in the different regimes we can again obtain simple analytic forms. To this end, we note the expansion for $M>0$ and $\myim z>0$:
\begin{align}\label{eq:arcsecExp}
\begin{split}
& \myim \cfrac{\operatorname{arcosh}\left(-i \frac{M}{z} \right)}{z \sqrt{1+\cfrac{M^2}{z^2}} } \to
\begin{cases}
\myim \dfrac{2\log\left(\frac{2z}{M} \right)-i \pi}{2z} \, , & \!\!\text{if } M \ll |z|\\
-\dfrac{\pi}{2M} \, , & \!\! \text{if } M \gg |z|
\end{cases} .
\end{split}
\end{align}
Let us focus first on the fermionic damping rate by setting $\omega_k=0$ in Eq.~\eqref{eq:SFTfin} and neglecting the negative contribution from the quantum component~\eqref{eq:resSFQw0}, which will turn out to be irrelevant below. With the given expansion we find
\begin{widetext}
\begin{align}\label{eq:resGammaF}
\begin{split}
\Gamma_{F,\mathbf k_F}(T) & = -\frac{ \alpha \dkFq T}{4 A^2 \eF} \myim \frac{\operatorname{arcosh}\left(\dfrac{\Gamma_{F, \mathbf k_F(T)}}{M/A}\right)}{i \Gamma_{F, \mathbf k_F}(T) \sqrt{1 - \dfrac{M^2/A^2}{\Gamma^2_{F \mathbf k_F}(T)}}}
\rightarrow
\frac{ \alpha \dkFq T}{A^2 \eF } 
\begin{cases}\dfrac{1}{4 \Gamma_{F,\mathbf k_F}(T)}
\log\left(\dfrac{2 \Gamma_{F, \mathbf{k}_F}(T)}{M/A} \right)\! &\text{if, } M/A \ll \Gamma_{F, \mathbf{k}_F}(T) \\[.5 cm]
 \dfrac{\pi}{8} \dfrac{1}{M/A}\!  &\text{if, } M/A \gg \Gamma_{F, \mathbf{k}_F}(T)
\end{cases}
\end{split} \, ,
\end{align} 
\end{widetext}
where we have used $\myre \Sigma^R_{F,T} (\mathbf k_F,\omega_k=0) = 0$ due to the symmetry arguments given in App.~\ref{sec:KK}.
The upper case is attained close to the classical critical line, where $M \to 0$, but also for large enough temperatures at the coupling strength corresponding to the QCP, as we will see below. The solution for $\Gamma_F(T)$ reads in this case~\cite{klein2020,damia2020}
\begin{align}\label{eq:resGammaFcc}
\Gamma_{F, \mathbf{k}_F} = \frac{|\dkF|}{2A}\sqrt{\frac{\alpha T}{\eF} \log \left(\frac{ \sqrt{ \alpha T} |\dkF|}{\sqrt{\eF} M/A^2} \right)}\, \, ,
\end{align} 
up to log(log...) terms.
Obviously, the scaling $\Gamma_F(T) \sim \sqrt{T \log T}$ renders the quantum component, which at most scales like $T^{2/3}$ for small masses, indeed irrelevant.
In particular, Klein et al.~\cite{klein2020} showed that this solution agrees very well with QMC simulations, however at temperatures well above the superfluid transition of the INM~\cite{metlitski2015} $T_c \sim \win$.
Interestingly, the fermionic damping rate diverges logarithmically upon approaching $T_c$.
As we will see in Sec.~\ref{sec:vertexCorr}, the vertex corrections contain similar logarithms which require a careful resumation to capture the real physical result in the limit $M \to 0$. In this regard, Damia et al.~\cite{damia2020} showed that including a bosonic self-interaction in the self-consistent equations cures the divergences.
In the opposite regime $M(T) \gg \Gamma_F(T)$, one finds
\begin{align}\label{eq:resGammaFQCR}
\Gamma_{F,\mathbf k_F}(T) =  \frac{\pi \alpha \dkFq}{8 \eF A} \frac{T}{M(T)} \, ,
\end{align} 
which has been obtained previously by Dell'Anna and Metzner~\cite{dell2006}, as well as by Punk~\cite{punk2016}. 
In fact, this form turns out to govern the Eliashberg equations at the onset of finite temperatures (see also the arguments for the quantum critical scaling, given in Sec.~\ref{sec:keyRes} and Sec.~\ref{sec:bosMass}). Fig.~\ref{fig:SFT} shows that this result indeed describes the $\omega_k \to 0$ limit of $\myim \Sigma^R_{F,T}$ at the smallest temperatures. 

In addition to the plateau of $
\myim \Sigma^R_{F,T}(\mathbf k_F, \omega_k)$, we can also extract the large-frequency behavior from Eq.~\eqref{eq:SFTfin}. More precisely, we refer to the regime $|\omega_k-\Sigma^R_{F}(\mathbf k_F,\omega_k)| \gg M$, which is formally equivalent to the $M \to 0$ limit in Eq.~\eqref{eq:arcsecExp}, In this regime we find:
\begin{align}\label{eq:SFTasy}
\begin{split}
\myim \Sigma^R_{F,T}&(\mathbf k_F, \omega_k)  \simeq  \\
&\!\!\frac{\alpha \dkFq T}{4 A^2 \varepsilon_F}\myim  \left[\dfrac{2 \log \left(2\dfrac{\omega_k- \Sigma^R_{F} (\omega_k, \varphi_{\mathbf k_F})}{M/A} \right)-i \pi}{\omega_k- \Sigma^R_F(\omega_k,\varphi_{\mathbf k_F})} \right] \, .
\end{split}
\end{align} 
Moreover, for asymptotically large arguments the solution of the self-consistent problem converges to the solution of the ordinary equation where $\Sigma^R_F$ on the right-hand side is replaced by $\Sigma^R_{F,Q}$. This allows to distinguish two scenarios: At the highest frequencies $|\omega_k| \gg \win$ the frequency dependence of the right-hand-side is determined by the linear frequency terms which implies $\myim \Sigma^R_F(\mathbf k_F, \omega_k) \sim |\omega_k|^{-1}$. However, if the asymptotic regime is encountered already for $|\omega_k| \lesssim \win$, the right-hand-side is dominated by the asymptotics $\myim\Sigma^R_{F,Q}(\mathbf k_F, \omega_k)\to -|\dkF|^{4/3}\win^{1/3} |\omega_k|^{2/3}$. Then, the non-FL correlations give rise to an intermediate regime characterized by an $|\omega_k|^{-2/3}$ tail with logarithmic corrections. If present, this regime emerges between the plateau in the limit $\omega_k \to 0$ and the high-frequency $|\omega_k|^{-1}$ power-law. Such behavior is indeed observed in Fig.~\ref{fig:SFT} for the lowest temperatures. We state the functional forms more precisely in Eq.~\eqref{eq:resSFTsum} below, but we first extract the crossovers to complete the analysis. By comparing $\Gamma_F(T)$ from Eq.~\eqref{eq:resGammaFQCR} to the high-frequency regime we obtain a simple estimate for the crossover scale between the low- and high-frequency asymtptotics of $\myim \Sigma^R_{F,T}$
\begin{align}\label{eq:SFTcross}
\begin{split}
\frac{1}{M}\sim\left|\myim \left[\dfrac{2\log \left(2\dfrac{\omega_k- \Sigma^R_{F,Q} (\omega_k, \varphi_{\mathbf k_F})}{M} \right)-i \pi}{\omega_k- \Sigma^R_{F,Q}(\omega_k,\varphi_{\mathbf k_F})} \right]\right| \, ,
\end{split}
\end{align} 
where we have omitted numerical factors of order one.
Based on the previous arguments we can further simplify this relation: 
At the smallest temperatures we expect $M(T) \lesssim \win$ such that the condition $|\omega_k-\Sigma^R_{F,Q}(\omega_k,\varphi_{\mathbf k_F})| \gg M(T)$ is satisfied for frequencies comparable or even below $\win$ and the intermediate regime governed by the non-FL correlations is present. 
On the other hand, 
as soon as $M(T)$ increases with temperature to $\mathcal{O} (\win)$, or even beyond, the condition $|\omega_k-\Sigma^R_{F,Q}(\omega_k,\varphi_{\mathbf k_F})| \gg M(T)$ entails a direct crossover from $\Gamma_F(T)$ to the $1/|\omega_k|$ tail. In summary, we have the following asymptotic scaling behavior:
\begin{widetext}
 \begin{align}\label{eq:resSFTsum}
 \begin{array}{l  }
 M(T) \ll \win:  \qquad\\
 \myim \Sigma^R_{F,T} (\mathbf k_F, \omega_k)\simeq
  - \dfrac{\alpha \dkFq T}{4 \varepsilon_F A^2} 
\begin{cases}
 \dfrac{\pi}{2 M} \, , \quad \text{if } |\omega_k| \ll \dfrac{M(T)^{3/2}}{\win^{1/2}} \\[.5 cm]
 \dfrac{\pi}{|\omega_k|}\, , \quad \text{if } \win \ll|\omega_k| \\[.5 cm]
 \dfrac{1}{|d_{\mathbf k_F}|^{4/3}\win^{1/3} |\omega_k|^{2/3}}\myim \!\left[\!\dfrac{2\log\left(\dfrac{2}{M/A}(\sqrt{3}\sgn(\omega_k) +i)|d_{\mathbf k_F}|^{4/3}\win^{1/3} |\omega_k|^{2/3} \right)- i\pi}{(\sqrt{3} \sgn(\omega_k) +i)}\!\right]\! , \text{else }
 \\[.5 cm]
 \end{cases} \\[2.5 cm]
 M(T) \gg \win: 
\qquad \myim \Sigma^R_{F,T} (\mathbf k_F,\omega_k) \simeq -\dfrac{ \alpha \dkFq T}{4\varepsilon_F A^2 }  
 \begin{cases}
 \dfrac{\pi}{2M} \, , & \text{if } |\omega_k| \ll M \\[.5 cm]
 \dfrac{\pi}{|\omega_k|}\, , &\text{if } M \ll|\omega_k| \, ,
 \end{cases}
 \end{array}
\end{align}   
\end{widetext}
where the crossover scales have been calculated up to logarithmic corrections. The result for the largest $|\omega_k|$ agrees again with the analytic continuation of the Matsubara expressions of Ref.~\cite{klein2020}.  We also notice that the $T/|\omega_k|$ tail is not restricted by the non-FL scaling arguments and thus extends beyond the frequency cut-off $\wmax$.
As already mentioned, the numerically obtained self-consistent solution for $\myim \Sigma^R_{F,T}$ indeed follows these asymptotic expression as is shown in Fig.~\ref{fig:SFT}.  
Regarding the numerical evaluation we have not properly considered the nonzero real part of the thermal component. However, in App.~\ref{sec:KK} we argue that these corrections vanish for small $\alpha, T$ in the non-FL regime and that neglecting them is consistent with the structure of the Eliashberg equations. In addition, we have checked that the numerical stability against small perturbations that mimic corrections to the real part.

Eq.~\eqref{eq:resSFTsum} reveals that a full understanding of the thermal component requires the bosonic mass $M(T)$ as input parameter.
Before embarking on the computation of $M(T)$, we briefly consider the momentum dependence of $\Sigma^R_{F,T}$. This is necessary to estimate the upcoming integrals. We can parametrize any momentum $\mathbf k$ close but not exactly on the FS by a small deviation $\delta\mathbf{k} = \mathbf k - \mathbf k_F $ that is parallel to $\mathbf k_F$ (i.e. $\varphi_{\mathbf k} = \varphi_{\mathbf k_F}$) and satisfies $|\delta \mathbf k | \ll k_F$. To compute $\myim \Sigma^R_{F,T}(\mathbf k ,\omega)$ in analogy to Eq.~\eqref{eq:SFTint3} we have to insert the fermionic Green's function
\begin{align}\label{eq:GRp}
\begin{split}
& G^R(\mathbf k - \mathbf p,\omega_k-\omega_p ) \simeq\\
& \frac{1}{\omega_k-\omega_p-\Sigma^R_F(k-p)-v_F p \cos(\varphi_{\mathbf k}-\varphi_{\mathbf p})-v_F \delta k } \, ,
\end{split}
\end{align}  
where we have linearized the dispersion as usual. Following the same steps for the $\omega_p$ integration as above and using that $G^R(k-p)$ remains an analytic function of $\omega_p$ in the lower complex half plane, we find 
\begin{align}\label{eq:SFTint4}
\begin{split}
\myim \Sigma^R_{F,T}(\mathbf k,\omega_k) =  g^2  \dkFq T \myim \int \frac{d^2 p}{(2\pi)^2}\frac{1}{c_B^2 p^2 + M^2} \\
 G^R_F(\mathbf k- \mathbf p,\omega_k) \, .
\end{split}
\end{align}
Here, we have approximated the d-wave form factor by neglecting corrections from both $\delta\mathbf{k}$ and the bosonic momentum $\mathbf p$. The determination of $\myim \Sigma^R_{F,T}(\mathbf k, \omega_k)$ for arbitrary momenta and frequencies actually requires a full numerical solution of Eq.~\eqref{eq:SFTint4} with the Green's function~\eqref{eq:GRp}.
However, the value $\Sigma^R_{F,T}(\mathbf k_F , \omega \to 0) = - \Gamma_{F,\mathbf k_F}(T)$ must still be approached in the limit $\delta k \to 0, \omega_k \to 0$.
Moreover, asymptotically large arguments allow to replace $\Sigma^R_F(k-p) \to \Sigma^R_{F,Q}(k-p) \simeq \Sigma^R_{F,Q}(\omega_p - \omega_k,\varphi_{\mathbf k_F})$
in analogy to the arguments given below Eq.~\eqref{eq:SFTasy}. As a result of the momentum-independent quantum self-energy, we can solve the $\varphi_{\mathbf p}$ integral just as in Eq.~\eqref{eq:SFTint3}. Afterwards, we perform the $p$ integration like in Eq.~\eqref{eq:SFTfin}. This leads to the asymptotic expression~\eqref{eq:SFTasy}, yet with the dispersive shift $\omega_k \to \omega_k- v_F \delta k$:
\begin{align}\label{eq:SFTasyfull}
\begin{split}
&\myim \Sigma^R_{F,T}(\mathbf k, \omega_k)  \simeq  \\
&\frac{\alpha \dkFq T}{4 A^2 \varepsilon_F}\myim \! \left[\dfrac{2 \log \left(\! 2\dfrac{\omega_k- \Sigma^R_{F,Q} (\omega_k, \varphi_{\mathbf k_F})-v_F \delta k}{M/A} \right)\!-i \pi}{\omega_k- \Sigma^R_{F,Q}(\omega_k,\varphi_{\mathbf k_F})-v_F \delta k} \right] \! .
\end{split}
\end{align} 
At $\delta \mathbf k = 0 $ we have seen that
$\myim \Sigma^R_{F,T}$ approaches the constant value $\Gamma_{F,\mathbf{k}_F}(T)$ for small frequencies whereas at higher frequencies this plateau crosses quickly over to algebraic tails. For finite deviations away from $\mathbf k_F$, we find the crossover condition~\eqref{eq:SFTcross} but again shifted by $\omega_k \to \omega_k -v_F \delta k$:
\begin{align}\label{eq:SFTcrossfull}
\begin{split}
\frac{1}{M}\sim\left|\myim\dfrac{2\log \left(2\dfrac{\omega_k- \Sigma^R_{F,Q} (\omega_k, \varphi_{\mathbf k_F})-v_F \delta k}{M/A} \right)-i \pi}{\omega_k- \Sigma^R_{F,Q}(\omega_k,\varphi_{\mathbf k_F})-v_F \delta k}\right| .
\end{split}
\end{align}  
Regarding the range-of-validity of the asymptotic forms of $\myim \Sigma^R_{F,T}(\mathbf k, \omega_k)$ as function of the momentum, we expect that $\myim \Sigma^R_{F,T}(\mathbf k, \omega_k)$ decays quickly if the linearization $\ek -\mu \simeq v_F \delta k  $ fails to be accurate. As discussed below Eq.~\eqref{eq:defcut}, the corresponding scale exceeds the cut-off $\Lambda$ from the interaction-based INM scaling relations for small enough $\alpha$.
\subsection{Nematic susceptibility}\label{sec:bosMass}
The mass parameter $M^2(T)$ introduced in Eqs.~\cref{eq:ansDR,eq:m2def} is encoded in $\myre \Sigma^R_B(0)$. 
To obtain an explicit expression for the real part of the self-energy, we insert $\myim \Sigma^R_B$ from Eq.~\eqref{eq:defSB} into the Kramers-Kronig relation~\eqref{eq:defKK} and apply the latter to $\myim G^R$. This yields
\begin{align}\label{eq:SBRe1}
\begin{split}
\myre \Sigma^R_B(0)   = -8g^2\! \!\int\!\! \frac{d^2k \,d\omega_k}{(2\pi)^3} \dkq n_F(\omega_k)\myim G^R(k)\! \myre G^R(k) .
\end{split}
\end{align}
Given this expression, we can consider the connection between $M^2(T)$ and the compressibility $(\partial n/\partial \mu)_T$, introduced below Eq.~\eqref{eq:m2def}, in further detail. 
First, we note that $\myre \Sigma^R_B(0)$ can be rewritten as 
\begin{align}\label{eq:SBRRe2}
\begin{split}
\myre \Sigma^R_B(0) 
= 4 g^2 \int \frac{d^2k \,d\omega_k}{(2\pi)^3} \dkq \, n_F(\omega_k) \frac{\partial \myim G^R(k)}{\partial\mu}
 \, ,
\end{split}
\end{align}
when the dressed Green's function~\eqref{eq:GRex} is inserted, since $\Sigma^R_F$ is independent of $\mu$ within ET. 
The total density $n$ reads quite generally~\cite{fetter2003Book}
\begin{align}
n= -2\int \frac{d^2 k}{(2\pi)^2} \int \frac{d \omega_k}{\pi} n_F(\omega_k) \myim G^R(k) \, , 
\end{align}
with the factor 2 for spin.
This implies
\begin{align}\label{eq:comp}
\left(\frac{\partial n}{\partial \mu}\right)_T =-4 \int \frac{d^3 k}{(2\pi)^3} \int \frac{d \omega_k}{(2\pi)} n_F(\omega_k)\frac{\partial \myim G^R(k)}{\partial \mu}  \, ,
\end{align}
which differs from Eq.~\eqref{eq:SBRRe2} only by factors of the coupling constant and the accompanying angular weight from the d-wave form factor. The opposite sign arises from the Fermion loop that is taken into account in $\Sigma^R_B$.
Consequently, Eliashberg theory incorporates the thermodynamic relation between the fluctuations of the conserved electron density and the compressibily correctly, as is expected for a self-consistent quantum field theory on general grounds~\cite{baym1961}. Moreover, both the invers order-parameter susceptibility $M^2(T)$ and the compressibility $(\partial n/\partial \mu)_T-(\partial n/\partial \mu)_{T=0}$ scale in the same way with $T$, since the nematic form factor does not introduce singularities. Note that we have shown this relation for the dressed Green's function but the same statements hold if $G^R_0=(\omega-\ek + \mu +i 0^+)$ is used instead.
However, there is a marked difference between evaluating $(\partial n/\partial \mu)_T$ with bare or interacting Green's functions: 
For the spherical FS of noninteracting electrons one finds in $d=2$
\begin{align}\label{eq:compBare}
\begin{split}
\left(\frac{\partial n^{(0)}}{\partial \mu}\right)_T & =2\frac{\partial}{\partial \mu} \int \frac{d^2 k\, d\omega_k}{(2\pi)^2} n_F(\omega_k)\delta(\omega_p -\ek +\mu)\\& = 2\frac{\partial}{\partial \mu} \int \frac{d^2 k}{(2\pi)^2} n_F(\ek-\mu) \\
& = 2\frac{\partial}{\partial \mu} \frac{m T}{2\pi} \log(1+e^{\beta \mu})=\frac{m}{\pi} \frac{1}{1+e^{-\beta \mu}} \, ,
\end{split}
\end{align} 
which is equivalent to the constant density of states $D(\varepsilon) = m/\pi$ and exponentially small thermal corrections. The arguments from the previous paragraph entail the that the same behavior arises from the perturbative evaluation of the one-loop diagram for $M^2(T)$.  
When the band structure of the underlying lattice is taken into account, the Sommerfeld expansion for small temperatures gives rise to analytic corrections $\sim D'(\eF) T^2$ to the ground-state result. Even in a FL one finds the same temperature dependence since the finite decay rate of the low-energy quasiparticles, which scales like~\cite{baym2008landau} $\sim T^2$ with logarithmic corrections~\cite{bloom1975,chubukov2005FL} in $d=2$, vanishes therefore much faster than thermal broadening $\sim T$ of the distribution function $n_F$. The effect of interactions merely introduces a renormalization of the density of states. In contrast, the non-quasiparticle character of the excitations in the QCR, which is incorporated by the branch cuts of $G^R$, changes the structrue of the integrals Eq.~\cref{eq:SBRRe2,eq:comp} substantially: In the non-FL one faces broad spectral features of a finite width around the FS, instead of sharp spectral functions that essentially focus the evaluation on the FS. As a result, in the quantum critical case this spectral width $\wmax$ and the associated range in momentum space $\Lambda$, which correspond to a new, emergent energy and lenght scales, explicitly appear in the final results. 
Therefore, UV/IR mixing cannot be avoided at finite temperatures, in contrast to the ground state where the single parameter $\myre \Sigma^R_B(0)$ is absorbed in the definition of the critical point (see Eq.~\eqref{eq:renorm}). Similar effects of UV/IR mixing appear also at $T=0$, e.g. in the case of higher dimensions when the extended structure of the hot parts of the FS has to be taken into account to describe the interactions properly~\cite{mandal15}. 
Considering both the low- and high-energy degrees of freedom on equal footing is certainly beyond the scope of an analytic apporach. 
However, the estimates for $\wmax$ and $\Lambda$ provided by ET allow to obtain qualitative results for $M^2(T)$. Furthermore, the scaling with the UV parameters can be compared to numerical simulations.
 
To compute $M^2(T)$, we return now to Eq.~\eqref{eq:SBRe1}. Using the identity $n_F(-\omega) = 1- n_F(\omega)$ we can decompose the bosonic mass~\eqref{eq:m2def} into contributions from positive and negative frequencies
\begin{widetext}
\begin{align}\label{eq:defm12}
\begin{split}
M^2(T) &= M_-^2(T) + M^2_+(T)\\
&= - 8 g^2 \int \frac{d^2 k }{(2\pi)^2} \dkq 
\left[\int_{-\infty}^0 \frac{d \omega_k}{2 \pi} \bigg(\left[\myim G^R(k) \myre G^R(k)\right]_T -  \left[\myim G^R(k) \myre G^R(k)\right]_{T=0} \bigg) \right.
\\
&\qquad \qquad\qquad\qquad\quad\, \left.  + \int_0^{\infty} \frac{d\omega_k}{2\pi}  n_F(\omega_k)\bigg(\left[\myim G^R(k) \myre G^R(k)\right]_T  - \left[\myim G^R(\mathbf k,-\omega_k) \myre G^R(\mathbf k,-\omega_k)\right]_T \bigg) \right] \, .
\end{split}
\end{align}
\end{widetext}
In the absence of well-defined quasiparticles, one expects 
the spectral width of the excitations to exceed the temperature, which is indicated by the thermal damping rate $\Gamma_F(T)$ that vanishes more slowly than $T$, too. As a consequence, the major contribution to the inverse order parameter susceptibility will arise from $M_-^2(T)$ since in $M_+^2(T)$ frequencies above $T$ are exponentially supressed by $n_F$. 

For the explicit calculation, we will assume that the non-FL correlations are restricted to the maximal range in frequencies $|\omega| \lesssim \wmax$ and the corresponding maximal momentum cut-off $v_F \Lambda \leq h_{\max} \eF$, given in Eqs.~\cref{eq:defcut,eq:deffullcut}. The analogous computation within the scheme of minimal cutoffs is presented in App.~\ref{sec:minCut}. However, we emphasize that the physical picture relies only on the existence of non-FL correlations at finite temperatures. In particular, the choice of the cutoffs does not affect the scaling with $T$. 
The following procedure consists of three steps: In Sec.~\ref{sec:MassQ} we compute $M^2(T)$ by considering only the quantum part $\Sigma^R_{F,Q}$ to find a preliminary result for the bosonic mass. Afterwards in Sec.~\ref{sec:domScal}, we use the latter to obtain the corresponding thermal part $\Sigma^R_{F,T}$ and determine the regime of $\omega/T$ scaling. Finally, we feed the total self-energy back into the Eliashberg equations in Sec.~\ref{sec:feedback} and show that the inclusion of $\Sigma^R_{F,T}$ yields only subleading corrections to $M^2(T)$ at the onset of finite temperatures. For larger $T$, however, these eventually dominate the inverse susceptibility and provide the result $M^2(T) \sim T$ with logarithmic corrections.  

\subsubsection{Mass gap from $\Sigma^R_{F,Q}$}\label{sec:MassQ}
In this section we compute the bosonic mass by setting $\Sigma^R_F(k) \to \Sigma^R_{F,Q}(\omega_k,\varphi_{\mathbf k})$ that is to a very good approximation independent of the magnitude of the momentum (see Sec.~\ref{sec:SFQ}).
Labeling the corresponding bosonic self-energy as $\Sigma^R_{B,Q}$ and inserting the explicit form of the Green's functions into Eq.~\eqref{eq:SBRe1} yields
\begin{align}\label{eq:SBRRe}
\begin{split}
&\myre \Sigma^R_{B,Q}(0)\simeq -8 g^2 \!  \int_0^{2\pi} \!\frac{d\varphi_{\mathbf k}}{2\pi}\!\int_{-\wmax}^{\wmax} \frac{d\omega_k}{2\pi}\! \int^{k_F+\Lambda}_{k_F - \Lambda} \frac{dk}{2 \pi} k \, \dkq \\
&\frac{n_F(\omega_k)\myim \Sigma^R_{F,Q}(\omega_k, \varphi_{\mathbf k})(\omega_k -\myre \Sigma^R_{F,Q}(\omega_k,\varphi_{\mathbf k})-\xi_k)}{\left[(\omega_k -\myre \Sigma^R_{F,Q}(\omega_k,\varphi_{\mathbf k})-\xi_k)^2+(\myim \Sigma^R_{F,Q}(\omega_k, \varphi_{\mathbf k}))^2\right]^2} \, ,
\end{split}
\end{align}
with $\xi_k = \ek -\mu$.
After linearizing the fermionic dispersion in the variable $u=v_F(k-k_F)$ as usual, we can integrate $u$ exactly
\begin{align}\label{eq:SBRRe3}
\begin{split}
&\myre \Sigma^R_{B,Q}(0)= 
-2 \alpha \!\int_0^{2\pi} \!\frac{d\varphi_{\mathbf k}}{2\pi}\int_{-\wmax}^{\wmax} \!\frac{d\omega_k}{2\pi} \sum_{\sigma =\pm 1}\dkq  n_F(\omega_k)\\
 & \frac{ \sigma \myim \Sigma^R_{F,Q}(\omega_k, \varphi_{\mathbf k})}{(\omega_k -\myre \Sigma^R_{F,Q}(\omega_k,\varphi_{\mathbf k})-\sigma v_F \Lambda)^2+(\myim \Sigma^R_{F,Q}(\omega_k, \varphi_{\mathbf k}))^2} .
\end{split}
\end{align}
Notice that sending $\Lambda \to \infty$ implies $\myre \Sigma^R_B(0)=0$ at all $T$ and thus $M^2(T) \equiv 0$. Taking this limit means that the effective low-energy theory with the linearized dispersion is extended over all energy scales. 
Therefore, ET captures the constraints by the emergent gauge symmetry of the two-patch model~\cite{metl10,mross2010,dalidovich2013}, which imply that the bosonic mass vanishes, unless the UV physics is taken into account~\cite{hartnoll2014}.

Let us focus now on $M_{-,Q}(T)^2$, obtained by evaluating the definition of the bosonic mass~\eqref{eq:defm12} with $\myre\Sigma^R_B$ from Eq.~\eqref{eq:SBRRe3}:
\begin{widetext}
\begin{align}\label{eq:m-q}
& M^2_{-,Q}(T) =-2 \alpha \!\int \!\frac{d\varphi_{\mathbf k}}{2\pi}
 \int_{-\wmax}^0 \!\frac{d\omega_k}{2\pi} \sum_{\sigma =\pm 1} \left[ \left. \dkq\frac{ \sigma \,\myim \Sigma^R_{F,Q}(\omega_k, \varphi_{\mathbf k})}{(\omega_k -\myre \Sigma^R_{F,Q}(\omega_k,\varphi_{\mathbf k})-\sigma v_F \Lambda)^2+(\myim \Sigma^R_{F,Q}(\omega_k, \varphi_{\mathbf k}))^2}\right|_T\!\! - (T \to 0) \right]\, .
\end{align}
\end{widetext}
In the following, we will assume that $\Sigma^R_{F,Q}$ can be evaluated in the limit $M \to 0$, or more precisely $\bar M \ll 1$, see Eq.~\eqref{eq:defMbar}. This simplification will be justified within our self-consistent solution below. 
To solve the integral over frequencies, we split it in two parts $|\omega_k| \lesssim T$ and $|\omega_k| \gtrsim T$, where the precise value of the prefactor is irrelevant in the limit $T \to 0$. For $|\omega_k| \lesssim T$, we rescale $\bar \omega_k = \omega_k / T$ and expand to linear order in $\omega_k - \myre \Sigma^R_{F,Q}$ due to the prerequiste $|\omega_k - \myre \Sigma^R_{F,Q}| \ll v_F \Lambda$ for non-FL correlations. This results in a contribution that scales like $T^{7/3}$. 
Instead, the more important contribution arises from $|\omega_k| \gtrsim T$, where the self-energy acquires the form
$$\myim\Sigma^R_{F,Q}  = \myim \left. \Sigma^R_{F,Q}\right|_{T=0} + \delta \myim \Sigma^R_{F,Q}\, , $$
see also Eq.~\eqref{eq:defSFQdelta}. In this regime $\delta \myim \Sigma^R_{F,Q}(\varphi_{\mathbf k}) = 0.318919...|\dkF|^{4/3}  \alpha^{2/3}T^{2/3}/(A^{4/3}\eF)$ can be considered as a small correction to the ground state self-energy (cf. Eqs.~\eqref{eq:resSFQ1}, \eqref{eq:resSFQdelta} and Fig.~\ref{fig:deltaSigma} for details). As a result, we can expand the integrand both in $ \myim \delta\Sigma^R_{F,Q}$ and then  in $\omega_k - \myre \Sigma^R_{F,Q}$. Retaining only the most important term yields
\begin{align}\label{eq:MQdom}
\begin{split}
M^2_{-,Q}(T) =
-2 \alpha \!\int_0^{2\pi} \!\frac{d\varphi_{\mathbf k}}{2\pi}\int_{-\wmax}^{-T} \!\frac{d\omega_p}{2\pi} \myim \delta \Sigma^R_{F,Q}(\varphi_{\mathbf k}) \\
\times \sum_{\sigma =\pm 1} \frac{ 2 \sigma d^2_{\mathbf k} \omega_p}{(\sigma v_F \Lambda)^3}\, .
\end{split}
\end{align}  
After inserting $\myim \delta\Sigma^R_{F,Q}$ and
the values $\wmax = h_\omega \sqrt{\alpha}$ and $v_F \Lambda = h_{\max} \eF=h_\omega^{1/3} \alpha^{1/2}$, defined in Eq.~\eqref{eq:deffullcut}, we obtain
\begin{align}\label{eq:resM2-}
M^2_{-,Q}(T) \simeq 0.082417... h_\omega \frac{\alpha^{7/6} T^{2/3}}{\eF} \, .
\end{align}
where the angular integral has been computed numerically. Note that the leading contribution to the $\omega_p$ integral arises from the lower boundary at $-\wmax$ such that the precise choice of the upper boundary is indeed irrelevant. In the corresponding expression for $M^2_{+,Q}$ from Eq.~\eqref{eq:defm12} the integrand  is exponentially supressed by the Fermi function at frequencies of order $\wmax$. Furthermore, the regime $\omega_k \leq T $ merely yields a negligible contribution analogously to the case for $M^2_{-,Q}$. In agreement with our previous statement the leading behavior is given by
\begin{align}
M^2_Q(T) = M^2_{-,Q} (T) \, .
\end{align} 
Several commments on the last two equations are in order: First of all, we note that the scaling with $T^{2/3}$ has been generated by the non-FL correlations in the quantum component. They are intimately connected to the $\omega^{2/3}$ power law which incorporates the nonquasiparticle excitations of the non-FL. As a result, including the latter in the Eliashberg equations gives rise to a scaling law that supersedes the estimate $T \log T$ at small $T$. According to the connection of $M^2(T)$ and the thermal contribution to the compressibility, we have $(\partial n/\partial_\mu)_T -(\partial n/\partial_\mu)_{T=0} \sim T^{2/3}$, too. 
Finally, the nonuniversal character of $M_Q^2(T)$ requires to extract the material parameters $h_\Lambda $ and $h_\omega$ from the interplay of the non-FL correlations and the band structure for each system individually, in order to obtain a more quantitative results. Nevertheless, our approach provides a first estimate for the static nematic susceptibility and in particular allows to test the scaling with $\alpha$. In this regard, we also point out that 
despite the condidition $h_\omega \ll 1$, the numerical prefactor of $M$ is not necessarily extremely small: for instance $h_\omega =10^{-1}$ implies $\sqrt{0.082417... h_\omega} \simeq 0.09$.
\subsubsection{Scaling relations from $\myim \Sigma^R_{F,Q}$}\label{sec:domScal}
\begin{figure}[t]
\begin{center}
\includegraphics[width=\columnwidth]{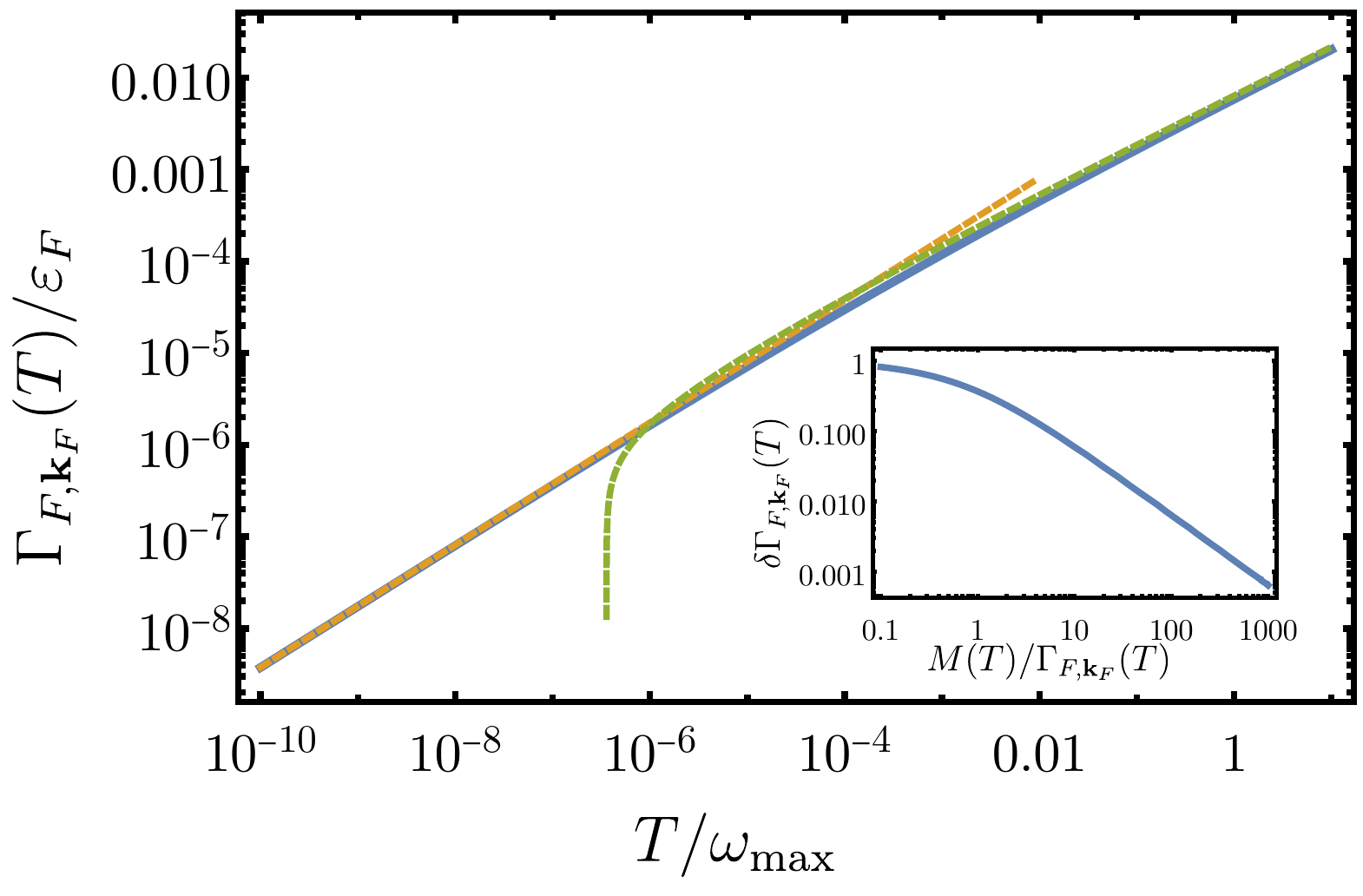}
\caption{Blue: Self-consistent thermal damping rate from Eq.~\eqref{eq:resGammaF} with parameters $\alpha/\eF =10^{-2}$, $h_\omega=10^{-1}$, $A=1$ and $\varphi_{\mathbf k_F}= \pi/10$; Yellow dashed: Expansion for $M(T) \gg \Gamma_{F, \mathbf k_F}(T)$ from Eq.~\eqref{eq:resGammaFQCR} that becomes exact at the lowest temperatures; Green dashed: Expansion around the opposite limit $M(T) \ll \Gamma_{F, \mathbf k_F}(T)$ from Eq.~\eqref{eq:resGammaFcc} for the behavior at larger temperatures;
Inset: relative deviation between the full result and the low-temperature asymtptotics: For $\delta \Gamma_{F, \mathbf k_F} \simeq 0.1$ a ratio $M(T)/\Gamma_{\mathbf k_F} \geq 10$ is required while an error below one percent is obtained only if  $M(T)/\Gamma_{F,\mathbf k_F} \simeq 100$.}
\label{fig:MvsGamma}
\end{center}
\end{figure}

Let us now determine on the basis of the previous section how the thermal decay rate and the various crossovers of the INM scale with $T$.
Calculating the numerical prefactors of order one with high precision is beyond the scope of our ansatz, which is rather focused on the dominant scaling behavior, Therefore, we will omit them whenever possible. Instead, we concentrate on the dependence on $T, \alpha$ and $\eF$ but keep $h_\omega \ll 1$.
First, we find from Eq.~\eqref{eq:resM2-}
\begin{align}\label{eq:resMfin}
M(T) \sim h^{1/2}_\omega\frac{\alpha^{7/12} T^{1/3}}{\eF^{1/2}} \, ,
\end{align} 
where we supress the index $Q$ because it will turn out to be equivalent to the total result at the onset of finite temperatures. The corresponding dimensionless mass $\bar M \sim  M/(\alpha T)^{1/3}$ from Eq.~\eqref{eq:defMbar} satisfies $\bar M \sim h^{1/2}_\omega (\alpha^{1/2}/\eF)^{1/2}\ll 1$ such that FL correlations do not play a role, according to Sec.~\ref{sec:SFQ}. The numerically evaluated, self-consistent thermal damping rate from Eq.~\eqref{eq:resGammaF} is presented in Fig.~\ref{fig:MvsGamma}. In particular, for the smallest temperatures the expansion of Eq.~\eqref{eq:resGammaFQCR} holds and $\Gamma_F$ becomes
\begin{align}\label{eq:resGammaFfin}
\Gamma_F(T) \sim h^{-1/2}_\omega\frac{\alpha^{5/12} }{\eF^{1/2}} T^{2/3} \, .
\end{align}
Thus $M(T) \gg \Gamma_F(T)$ in the limit $T \to 0$, such that the necessary condition for the application of Eq.~\eqref{eq:resGammaFQCR} is fulfilled. Note that the temperature dependence $\Gamma_F(T) \sim T^{2/3}$ is a prerequiste for $\omega/T$ scaling of the fermionic self-energy. Here, it originates from the ratio $\Gamma_F(T) \sim T/M(T)$ and $M^2(T) \sim T^{2/3}$ which is generated by the thermal corrections to the non-FL correlations as we have seen in the previous section.
However, the inset of Fig.~\ref{fig:MvsGamma} shows that a rather large rato $M/\Gamma_F \geq 10 -100$ is required to describe the exact result with an error between $10\%$ and $1\%$, respectively. Below we see that this fact limits the range-of-validity of this scaling relation.
As a consequence of Eq.~\eqref{eq:resGammaFfin}, the damping rate overcomes the positive contribution from the quantum part $\myim \Sigma^R_{F,Q}(\mathbf k_F ,0) \sim \alpha^{2/3} T^{2/3}/\eF$, in the limit $\alpha/\eF^2 \ll 1$ such that the positive-definiteness of the spectral function is restored. Furthermore, Fig.~\ref{fig:sumSF} shows that the dominant behavior of the total self-energy $\myim \Sigma^R_{F} = \myim \Sigma^R_{F,Q} + \myim \Sigma^R_{F,T}$ is very well approximated by the simple form
\begin{align}\label{eq:resSFtot1}
\myim \Sigma^R_{F} (\mathbf{k}_F, \omega_k,T) \simeq - \Gamma_{F, \mathbf k_F}(T) - \win^{1/3} |\dkF|^{4/3} |\omega_k|^{2/3}    
\end{align} 
at sufficiently small temperatures: As can be seen in Fig.~\ref{fig:sumSF}, this expression holds if the plateau $\myim \Sigma^R_{F,T} (\mathbf k_F,\omega_k \to 0) \sim -\Gamma_F$ crosses first the non-FL tail $\sim |\omega_k|^{2/3}$ of $\myim \Sigma^R_{F,Q}$ with increasing $\omega_k$, before the power-law asymptotics of $\myim \Sigma^R_{F,T} (\mathbf k_F, \omega_k \to \infty)$ sets in. Provided that this is the case, we can express $\myim \Sigma^R_{F}$ in terms of a scaling function by combining the last two equations with Eqs.~\cref{eq:resGammaFQCR,eq:resM2-}:
\begin{figure}[t]
\begin{center}
\includegraphics[width=\columnwidth]{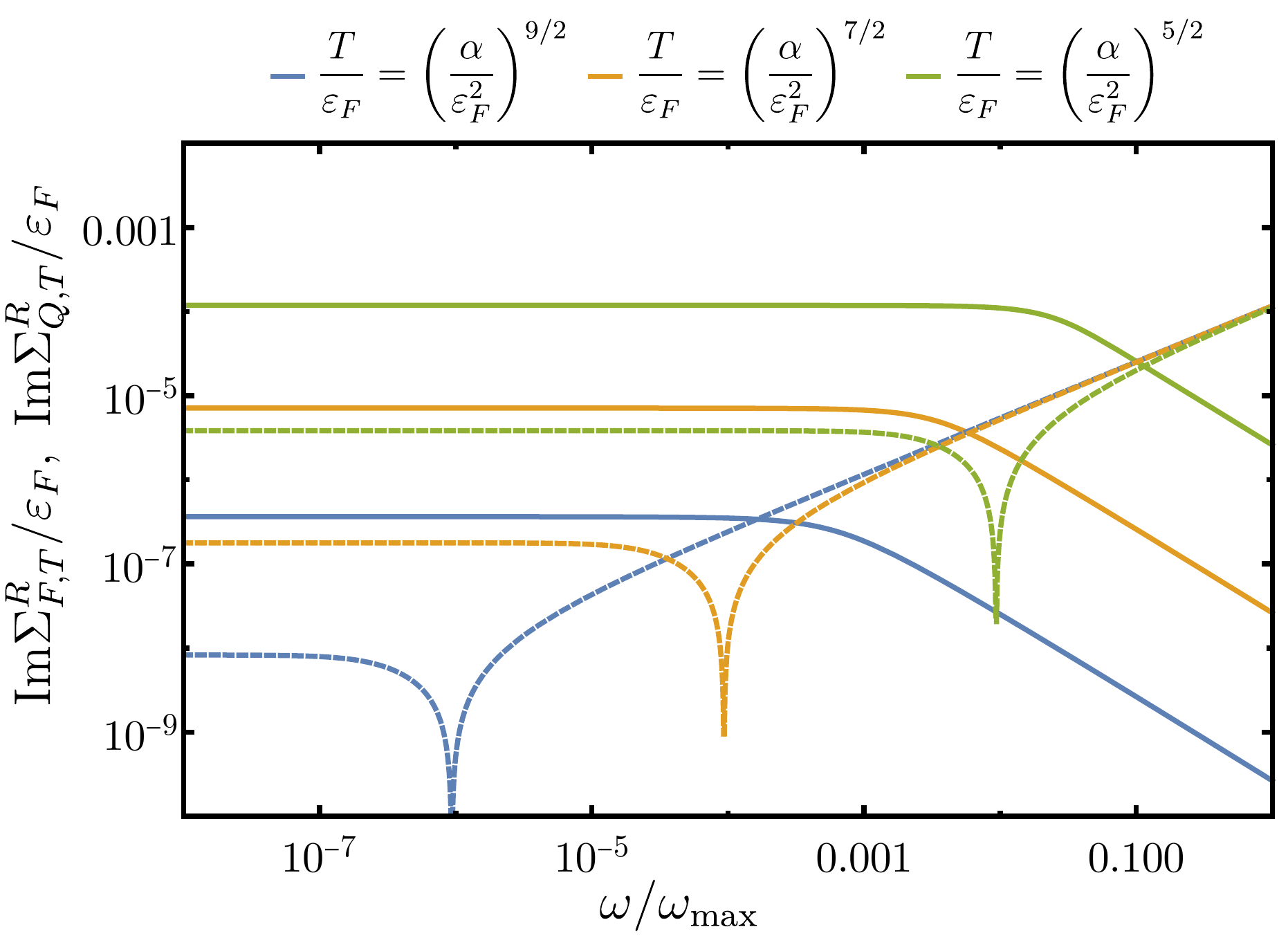}
\caption{Thermal component $\myim \Sigma^R_{F,T}(\mathbf k_F,\omega_k)$ (solid lines) and quantum component $\myim \Sigma^R_{F,Q}(\mathbf k_F,\omega_k)$ (dashed lines) for three different temperatures. The other parameters are $\alpha/\eF^2 = 10^{-2}$, $h_\omega = 10^{-1}$, $A=1$ and $\varphi_{\mathbf k_F} = \pi/10$. In case of the smallest temperature, the plateau of $\myim \Sigma^R_{F,T}$ at small frequencies crosses the $|\omega_k|^{2/3}$ non-FL asymptotics of $\myim \Sigma^R_{F,Q}$ at a frequency smaller than the threshold for the aysmptotic $|\omega_k|^{-1}$ power-law of the thermal component. In contrast, at higher temperatures an intermediate frequency regime, which is dominated by this asymptotic tail, emerges.}
\label{fig:sumSF}
\end{center}
\end{figure} 
\begin{widetext}
\begin{align}\label{eq:resSFtot2}
\begin{split}
\myim \Sigma^R_{F} (\mathbf{k}_F, \omega_k,T,h_\omega)  & =  - |\dkF|^{4/3} \frac{\alpha^{5/12 }}{\eF^{1/2}}T^{2/3} \tilde{\Sigma}^R_F\left(\frac{\omega_k}{T},\varphi_k,h_\omega\right) \\
& = - |\dkF|^{4/3} \frac{\alpha^{5/12 }}{\eF^{1/2}}T^{2/3}\left(1.367... h^{-1/2}_{\omega} |\dkF|^{2/3} + \frac{ \alpha^{1/4}}{8 \cdot 3^{1/2} A^{4/3} \eF^{1/2}}\left(\frac{|\omega_k|}{T} \right)^{2/3}\right) \, .
\end{split}
\end{align}
\end{widetext}
$\tilde{\Sigma}^R_F$ is represented in Fig.~\ref{fig:scalCol} where we indeed observe a scaling collapse at the smallest temperatures. 
We emphasize again that the prefactors are nonuniversal but depend on the spectral width of the non-FL excitations which is encoded in the UV-scale $h_\omega$. 
Furthermore, the $\omega_k/T$-scaling form $\tilde{\Sigma}^R_F$ admits only a single crossover between the low- and the high-frequency regime.
As a result, $\omega_{\lessgtr}$, defined in Eq.~\eqref{eq:defSFQ}, coincide and can be estimated as follows
\begin{align}\label{eq:wcross}
\Gamma_F(T) \sim \win^{1/3} |\omega_\lessgtr|^{2/3} \Rightarrow \omega_\lessgtr \sim  h_{\omega}^{-3/4}\frac{\eF^{3/4} }{\alpha^{3/8}} T   \, .
\end{align}  
As expected from $\omega/T$ scaling, the relation is linear in $T$ but acquires a nonuniversal prefactor, which is much larger than one, as a consequence of Eq.~\eqref{eq:resSFtot2}. In fact, the small numerical prefactor of $\win$ increases $\omega_{\lessgtr}$ even further as can be seen in Fig.~\ref{fig:scalCol}. 

Finally, these results allow to elaborate on the boundaries for the existence of $\omega/T$ scaling. First of all, it can only emerge if $\Gamma_F \ll M(T)$ since otherwise $\Gamma_F(T)$ follows either from Eq.~\eqref{eq:resGammaFcc} or from a more complicated crossover behavior of the full equation~\eqref{eq:resGammaF}. Then $\Gamma_F(T)$ does not scale in the same way with temperature as the ground-state self-energy does with frequencies and $\omega/T$ scaling is ruled out. However, the expansion~\eqref{eq:arcsecExp} around $z=0$ has only a small radius of convergence such that the first condition for $\omega/T$ scaling reads 
\begin{align}\label{eq:ovT1}
\frac{M(T)}{\Gamma_F(T)} \gtrsim B \, ,
\end{align}
with $B \simeq 100$. This value is required to obtain an error of approximately one percent in the determination of $\Gamma_F(T)$ via Eq.~\eqref{eq:resGammaFQCR}, as can be seen in the inset of Fig.~\ref{fig:MvsGamma}. 

In addition, there is another mechanism that leads to the destruction of $\omega/T$ scaling: As is presented in Fig.~\ref{fig:sumSF}, an intermediate regime that is governed by the asymptotic $|\omega_k|^{-1}$ tail~\footnote{The asymptotics $\sim |\omega_k|^{-2/3}$ 
(see Eq. \cref{eq:resSFTsum}) 
caused by the non-FL correlations is always preempted by the crossover to 
$\myim \Sigma^R_{F,Q}$.
}
of $\myim \Sigma^R_{F,T} $ emerges at sufficiently high temperatures.
In particular, the crossover from the $\Gamma_F(T)$ to this tail is determined by $M^{-1}(T) \sim |\omega_k|^{-1}$, as follows from Eq.~\eqref{eq:SFTcross}. Consequently, the intermediate frequency regime is visible in the total self-energy if the resulting crossover preempts the crossover to the non-FL asymptotics, which is encountered at the scale $\omega_{\lessgtr}$ from Eq.~\eqref{eq:wcross}. This allows to formulate a second criterion for the existence of $\omega/T$ scaling:  
\begin{align}\label{eq:ovT2}
\frac{1}{\omega_{\lessgtr}} \gg \frac{1}{M(T)} \, .
\end{align}

Inserting the results for $M^2(T)$ from Eq.~\eqref{eq:resM2-}, $\Gamma_F(T)$ from Eq.~\eqref{eq:resGammaFfin} and $\omega_{\lessgtr}$ from Eq.~\eqref{eq:wcross} into the last two equations, allows to derive two upper temperature boundaries for the existence of $\omega/T$ scaling. Finally, their minimum sets the scale $T_{\text{scal}}$ below which $\Sigma^R_{F}$ is given by the scaling form of Eq.~\eqref{eq:resSFtot2}
\begin{align}\label{eq:defTscal}
T_{\text{scal}} = 
\begin{cases}
h_\omega^{15/8} \left(\dfrac{\alpha^{1/2}}{\eF}\right)^{23/8} \eF \, ,&\text{if } \dfrac{\alpha}{\eF^2} \ll \dfrac{h_\omega^{6/5}}{B^{16/5}} \\
B^{-3} h_\omega^3 \alpha^{1/2}\, ,&\text{if } \dfrac{\alpha}{\eF^2} \gg \dfrac{h_\omega^{6/5}}{B^{16/5}} 
\end{cases} \, .
\end{align}
The first form, which applies to the smallest couplings, results from Eq.~\eqref{eq:ovT2} while the second one is obtained from Eq.~\eqref{eq:ovT1}.
Due to the (possibly) small prefactor, it is quite likely in both cases that $T_{\max}$
is comparable or even below the critical temperature~\cite{metlitski2015} $T_c \sim \win$ for the transition to the superconducting state. In spite of completely discarding pairing fluctuations in ET, our analysis may still contain relevant physical information at these small temperatures: In the presence of a finite superconducting gap $\Delta$ one expects that the Matsubara self-energy $\Sigma_F(i \omega_n)$ aquires a contribution $\Delta^2/(i \omega_n) \to \Delta^2/(\omega + i 0^+)$, which gives rise to a pronounced upturn of the self-energy at the lowest frequencies. Such an upturn is indeed observed in QMC simulations~\cite{lederer17QMC,berg19QMCrev,xu2020}. As a consequence, it is very plausible that the spectral functions at larger frequencies remain governed by the non-FL correlations considered by ET. 

We emphasize again that this analysis has been presented under the assumption that the thermal component $\Sigma^R_{F,T}$ does not alter the dominant scaling behavior of $M^2(T)$. This is confirmed in the next section.  

\subsubsection{Feedback of $\Sigma^R_{F,T}$}\label{sec:feedback}
Finally, we have to include the thermal component of the self-energy in the self-consistent computation of $M^2(T)$. To show that its effect is negligible, it suffices to insert the results from the previous section to check that only subleading corrections are generated. In contrast, if the dominant scaling behavior was changed, we would have to restart the calculation with both components treated on equal footing. Before embarking on the detailed calculations to confirm that this is not necessary, we give a simple argument: The corrections $\myim \delta \Sigma^R_{F,Q}$ extend over the same region in $(\omega_k ,\mathbf k)$ space as the non-FL correlations in the ground state, except for very small frequencies $|\omega_k| \lesssim T$. As a result, $M^2_{-,Q}$ essentially inherits the temperature dependence of $\myim \delta \Sigma^R_{F,Q}$ times a prefactor that is given by an integral over the entire non-FL regime.  The latter is determined in the ground state  and the leading behavior of the prefactor is, therefore, independent of $T$, see for instance Eq.~\eqref{eq:MQdom}. As described in Sec.~\ref{sec:SFT}, the thermal component $\myim\Sigma^R_{F,T}$ is instead characterized by the plateau $-\Gamma_F(T)$ at small momentum and frequency arguments and crosses over to algebraic tails with log corrections for large arguments. Quite importantly, both the crossover scales and the prefactors of the tails depend on $T$.   
Therefore, the plateau, which sets the maximum of $|\myim\Sigma^R_{F,T}|$ according to Fig.~\ref{fig:SFT} does not extend over the complete non-FL regime in $(\omega_k, \mathbf k)$ but is restricted to a smaller region with boundaries determined by the temperature. Since the $T$-dependent algebraic tails vanish fast enough, too, the contributions form $\Sigma^R_{F,T}$ scale to zero faster than the leading $M^2(T)\sim T^{2/3}$ term originating from $\Sigma^R_{F,Q}$. 
  
To show this in detail, let us again begin with $M^2_-(T)$, defined in Eq.~\eqref{eq:defm12} where now the total self-energy $\Sigma^R_{F,Q} + \Sigma^R_{F,T}$ is inserted into $G^R$. 
As we have argued in Sec.~\ref{sec:SFT}, the tails of $\myim \Sigma^R_{F,T}(\mathbf k, \omega_k) $ can extend beyond the frequency scale $\wmax$ and the momentunm cut-off $\Lambda$. Nevertheless, we focus first on the regime $-\wmax \leq \omega \leq 0$ and $|\mathbf k -\mathbf k_F| \leq \Lambda$ as before and return to the remaining frequencies and momenta at the end of the section.
According to Eq.~\eqref{eq:wcross} the crossover scale $\omega_\lessgtr$ is on order $T$. Following the same line of arguments as below Eq.~\eqref{eq:m-q}, shows that the contribution from $-\omega_\lessgtr \leq \omega_k \leq 0$ still scales like $T^{7/3}$ and can be neglected. In the regime $-\wmax \leq \omega \leq - \omega_\lessgtr$ we can again expand the corresponding integrand of Eq.~\eqref{eq:defm12} to first order in the deviation of the self-energy from its 
its ground state form: $\myim \Sigma^R_F - \myim \Sigma^R_{F}|_{T=0} = \myim \delta \Sigma^R_{F,Q} +\myim \Sigma^R_{F,T}$. Since $\myim \delta \Sigma^R_{F,Q}$ has already been discussed in the previous section, it remains to consider $\myim \Sigma^R_{F,T}$, which gives rise to the mass $M^2_{-,T}(T)$. After linearizing in $k' =k-k_F$, the corresponding expression reads
\begin{widetext}
\begin{align}\label{eq:SBRRe}
\begin{split}
M^2_{-,T}(T) & \simeq 8 g^2 k_F \int_0^{2\pi} \frac{d\varphi_{\mathbf k}}{2\pi}\int_{-\wmax}^{-\omega_\lessgtr} \frac{d\omega_k}{2\pi} \int^{+\Lambda}_{- \Lambda} \frac{dk'}{2 \pi} \dkq \, \myim \Sigma^R_{F,T}( k',\omega_k, \varphi_{\mathbf k})(\omega_k -\myre \Sigma^R_{F}|_{T=0}(\omega_k,\varphi_{\mathbf k})-v_F k') \\
&\qquad\qquad\qquad\qquad\qquad\times \bigg[  \frac{1}{\left[(\omega_k -\myre \Sigma^R_{F}|_{T=0}(\omega_k,\varphi_{\mathbf k})-v_F  k')^2+(\myim \Sigma^R_{F}|_{T=0}(\omega_k, \varphi_{\mathbf k}))^2\right]^2} \\
&\qquad\qquad\qquad\qquad \qquad \qquad- \frac{2 \myim \Sigma^R_{F}|_{T=0}(\omega_k, \varphi_{\mathbf k})^2}{\left[(\omega_k -\myre \Sigma^R_{F}|_{T=0}(\omega_k,\varphi_{\mathbf k})-v_F  k')^2+(\myim \Sigma^R_{F}|_{T=0}(\omega_k, \varphi_{\mathbf k}))^2\right]^3}\bigg] \, .
\end{split}
\end{align}
\end{widetext}
To proceed, we have to recall the properties of $\Sigma^R_{F,T}(\mathbf k, \omega_k)$ at finite momentum deviation $\delta k \equiv k'$ away from the FS, discussed at the end of Sec.~\ref{sec:SFT}. First of all, the criterion~\eqref{eq:SFTcrossfull} sets the crossover scale between the asymptotics of small and large arguments. It can be simplified by introducing the variable $u = v_F k' -\omega_k + \myre \Sigma^R_{F}|_{T=0}(\omega_k,\varphi_{\mathbf k})$. After neglecting the logarithmic terms, which cannot change the leading scaling behavior, and also the correction $\myim \delta \Sigma^R_{F,Q}$, the condition for the crossing of the two regimes reads as
\begin{align}\label{eq:asySimp}
1 \sim \frac{M |u|}{u^2 + (\myim \Sigma^R_{F}|_{T=0}(\omega_k, \varphi_{\mathbf k}))^2} \, .
\end{align}
The crossover scale is determined by approaching the value of $|u|$ that equates both sides from large $u$.
If the right-hand side is larger than one, we can replace $\myim \Sigma^R_{F,T} \simeq - \Gamma_{F,\mathbf k_F}(T)$ with the scattering rate from Eq.~\eqref{eq:resGammaFfin}. In the opposite case, one encounters the asymptotic behavior of $\myim \Sigma^R_{F,T}$ from Eq.~\eqref{eq:SFTasyfull}:
\begin{align}\label{eq:alg1}
\myim \Sigma^R_{F,T} (\mathbf k, \omega_k) \simeq - \frac{\pi \alpha \dkFq T}{4A^2 \varepsilon_F} \frac{|u|}{u^2+ (\myim\Sigma^R_F|_{T=0}(\omega_k,\varphi_{\mathbf k}))^2} \, ,
\end{align}
up to logarithmic corrections.
From the relevant roots of Eq.~\eqref{eq:asySimp}, we obtain the curves $u(\omega_k, \varphi_{\mathbf k})$ 
\begin{align}
u_{\pm}(\omega_k,\varphi_{\mathbf k}) = \frac{\pm M \pm \sqrt{M^2-4(\myim\Sigma^R_F|_{T=0}(\omega_k,\varphi_{\mathbf k}))^2}}{2}\, ,
\end{align} 
that indicate the crossover in terms of the variable $u$ at given $(\omega_k,\varphi_\mathbf{k})$.
However, these two solutions are only real provided that $M(T) \geq 2 |\myim\Sigma^R_F|_{T=0}(\omega_k,\varphi_{\mathbf k})|$, which requires $|\omega_k|$ to be smaller than
\begin{align}\label{eq:defMT}
\omega_{M(T)} =\frac{1}{\dkq \win^{1/2}} \left(\frac{M(T)}{2}\right)^{3/2} \, .
\end{align} 
If $|\omega_k|>\omega_{M(T)}$, only the asymptotic tails of Eq.~\eqref{eq:alg1} exist for all relevant momenta.
For temperatures below $T_{\text{scal}}$ from Eq.~\eqref{eq:defTscal}, we have $\omega_{M(T)} \gg \omega_{\lessgtr}$. On the other hand, in this temperature regime $\omega_{M(T)}$ never exceeds $\omega_{\max}$. This means that the plateau of $\myim \Sigma^R_{F,T}$ never extends over the complete $(\omega_k, \mathbf k)$ region that hosts the non-FL correlations. This observation gives a first hint that the thermal component of the fermionic self-energy does not change the scaling. To confirm this mathematically, it remains to further subdivide the integral of Eq.~\eqref{eq:SBRRe} in three subparts: 
\begin{widetext}
\begin{align}
\arraycolsep=10.4pt\def\arraystretch{2.2}
\begin{array}{c|c|c|c}
          &  \omega_k  & u & \myim \Sigma^R_{F,T} \\
\hline
M_{-,1,T} &   -\omega_{M(T)} \leq \omega_k \leq -\omega_{\lessgtr} & -u_+ \leq u \leq u_+ & -\Gamma_{F, \mathbf k_F} \\ 
M_{-,2,T}  & -\omega_{M(T)} \leq \omega_k \leq -\omega_{\lessgtr} & u_{-\Lambda} \leq - u_+ \cup u_+ \leq u \leq  u_{\Lambda} & \text{cf. Eq.~\eqref{eq:alg1}} \\
M_{-,3,T} & -\wmax \leq \omega_k \leq \omega_{M(T)} & 
u_{-\Lambda} \leq u\leq u_\Lambda &\text{cf. Eq.~\eqref{eq:alg1}}
\end{array}
\end{align}
\end{widetext}
Here, $u_{\pm \Lambda} = \pm v_F \Lambda -\omega_k + \myre \Sigma^R_{F}|_{T=0} (\omega_k, \varphi_{\mathbf{k}})$ is introduced as shorthand notation. Furthermore, we have used $u_- = -u_+$. The last column indicates the corresponding form of $\myim \Sigma^R_{F,T}$.
Written out, the first term becomes
\begin{widetext}
\begin{align}
\begin{split}
 M^2_{-,T,1}(T) & \simeq
8 g^2 m \int_0^{2\pi} \frac{d\varphi_{\mathbf k} }{2\pi}\int_{-\omega_{M(T)}}^{-\omega_{\lessgtr}} \frac{d\omega_k}{2\pi} \int^{u_+(\omega_k,\varphi_\mathbf{k})}_{-u_+(\omega_k,\varphi_\mathbf{k})} \frac{du}{2 \pi} \dkFq\,\Gamma_{F,\mathbf{k}_F}(T)\, u 
\\
&\qquad\qquad\times\left[  \frac{1}{\left[u^2+(\myim \Sigma^R_{F}|_{T=0}(\omega_k, \varphi_{\mathbf k}))^2\right]^2}
- \frac{2 \myim \Sigma^R_{F}|_{T=0}(\omega_k, \varphi_{\mathbf k})^2}{\left[u^2+(\myim \Sigma^R_{F}|_{T=0}(\omega_k, \varphi_{\mathbf k}))^2\right]^3}\right] =0\, ,
\end{split}
\end{align}
\end{widetext}
which vanishes by symmetry. Next, we have 
\begin{widetext}
\begin{align}\label{eq:Mm1T}
\begin{split}
& M^2_{-,T,2}(T) =
-8 g^2 m \int_0^{2\pi} \frac{d\varphi_{\mathbf k}}{2\pi}\int_{-\omega_{M(T)}}^{-\omega_{\lessgtr}} \frac{d\omega_k}{2\pi} \left(\int_{u_{-\Lambda}}^{-u_+(\omega_k,\varphi_\mathbf{k})}+\int_{u_+(\omega_k,\varphi_\mathbf{k})}^{u_\Lambda} \right) \frac{du}{2 \pi} \dkFv\frac{\pi \alpha T}{4 A^2 \eF} \\
&\qquad\qquad\qquad
\times \frac{|u| u}{u^2+(\myim \Sigma^R_{F}|_{T=0}(\omega_k, \varphi_{\mathbf k}))^2}\left[  \frac{1}{\left[u^2+(\myim \Sigma^R_{F}|_{T=0}(\omega_k, \varphi_{\mathbf k}))^2\right]^2}
- \frac{2 \myim \Sigma^R_{F}|_{T=0}(\omega_k, \varphi_{\mathbf k})^2}{\left[u^2+(\myim \Sigma^R_{F}|_{T=0}(\omega_k, \varphi_{\mathbf k}))^2\right]^3}\right]
\end{split}
\end{align}
\end{widetext}
Due to the antisymmetry of the integrand with respect to $u$, it is equivalent to integrate $u$ only between $v_F \Lambda \mp (\omega_k - \myre \Sigma^R_{F}|_{T=0})$. With the non-FL condition $v_F \Lambda \gg |\omega_k - \myre \Sigma^R_{F}|_{T=0}|$ from Eq.~\eqref{eq:ETcond1}, we can simplify the expression by replacing $u \to v_F \Lambda$ in the integrand while multiplying with the directed integration range $-2 (\omega_k-\myre \Sigma^R_F|_{T=0})$. In addition, we also have $v_F \Lambda \gg  \myim \Sigma^R_{F}|_{T=0}$ such that we obtain
\begin{widetext}
\begin{align}\label{eq:Mm2T}
\begin{split}
& M^2_{-,T,2}(T) =
\frac{2\pi \alpha^2 T }{A^2 \eF}  \int_0^{2\pi} \frac{d\varphi_{\mathbf k}}{2\pi}\int_{-\omega_{M(T)}}^{-\omega_{\lessgtr}} \frac{d\omega_k}{2\pi}  \dkFv
 \bigg[  \frac{1}{(v_F \Lambda)^4}
- \frac{2 \myim \Sigma^R_{F}|_{T=0}(\omega_k, \varphi_{\mathbf k})^2}{(v_F \Lambda)^6}\bigg](\omega_k -\myre \Sigma^R_F|_{T=0}(\omega_k, \varphi_{\mathbf k})) \, .
\end{split}
\end{align}
\end{widetext}
The frequency integral is now trivial and implies the leading temperature scaling
\begin{align}\label{eq:M2T}
M^2_{-,T,2}(T) \sim \frac{\alpha^2 T }{\eF} \frac{\win^{1/3} \omega^{5/3}_{M(T)} }{h^{4/3}_\omega\alpha^2} \sim \frac{\alpha^{11/24} T^{11/6}}{h^{1/12}_\omega\eF^{3/4}} \, ,
\end{align}
where we have inserted $M^2(T) = M^2_{-,Q}(T)$ from Eq.~\eqref{eq:resM2-} and $v_F \Lambda = h_{\max} \eF = h_\omega^{1/3}\sqrt{\alpha}$. In particular, this result arises from the combination of $\myre \Sigma^R_F|_{T=0}$ with the $(v_F \Lambda)^{-4}$ term. Comparing $M^2_{-,T,2}(T)$ with $M^2_{-,Q}(T) $ shows that they become comparable only for temperatures larger than $T \sim h_\omega^{13/14}(\alpha/\eF^2)^{17/28} \eF   $ which is almost comparable to $\wmax \sim h_\omega \alpha^{1/2}$ and certainly exceeds $T_{\text{scal}}$. Finally, the last contribution reads as $\omega_{M(T)}$
\begin{widetext}
\begin{align}
\begin{split}\label{eq:Mm3T}
& M^2_{-,T,3}(T) = -8 g^2 m \int_0^{2\pi} \frac{d\varphi_{\mathbf k}}{2\pi}\int_{-\wmax}^{-\omega_{M(T)}} \frac{d\omega_k}{2\pi} \int_{u_{-\Lambda}}^{u_\Lambda}  \frac{du}{2 \pi} \dkFv \frac{\pi \alpha T}{4 A^2 \eF}  \frac{|u| u}{u^2+(\myim \Sigma^R_{F}|_{T=0}(\omega_k, \varphi_{\mathbf k}))^2}\\
&\qquad\qquad\qquad\qquad\qquad\times\bigg[  \frac{1}{\left[u^2+(\myim \Sigma^R_{F}|_{T=0}(\omega_k, \varphi_{\mathbf k}))^2\right]^2}
- \frac{2 \myim \Sigma^R_{F}|_{T=0}(\omega_k, \varphi_{\mathbf k})^2}{\left[u^2+(\myim \Sigma^R_{F}|_{T=0}(\omega_k, \varphi_{\mathbf k}))^2\right]^3}\bigg]\, .
\end{split}
\end{align}
\end{widetext}
Since the $u$ integration runs again between $v_F \Lambda \mp (\omega_k - \myre \Sigma^R_F|_{T=0}) $, we can employ the very same steps as in case of the previous integral. 
The frequency integral is now dominated by the boundary at $-\wmax$, because of $\wmax \gg \omega_{M(T)}$. In this case the most important contribution arises from the product $\omega_k$ times the $(v_F \Lambda)^{-4}$ term.
For the maximal cut-off, this results in 
\begin{align}\label{eq:MT3}
M^2_{-,T,3}(T) \sim \frac{\alpha^2 T }{\eF} \frac{\wmax^2}{h^{4/3}_\omega\alpha^2} \sim h^{2/3}_\omega \frac{\alpha T}{\eF} \, ,
\end{align}
which also scales less important than $M^{2}_{-,Q}(T)$ in the limit $T \to 0$. Here, the crossover between the two appears at very high temperatures of $T_{\max} \sim  h_\omega \alpha^{1/2} =\wmax$. 
Note that the linearization of the self-energy around $\Sigma^R_F|_{T=0}$ at frequencies of order $\wmax$ and momenta $|\mathbf k - \mathbf{k}_F| \sim \Lambda$ remains possible up to $T_{\max}$, too.
We will discuss the implications of this result below but first we complete the analysis of the feedback of $\myim \Sigma^R_{F,T}$.

For temperatures below $T_{\max}$, the term $M^2_{+,T}(T)$ can be discarded for the same reason as above: The Fermi function affects only the regime $\omega_k \lesssim T$ which does not contribute substantially.
Another contribution arises from the fact that, the thermal component extends beyond the momentum and frequency cut-offs of the non-FL regime. However, outside of this regime the self-energy at finite $T$ can be approximated as $\Sigma^R_F \simeq \Sigma^R_{F,T}$ since $\Sigma^R_{F,Q}$ becomes negligible. Analogously, $\Sigma^R_{F,Q}$ does not appear in the self-consistent equation~\eqref{eq:SFTfin}. Consequently, all the information specific to the non-FL state has dropped out of $\Sigma^R_{F}$. Therefore, it has to acquire a form identical to the high-energy asymptotics in a Fermi liquid, which is irrelevant for the low-energy physics. 

All in all, we have shown that the thermal component of the self-energy does not change the leading scaling behavior of the bosonic mass gap when the thermal self-energy $\Sigma^R_{F,T}$ generated by $M^2_{-,Q}$ is inserted into the self-consistent loop. As a result, we can write
\begin{align}
M^2(T) = M_{-,Q}^{2}(T)
\end{align}
up to subleading corrections. Therefore, the scaling relations presented at the end of the previous section provide the full solution for the Eliashberg equations. 

Let us return to the result~\eqref{eq:MT3}. It indicates that the quantum critical scaling $M^2(T) \sim T^{2/3}$ crosses over at $T_{\max} \sim \wmax$ to a linear temperature dependence, which strongly resembles the analytic results of Refs.~\cite{millis1993,hartnoll2014} and the QMC-observation~\cite{schattner16QMC}. Formally, we miss the logarithmic corrections, which have not been considered in detail within the calculation. However, restoring them properly at temperatures close to $T_{\max}$ is not straightforward: For instance, the crossover condition Eq.~\eqref{eq:SFTasyfull} changes because $\Gamma_F(T)$ is given by Eq.~\eqref{eq:resGammaFcc} instead of Eq.~\eqref{eq:resGammaFQCR}. Moreover, the calculation of $M^2(T)_{-,T,2}$ in Eq.~\eqref{eq:M2T} turns into a genuinely self-consistent problem as $\omega_{M(T)}$ depends itself on $M(T)$. This suppresses the increase of $M(T) \sim T^{11/6}$ suggested by Eq.~\eqref{eq:M2T}.
Furthermore, for $T \simeq T_{\max}$ the contribution of $M^2_+(T)$ has to be taken into account, too, because $n_F(\omega_k)$ ceases to suppress the largest frequency scales. 
Nevertheless, Eq.~\eqref{eq:MT3} and the associated crossover appear robust for two physical reasons: First of all, we note that for the integration in Eq.~\eqref{eq:Mm3T} only the universal asymptotic tails of $\Sigma^R_{F,T}$ and the estimate of the UV-cut-offs from the ET scaling arguments matter. Both of them are not affected by the previously described effects. Moreover, the linear temperature dependence arises only from the asymptotics of $\Sigma^R_{F,T}$, which does not contain any information specific to the non-qp but incorporates the presence of the almost critical bosonic fluctuations. Furthermore, we note that the thermal width of the Fermi-Dirac distribution $n_F$ starts to exceed the maximal spectral width $\wmax$ of the non-FL excitations for temperatures $T \gtrsim T_{\max}$. As a result, a qp-picture is restored with \emph{growing} temperature. These arguments indicate that the nematic susceptibility obtained in ET indeed approaches the form $M(T) \sim (T \log T)^{1/2}$ from Eq.~\eqref{eq:MTqp} at temperatures larger than $T_{\max}$. 

Finally, we point out that the crossover temperature between $M(T) \sim T^{1/3}$ and $M(T) \sim (T \log T)^{1/2}$ is always given by the UV-frequency cut-off provided that the cut-off scheme is constructed along the lines of Sec.~\ref{sec:ansatz}: For a given $\Lambda$ one chooses the maximal frequency as $\omega_\Lambda \sim (v_F \Lambda)^3/\alpha$ according to Eq.~\eqref{eq:ETcond2}. The dominant contribution to the bosonic mass from the integral~\eqref{eq:MQdom} reads then $$M^2_{-,Q}(T)\sim \alpha^{5/3} T^{2/3} \omega_\Lambda^2/(v_F \Lambda)^3.$$ In addition, the contribution containing the asymptotic tails of $\myim \Sigma^R_{F,T}$ from Eq.~{\eqref{eq:Mm3T}} scales like $$M^2_{-,T,3}(T) \sim \alpha^2 T \omega_\Lambda^2/(\eF (v_F \Lambda)^4).$$ As a result, the crossover temperature becomes 
\begin{align}\label{eq:Tcrossgen}
T \sim \frac{(v_F \Lambda)^3}{\alpha} \sim \omega_\Lambda\, .
\end{align}  

\section{Vertex corrections}\label{sec:vertexCorr}
As already mentioned in the introduction~\ref{sec:intro}, ET, like any self-consistent approximation used in quantum field theory, is uncontrolled in the sense that, by omitting diagrams, important low-energy contributions may get lost. In the context of the INM the vertex function has been studied both in the ground state~\cite{chubukov2005Ward,rech2006, metl10} and at finite temperatures~\cite{punk2016}. At $T=0$ the result depends  crucially on how the limit of vanishing external arguments is taken: In some cases the outcome is perturbative in $\alpha$ whereas in other ones even a divergent result is obtained. For $T>0$, the thermal contribution becomes comparable to the bare vertex and a resummation of the vertex correction is mandatory.

\begin{figure}[ht]
\begin{center}
\includegraphics[width=.9\columnwidth]{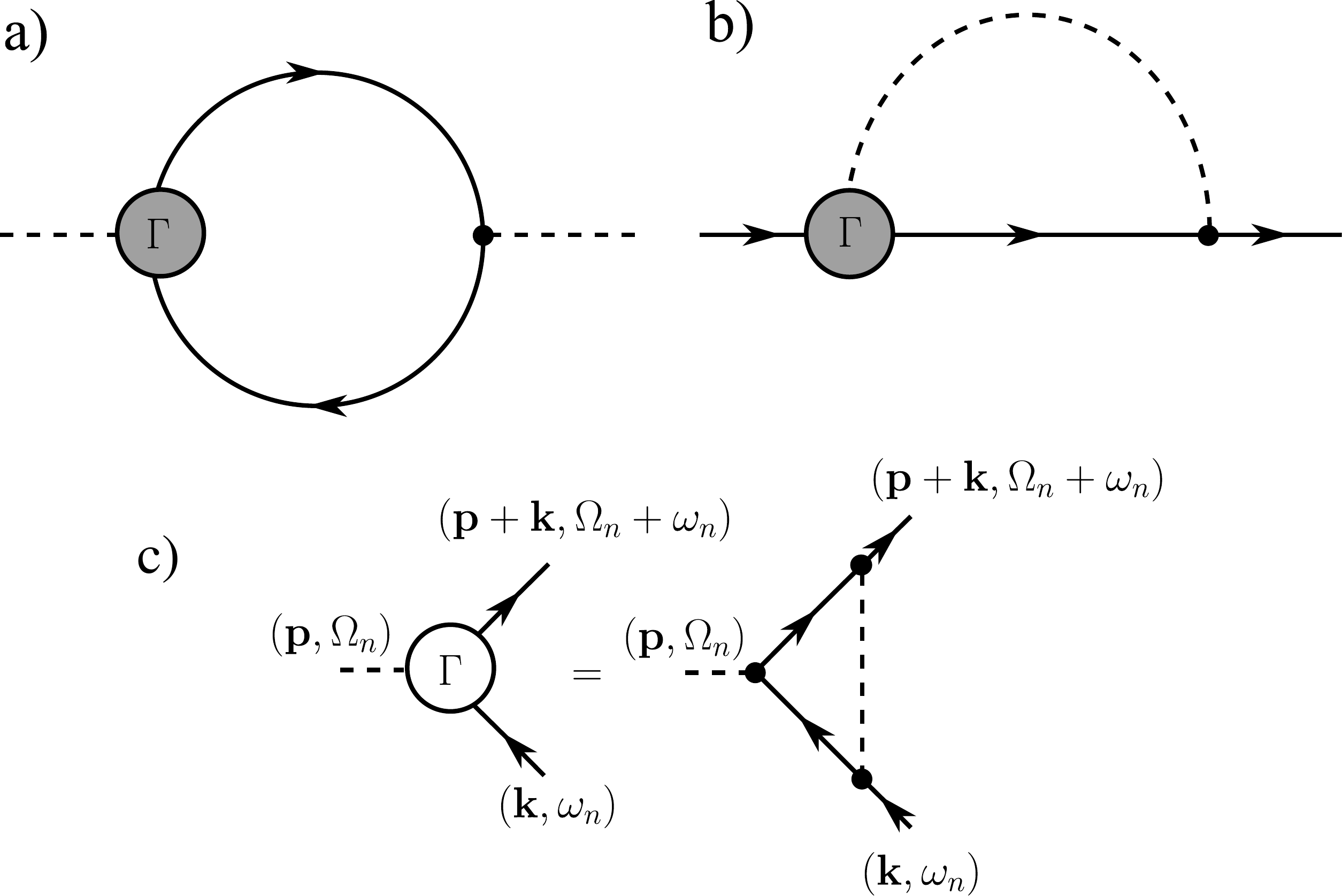}
\caption{
a) Bosonic self-energy dressed by the self-consistent vertex function depicted as grey circle; b) dressed fermionic self-energy; c) Lowest-order approximation to the vertex shown as white circle}
\label{fig:vertexCorr}
\end{center}
\end{figure}

Here, we use our Eliashberg results from the previous sections to give a more unified picture regarding the effect of the leading order vertex correction:
\begin{align}
\begin{split}
\Gamma(p, k) = \frac{g^3}{(\beta V)^{3/2}} \sum_{q}
& G(k+q) G(k+p+q) D(q) \\
& \times d_{\mathbf k+\mathbf q/2} d_{\mathbf k + \mathbf{p}/2 + \mathbf q} d_{\mathbf k+ \mathbf p + \mathbf q/2}
\end{split}
\end{align} on the self-energies $\Sigma^R_B(\mathbf p, \omega_p)$ and $\Sigma^R_F(\mathbf k, \omega_k)$. The corresponding diagrams are depicted in Fig.~\ref{fig:vertexCorr}.
   
To caclucate $\Gamma(\mathbf p , \omega_p; \mathbf k ,\omega_k)$ as function of the real frequencies $\omega_{p,k}$ with the dressed propagators $G^R(\mathbf k,\omega_k)$ and $D^R(\mathbf p, \omega_p)$ from above, we have to perform the analytic continuation according to Eq.~\eqref{eq:defHilbert}. This step, however, mixes retarded and advanced Green's functions $G^A=(G^R)^\ast ,D^A =(D^R)^\ast$ in the two-loop self-energies such that the vertex function is decomposed into several subparts. 
To obtain them correctly, we state the full expressions for the self-energies including the the lowest-order vertex correction and extract the vertex parts from the~\footnote{The two-loop self-energies can also be obtained within the equivalent Keldysh formalism~\cite{kame2011}. However, the individual terms of Eqs.~\cref{eq:SB2loop,eq:SF2loop} do not necessarily  represent single Keldysh diagrams but rather linear combinations of them.}.
In case of $\Sigma^R_B$ we find
\begin{widetext}
\begin{align}
\begin{split}\label{eq:SB2loop}
&\Sigma^R_B(\mathbf p, \omega_p) = \\
&- g^4 \!\int_{k,q}\bigg\{ 
\left[
\text{th}\!\left(\!\frac{\beta \omega_1}{2}\!\right)\! \myim G^R(1) G^R(2) D^R(3) 
+ \text{th}\!\left(\!\frac{\beta \omega_2}{2}\!\right)\!G^A(1) \myim G^R(2) D^A(3)
+G^A(1) G^R(2)\text{ct}\!\left(\!\frac{\beta \omega_3}{2}\!\right)\! \myim D^R(3)
\right] \\
&\qquad\qquad \!\! \times \left[
\text{th}\left(\frac{\beta \omega_4}{2}\! \right)\! \myim G^R(4) G^A(5)
+\text{th}\left(\frac{\beta \omega_5}{2}\! \right)\! \myim G^R(4) \myim G^R(5)
 \right]
\\
& \qquad\qquad \!\! +\left[ \text{th}\!\left(\!\frac{\beta \omega_1}{2}\!\right)\! \text{ct}\!\left(\!\frac{\beta \omega_3}{2}\!\right)\! -1 \right]\!\!\left[ G^A(1) \myim G^R(2) \myim D^R(3) G^A(4) G^A(5)\!
+\!\myim G^R(1) G^R(2) \myim D^R(3) G^R(4) G^R(5)\right]\!\!\bigg\},
\end{split}
\end{align}
\end{widetext}
with the abbreviations of the arguments: $1 = ( \mathbf{k} + \mathbf q, \omega_k + \omega_q) $, $2= (\mathbf k + \mathbf p + \mathbf q, \omega_k + \omega_p + \omega_q)$, $3 = (\mathbf q , \omega_q)$, $4= (\mathbf k + \mathbf p , \omega_k + \omega_p) $ and $5 =(\mathbf k, \omega_k)$ while $\omega_i$ refers to the frequency argument of $i= 1,...,5$. Furthermore, $\text{th, ct}$ denote $\tanh , \coth$, respectively, and $\int_q \equiv (2\pi)^{-3}\int d^2 q \, d\omega_q $.

In the following, we will identify those contributions that are not perturbative in the small coupling $\alpha /\eF^2 \ll 1$. To this end, we 
split the bosonic distribution functions $\text{ct}(\beta \omega/2)$ again in a thermal $2T/\omega_p$ and a quantum part $\text{ct}(\beta\omega/2)-2T/\omega$ like in case of the fermionic self-energy. 
By inspection of the two-loop result for $\Sigma^{R}_{B}$, we have to consider several thermal contributions to the vertex function that can be evaluated for an incoming Boson with $(\mathbf{p},\omega_p)=(0,0)$ and incoming Fermion with $(\mathbf k, \omega_k ) =(\mathbf k_F ,0)$. The most important one is found in the first line of Eq.~\eqref{eq:SB2loop} and reads 
\begin{align}
\begin{split}\label{eq:GammaTint1}
\Gamma_T (\mathbf 0, 0; \mathbf k_F, 0) =- g^3 & \dkF^3\int  \frac{d^2 q\, d\omega_q}{(2\pi)^3}  \frac{2T}{\omega_q} \myim D^R(\mathbf q, \omega_q)\\
& \times  G^A(\mathbf{k}_F + \mathbf q, \omega_q)
  G^R(\mathbf k_F + \mathbf q, \omega_q) \, .
\end{split}
\end{align}
The integral can be calculated in close analogy to $\myim \Sigma^R_F$ in Sec.~\ref{sec:SFT}. At first glance, we cannot avoid the singularities of the fermionic Green's functions due to the presence of $G^R \cdot G^A$, which entails non-analyticities in both the upper and lower half-plane of the complex frequency $\omega_q$. Nevertheless, the dominant contribution to the integral arises from the pole of $\myim D^R(\mathbf q,\omega_q)$ at $\omega= i \Omega_q$, defined in Eq.~\eqref{eq:scaleDR}: As discussed  above Eq.~\eqref{eq:SFTint2}, we have either $\Omega_q \sim T$ with a small prefactor or $\Omega_q$ vanishes faster than $T$. In contrast, the low-frequency behavior of $G^{A,R}$ is regularized by $\Gamma_F(T) \sim T^{2/3}$ and thus yields only subdominant terms.    
Therefore, we have
\begin{align}
\begin{split}\label{eq:GammaTint2}
\Gamma_T(\mathbf 0,0;\mathbf k_F,0)& \simeq  g^3 T\dkF^3 \int \frac{d^2 q}{(2\pi)^2} \frac{1}{c_B^2 q^2 +M^2(T)}
\\& \qquad \quad\times G^A(\mathbf k_F + \mathbf q,0) G^R(\mathbf k_F + \mathbf q,0) \, .
\end{split}
\end{align}
The angular integral over $\varphi_\mathbf{q}$ yields 
\begin{align}\label{eq:angInt1}
\begin{split}
\int_0^{2\pi} \frac{d\varphi_{\mathbf q}}{(2\pi)} \frac{1}{(v_F p)^2 \cos^2(\varphi_{\mathbf{q}} -\varphi_{\mathbf k_F}) + \Gamma_{F, \mathbf k_F}^2} =  \\
\frac{1}{\Gamma_{F, \mathbf{k}_F}} \frac{1}{
\sqrt{v_F^2 p^2 + \Gamma_{F, \mathbf{k}_F}^2}}\, .
\end{split}
\end{align}
Inserted into the thermal component of the vertex this results in 
\begin{align}
\begin{split}
\Gamma_T(\mathbf 0,0;\mathbf k_F,0)& =  \frac{g^3 T\dkF^3}{ v_F^2 A^2 \Gamma_{F, \mathbf k_F}} \int_0^\infty \frac{du}{2\pi}  \frac{u}{u^2+M^2/A^2} \\
& \qquad\qquad\qquad\qquad\times\frac{1}{\sqrt{u^2+\Gamma_{F,\mathbf k_F}^2}} \\
& = \frac{g \alpha T\dkF^3}{4 \eF A^2 \Gamma_{F, \mathbf k_F}} \frac{\arccos\left(\dfrac{\Gamma_{F, \mathbf k_F}}{M/A}\right)}{\sqrt{M^2/A^2 -\Gamma^2_{F,\mathbf k_F}}} \, .
\end{split}
\end{align}
This form admits an expansion equivalent to Eq.~\eqref{eq:arcsecExp}. The limit $\Gamma_{F, \mathbf k_F}(T) \gg M(T)$ implies logarithmic terms like in the thermal component of the self-energy. In the opposite regime $\Gamma_{F, \mathbf k}(T) \ll M(T)$,  relevant to the QCR at the lowest temperatures, we find however,
\begin{align}\label{eq:resVertexT}
\Gamma_T(\mathbf 0,0;\mathbf k_F,0) = g \dkF \, ,
\end{align} 
where we have inserted the Eliashberg Eq.~\eqref{eq:resGammaFQCR} for the fermionic damping rate. We observe that the thermal component of the vertex in the low-energy limit coincides exactly with the bare coupling constant. This has been found in Ref.~\cite{punk2016}, too. 
As a result, the thermal vertex corrections do not 
vanish in the limit $T \to 0$. Moreover, they do not
become negligible even in the weak-coupling limit and therefore have to be properly resummed. To this end, we first have to identify all nonperturbative contributions to the vertex. 

We first turn to the quantum part $\Gamma^{(1)}_q$ resulting from the first line of Eq.~\eqref{eq:SB2loop}. The antisymmetry of the bosonic propagator $\myim D^R$ with respect to $\omega_p$ implies the low-energy limit of the bosonic damping $\myim \Sigma^R_B(p \to 0) = 0$, irrespective of how the origin of the $(\mathbf p ,\omega_p)$ plane is approached. Therefore, we concentrate on the corrections to the bosonic mass $M^2(T)$, or equivalently $\myre \Sigma^R_B(0)$, and set the external bosonic frequency and momentum to zero. However, we keep a nonzero internal fermionic frequency $\omega_k$: In some terms this turns out to be necessary to obtain a well-defined result while, physically, $M^2(T)$ is dominated by finite frequencies $|\omega_k| \gg \omega_{\lessgtr} \sim T$, too (cf. Eq.~\eqref{eq:MQdom}). Thus, the corresponding vertex part reads as
\begin{align}\label{eq:GammaqStart}
\begin{split}
&\Gamma_q^{(1)}(\mathbf{0},0;\mathbf k_F, \omega_k)  = \\ 
& - g^3 \dkF^3 \!\!\int_q \! \left[\text{ct} \!\left(\!\frac{\beta \omega_3}{2} \!\right)\!-\! \frac{2 T}{\omega_3}\! -\! \text{th}\! \left(
\!\frac{\beta \omega_1}{2}\! \right)\! \right] \! |G^R(1)|^2 \!\myim \! D^R(3) \\
&- g^3 \dkF^3\! \int_q \frac{1}{2i} \text{th}\left(\frac{\beta\omega_1}{2}\right) [G^R(1)^2 D^R(3)  -  G^A(1)^2 D^A(3)] \, .
\end{split}
\end{align}
In App.~\ref{sec:irrVertexCorr}, we show that the second integral is perturbative in the coupling constant such that we can neglect it. In the first line the dominant contribution arises when $\mathbf{q} \perp \mathbf k_F$ and as usual $|\mathbf q| \ll k_F$, similarly to the situation encountered in the computation of $\myim \Sigma^R_{F,Q}$. Concentrating on this case, the angular integral is of the form~\eqref{eq:angInt1} and yields
\begin{align}
\begin{split}
\Gamma_q^{(1)}(\mathbf{0},0;\mathbf k_F, \omega_k)\! \simeq  \!
- g^3\!\!\int \! \!\frac{d\omega_q}{2\pi}\! \!\int_0^\infty \! \!\frac{d q }{2 \pi}  \frac{\dkF^3 q  }{\myim \!\Sigma^R_F(\omega_k+\omega_q, \varphi_{\mathbf{k}_F})} \\
\frac{\myim D^R(\mathbf q,\omega_q)|_{\varphi_{\mathbf q}=\varphi_{\mathbf k_F}}}{\sqrt{(v_F q)^2 + \myim \Sigma^R_F(\omega_k + \omega_q,\varphi_{\mathbf k_F})^2}} \\
 \left[\text{ct} \!\left(\!\frac{\beta \omega_q}{2} \!\right)\!- \frac{2 T}{\omega_q} - \text{th}\! \left(
\!\frac{\beta(\omega_k + \omega_q)}{2}\! \right)\! \right]\! .
\end{split}
\end{align}
Note that we have approximated $\omega_k +\omega_q - \Sigma^R_F(\omega_{k}+\omega_q,\varphi_{\mathbf k_F+\mathbf q}) \simeq -i \myim \Sigma^R_F(\omega_k +\omega_q ,\varphi_{\mathbf k_F})$ to capture the dominant behavior in the limit of small frequencies. Furthermore, the $q$ integration remains well-defined when we replace $\sqrt{(v_F q)^2 + \myim \Sigma^R_F(\omega_k + \omega_q,\varphi_{\mathbf k_F})^2} \simeq v_F q$, which corresponds to the relation~\eqref{eq:ETcond1} for the applicability of ET. Moreover, this approximation brings the $q$-dependence of the integrand to a form  identical to the one found in Sec.~\ref{sec:SFQ}, up to the momentum independent factor $1/\myim \Sigma^R_F(\omega_k +\omega_q ,\varphi_{\mathbf k_F})$. 
In case of the self-energy, it is the $q$ integration that establishes the characteristic relation $v_F q \sim (\alpha \omega_q)^{1/3}$ of Eq.~\eqref{eq:ETcond2}, which is responsible for the generation of the non-FL correlations at the Ising nematic QCP. As a result, we expect $\Gamma^{(1)}_q$ to be compatible with this scaling structure.
By rescaling like above $\bar\omega = \omega/T $ and $\bar q = v_F A^{2/3}/(\dkFq \alpha T)^{1/3}\cdot q $, we have:
\begin{align}\label{eq:interVertexQ1}
\begin{split}
&\Gamma_q^{(1)}(\mathbf{0},0;\mathbf k_F, \omega_k) \simeq 
- g \dkF \frac{ \alpha^{1/4} }{12 \sqrt{3} A^{4/3} \eF^{1/2}}\!\!\int \! d\bar \omega_q \frac{1}{|\bar \omega_q|^{1/3}} \\
&\! \times \! \frac{1 }{ \tilde\Sigma^R_F(\bar \omega_k+\bar \omega_q, \varphi_{\mathbf{k}_F},h_\omega)}
 \left[\text{ct} \!\left(\!\frac{\bar\omega_q}{2} \!\right)\!- \frac{2}{\bar\omega_q} - \text{th}\! \left(
\!\frac{\bar \omega_k + \bar \omega_q}{2}\! \right)\! \right]\! ,
\end{split}
\end{align}
where we have inserted the results from Eqs.~\cref{eq:SQFdefScal,eq:SFQexp} for $\bar M =0 $ and our scaling function $\tilde \Sigma^R_F$ for the fermionic self-energy from Eq.~\eqref{eq:resSFtot2}. We note that $\Gamma_q^{(1)}$ is independent of $T$ and thus does not scale to zero when the ground state is approached. To obtain the result in the relevant limit $|\omega_k|\gg \omega_{\lessgtr} \sim T$, we can evaluate the last equation with $T=0$ expressions. This leads to
\begin{align}\label{eq:resVertexQ1}
\begin{split}
\Gamma_q^{(1)}(\mathbf{0},0;\mathbf k_F, |\omega_k| \gg \omega_{\lessgtr})& = 
- g \! \int_0^{|\bar\omega_k|} \!\! \frac{8 \sqrt{3} A^{4/3}\dkF\, d\bar \omega_q }{|\bar \omega_q|^{1/3} ||\bar \omega_k| + \bar\omega_q|^{2/3}} \\
 &= -32 \pi A^{4/3} g \dkF  \, ,
\end{split}
\end{align}
which is \emph{negative} and \emph{independent} of $\omega_k$. This behavior results from two counteracting effects that cancel the dependence on $\omega_k $: On the one hand, the integrand becomes strongly enhanced at small $\omega_q$ when $\omega_k\to 0$. On the other hand, the integration range shrinks. By setting $\omega_k =0 $ in Eq.~\eqref{eq:GammaqStart} from the beginning one quite likely misses this contribution in the ground state erroneously such that only perturbative terms remain. Similar effects have been seen in Ref.~\cite{rech2006}. 

Let us return to the self-energy~\eqref{eq:SB2loop} now and consider the last line.
Performing first $\int_q$ by the methods used so far we obtain a result that is to a good approximation independent of $\omega_k$. As a result, $\int d\omega_k G^{A(R)}(4) G^{A(R)}(5) = 0 $ since the contour can always be closed without enclosing singularities.   
In other words, the important vertex corrections for the bosonic self-energy are given by Eqs.~\cref{eq:resVertexT,eq:resVertexQ1}. As was shown in Ref.~\cite{chubukov2005Ward}, the dominant higher-order vertex diagrams in ET at $T=0$ are typically the ladder diagrams. Assuming that this applies to finite temperatures, too, and that the low-energy limit of the resummed vertex function for $\Sigma^R_B$ is given by the geometric series, we find similar to Ref.~\cite{punk2016}
\begin{align}
\begin{split}
\lim_{\omega_k \to 0} \Gamma_{\text{ladder}}(\mathbf 0, 0; \mathbf k_F , \omega_k) & = \frac{g \dkF}{1-(g \dkF)^{-1}(\Gamma_T+\Gamma_q^{(1)})} \\
& = \frac{g \dkF}{32 \pi A^{4/3}}\, .
\end{split}
\end{align}
We emphasize that the quantum part is necessary to avoid an instability that would occur if only $\Gamma_T = g \dkF$ had been considered. However, by taking the limit $\omega_k \to 0$ properly into account, the total vertex acquires merely a numerical prefactor whereas no scaling with $T$ and $\alpha$ is introduced. 

Finally, we also consider the fermionic self-energy dressed with the vertex function:
\begin{widetext}
\begin{align}\label{eq:SF2loop}
\begin{split}
 \Sigma^R_F (\mathbf k, \omega_k) = \qquad \!\! & \\
 - g^4 \dkF^3\int_{q,p} \bigg\{& \left[\text{th}\!\left(\!\frac{\beta \omega_1}{2}
\!\right)\! \myim G^R(1) G^A(3) D^A(3) 
+ \text{th}\!\left(\!\frac{\beta \omega_2}{2}\!\right)
\! G^R(1) \myim G^R(2) D^A(3) 
+ \text{ct}\!\left(\!\frac{\beta \omega_3}{2}\!\right)\! G^R(1) G^R(2) \myim D^R(3) 
\right]\\
\times & \left[\text{th}\!\left(\!\frac{\beta \omega_4}{2}\!\right)\! \myim G^R(4) D^A(5) +\text{ct}\!\left(\!\frac{\beta \omega_5}{2}\!\right)\! G^R(4) \myim D^R(5) \right] \\
+ & \left[ \text{th}\!\left(\!\frac{\beta \omega_1}{2}\!\right)\! \text{th}\!\left(\!\frac{\beta \omega_2}{2}\!\right)\!-1\right] \myim G^R(1) \myim G^R(2) D^A(3) G^R(4) D^R(5) \\
+ & \left[ \text{th}\!\left(\!\frac{\beta \omega_2}{2}\!\right)\! \text{ct}\!\left(\!\frac{\beta \omega_3}{2}\!\right)\!-1\right] G^R(1) \myim G^R(2) \myim D^R(3) G^A(4) D^A(5) \bigg\} \, ,
\end{split}
\end{align}
\end{widetext}
with $5=(\mathbf p,\omega_p)$ while the other definitions remain as stated below Eq.~\eqref{eq:SB2loop}. The thermal components $\Gamma_T$ are all of the form $\myim D^R G^A G^A $ or $\myim D^R G^R G^R $, respectively, and give rise to negligible contributions (see Eq.~\eqref{eq:GammaTirr}). Thereby, a large contribution originating from combining the thermal poles of the two loops of the diagram (e.g from the product of the last terms of the first two lines) is avoided by the associated irrelevant scaling of the thermal vertex correction.  
Let us now consider the quantum parts. There are two interesting limits to be studied: First, we consider $\omega_k =0$ at finite temperatures to exclude large contributions to the positive value $\myim \Sigma^R_{F,Q} (\mathbf k_F, 0)$ that could give rise to an instability by introducing a negative total damping rate $\Gamma_F(T)$. Afterwards, we keep a finite $\omega_k$ at $T=0$ to check the corrections to the non-FL correlations. In any case, we use a finite $\omega_p $ to prevent spurious cancelations of actually large terms. 

The quantum component of the first line reads for $\omega_k=0$:
\begin{align}\label{eq:GammaQ2}
\begin{split}
&\Gamma^{(1)}_q ( \mathbf 0, \omega_p; \mathbf k_F, 0)|_T = \\
&-g^3 \dkF^3 \int_q \bigg\{\left[\text{ct}\!\left(\!\frac{\beta \omega_q}{2}\!\right)  -\frac{2T}{\omega_q}\right] G^R(1) G^R(2) \myim D^R(3)\\
&+\frac{1}{2 i} \left[ \text{th}\!\left(\!\frac{\beta \omega_q}{2}\!\right) -\text{th}\!\left(\!\frac{\beta (\omega_p +\omega_q)}{2}\!\right)\right]\!  G^A(1) G^R(2) D^A(3) \\
&+ \frac{1}{2 i}\left[ \text{th}\!\left(\!\frac{\beta (\omega_p +\omega_q)}{2}\!\right)\! -\text{th}\!\left(\!\frac{\beta \omega_q}{2}\!\right)\! \right]\! G^R(1) G^R(2) D^A(3)\! \bigg\} .
\end{split}
\end{align}
All but the second line are of the form~\eqref{eq:GammaQirr} and thus irrelevant at arbitrary temperatures. In the second line the angular integral is solved via Eq.~\eqref{eq:angInt1}. We emphasize that the resulting integrand is both IR and UV integrable even without the tanh functions. Because of their difference the limit $\omega_p \to 0$ vanishes, unlike in Eq.~\eqref{eq:interVertexQ1}. As a result, we do not obtain important vertex corrections. 

Next we consider $\Gamma^{(1)}_q ( \mathbf 0, \omega_p; \mathbf k_F, \omega_k)|_{T=0} $. The second line becomes then proportional to 
$$ \int_q\!\left[\sgn(\omega_k+\omega_q) - \sgn(\omega_k + \omega_p + \omega_q)\right]\! G^A(1) G^R(2) D^A(3).$$ 
Now taking the limit $\omega_p \to 0$ is possible since the frequency argument $\omega_2 \to \omega_1 = \omega_k+\omega_q$ of the fermionic propagators still differs from the bosonic frequency $\omega_3 =\omega_q$. In analogy to \cref{eq:interVertexQ1,eq:resVertexQ1} the expression is thus well-defined at $\omega_p=0$ but evaluates to zero due to the sign functions. 
Finally, the last two lines of Eq.~\eqref{eq:SF2loop} vanish, too, due to the structure $G^{R,A}(4) D^{R,A}(5)$: integrating first $\int_q$ gives a constant independent of $\omega_p$, such that one can close the contour for the integration over $\omega_p$ without encircling singularities in the complex plane.

As a result, we observe that the leading order vertex correction affects mostly the $\mathcal{O}(1)$ prefactor of $\Sigma^R_B$. The scaling with $T$ and $\alpha$ of both the fermionic and the bosonic self-energies obtained from ET remains unchanged. 

\section{Conclusion and Outlook}\label{sec:conclude}
We have presented a new solution to the Eliashberg equations of the INM at finite temperature. In particular, we have shown that the temperature dependence of the inverse nematic susceptibility and of the fermionic self-energies obeys the expectations of quantum critical scaling: $M(T) \sim T^{1/3}$ and $\Sigma^R_F(\omega, T) \sim T^{2/3} \tilde \Sigma^R_F(\omega/T)$ at the onset of finite temperature. However, the dimensionless coefficients exhibit  IR/UV mixing and are thus nonuniversal. Physically, this effect originates from the finite spectral width of the non-qp excitations which coincides with the energy scale up to which non-FL correlations persist. Furthermore, we have discussed several scenarios for the breakdown of the scaling theory by thermal fluctuations: While it is only stable for temperatures below $T_{\text{scal}} < \win$ the Hertz-Millis result $M(T) \sim (T \log T)^{1/2}$ is recovered for temperatures larger than $T_{\max}$ which marks the threshold when the thermal broadening of occupations around the FS becomes comparable to spectral width of the non-FL excitations. 
The various crossover scales are amenable to comparison with numerical simulations whereby our analytical expressions may help to identify the physical mechanism underlying data from more sophisticated methods. Moreover, our approach in real frequencies gives direct access to the spectral functions which can be observed in experiments. 
Although our scaling solution is quite likely restricted to temperatures below the critical temperature for superconductivity $T_c \sim \win$, our results can still comprise valuable information: While  
the presence of a superconducting gap is expected 
to give a sharp upturn of the self-energies at the smallest frequencies it is quite likely that larger frequencies remain governed by the Ising nematic fluctuations.
\section*{Acknowledgments}
We thank  S. Diehl, D. Pimenov, and M. Punk for fruitful discussions and valuable comments on the manuscript.
\appendix
\section{ Kramers-Kronig relations}\label{sec:KK}
The scaling solution~\eqref{eq:keyResSFtot2} for temperatures below $T_{\text{scal}}$ incorporates the thermal scattering rate of the Fermions at small frequencies in addition to the non-FL correlations at high frequencies. The latter were determined by analytic continuation from the imaginary frequency axis. Here, we show that this operation is consistent with the Kramers-Kronig relations and that omitting corrections to $\myre \Sigma^R_F(k)$ beyond the ground state term is consistent with our solution of the Eliashberg equations.
Starting out from the standard Kramers-Kronig relation~\eqref{eq:defKK} for $\Sigma^R_F$
we can separate the Cauchy principal value integral into a regime of small frequencies $|\omega_k| \lesssim \wmax$ and large frequencies $|\omega_k| \gtrsim \wmax$. Furthermore, we focus on momenta $\mathbf k_F$ on the FS and omit the irrelevant angle variable for brevity. In general, we have $\myre \Sigma_F^R(\mathbf{k}_F, -\omega_k)= -\myre \Sigma_F^R(\mathbf k_F,\omega_k)$ and $\myim \Sigma_F^R(\mathbf{k}_F, -\omega_k)= \myim \Sigma_F^R(\mathbf{k}_F,\omega_k)$. To confirm this we first note that the assumed ansatz~\eqref{eq:SFans} initially
satisfies these conditions. In addition, they are preserved in the self-consistent loop as the propagators of a real Boson obey the symmetry relations $\myre D^R(\mathbf p, -\omega_p)= \myre  D^R(\mathbf{p}, \omega_p)$ and $\myim D^R(\mathbf p, -\omega_p) =- 
\myim D^R(\mathbf{p}, \omega_p)$ generically. We also mention that initializing the iterative procedure with the bare Green's function $G_0^R(k)=(\omega_k -\ek + \mu + i0^+)^{-1}$ does not change the above statement since $G_0^R(\mathbf k_F , \omega_k)$ shares the same symmetry properties.  
Now, making use of the fact that $\myim \Sigma^R_F(\mathbf k_F, \omega_k)$ is even and inserting our result~\eqref{eq:resSFtot1} for the non-FL regime, we have
\begin{align}\label{eq:KKint1}
\begin{split}
\myre \Sigma^R_F(\mathbf k_F, \omega_k) \simeq 
  \fint_{-\wmax}^{\wmax} \frac{d \omega'}{\pi} \frac{-\Gamma_{F, \mathbf k_F} - \win^{1/3} |\omega'|^{2/3}}{\omega' - \omega_k}\\
+\int^{\infty}_{\wmax} \frac{d \omega'}{\pi} \frac{\myim \Sigma_F^R(\mathbf k_F,\omega')}{\omega'}\left(
\frac{1}{1-\frac{\omega_k}{\omega'}}-\frac{1}{1+\frac{\omega_k}{\omega'}}\right) \, .
\end{split}
\end{align}  
In the regime $|\omega_k| \lesssim \wmax$, relevant for the non-FL correlations, the second integrand can be simply Taylor expanded while we introduce $\omega' = u \cdot \omega_k$ in the first one
\begin{align}\label{eq:KKint1}
\begin{split}
&\myre \Sigma_F^R (\mathbf k_F, \omega_k \to 0) \simeq 2 
\omega_k \int_{\wmax}^\infty \frac{d \omega'}{\pi} \frac{\myim \Sigma^R_F(\mathbf k_F,
\omega')}{\omega'^2} \\ &\qquad - \frac{|\omega_k|}{\omega_k}
\fint_{-\wmax/|\omega_k|}^{\wmax/|\omega_k|} \frac{d u}{\pi} \frac{\Gamma_{F,\mathbf k_F} + \win^{1/3}|\omega_k|^{2/3} |u|^{2/3}}{u - 1} \\
& \qquad\qquad\qquad\qquad\quad=  2 
\omega_k \int_{\wmax}^\infty \frac{d \omega'}{\pi} \frac{\myim \Sigma^R_F(\mathbf k_F, 
\omega')}{\omega'^2} \\ 
& - 2\sgn(\omega_k)
\fint_{0}^{\wmax/|\omega_k|} \frac{d u}{\pi} \frac{\Gamma_{F,\mathbf k_F} \!+ \win^{1/3}|\omega_k|^{2/3} |u|^{2/3}}{u^2 - 1}.
\end{split} 
\end{align}
Beyond $\wmax$, the fermionic single-particle states are expected to approach the non-interacting ones. Therefore, the first term, which is linear in $\omega_k$ like $G^R_0(\mathbf k_F, \omega_k)^{-1}$, can be neglected since it acquires a negligible prefactor. Indeed, estimating the dominant contribution from the lower integration boundary yields
\begin{align}
\omega_k \!\int_{\wmax}^\infty\! \!\!\!\!d \omega' \frac{\myim \Sigma^R_F(\mathbf k_F,
\omega')}{\omega'^2} \sim \frac{\myim \Sigma^R_F(\mathbf k_F,
\wmax)}{\wmax} \sim \frac{\alpha^{1/2}}{\eF} \omega_k .
\end{align}
Next, the first term of the second line integrates to
\begin{align}
\begin{split}
\fint_{0}^{\wmax/|\omega_k|} \frac{d u}{\pi} \frac{\Gamma_{F,\mathbf k_F}}{u^2 - 1} & = 
\frac{\sgn(\omega_k) \Gamma_{F,\mathbf k_F}}{\pi}  \log\! \left[\frac{\wmax-|\omega_k|}{\wmax+ |\omega_k|}\right] \\
& \rightarrow \frac{2\Gamma_{F,\mathbf k_F} }{\pi \wmax} \omega_k \, .
\end{split}
\end{align}
Since the prefactor $\Gamma_F(T) /\wmax \ll 1 $, when our result~\eqref{eq:resGammaFfin} for the damping rate at the highest possible temperature $T_{\text{scal}}$ from Eq.~\eqref{eq:defTscal} are inserted, we obtain again only a negligible correction to the bare frequency dependence (under the condition $\alpha/\eF^2\ll 1$)~\footnote{If one uses the minimal cut-off scheme, one still obtains $\Gamma_F(T)/ \win \ll 1$ with the results of App.~\ref{sec:minCut}.}.
Finally, the leading contribution to $\myre \Sigma^R_F$ in the limit $\omega_k \to 0$ is obtained by sending the upper integration boundary to $+\infty$ in the remaing term of Eq.~\eqref{eq:KKint1}. This yields for arbitrary $0<a<1$
\begin{align}
\begin{split}
\frac{2 \win^{1-a} |\omega_k|^a  }{\pi} \fint_0^\infty \frac{du \, u^a}{u^2-1}  =  \win^{1-a} |\omega_k|^a \tan\left( \frac{a \pi}{2} \right) \\ \xrightarrow{a=2/3} \sqrt{3} \win^{1/3}  |\omega_k|^{2/3}\, .
\end{split}
\end{align}
Altogher, we have shown that the low-temperature scaling solution of the Eliashberg equations $\myim \Sigma^R_F(\mathbf k_F,\omega_k)=- \Gamma_{F,\mathbf k_F}(T)- \win^{1/3} |\omega_k|^{2/3}$ indeed leads to the real part
\begin{align}
\myre \Sigma^R_F (\mathbf k_F , \omega_k \to 0) 
\to \sqrt{3} \win^{1/3} \sgn(\omega_k) |\omega_k|^{2/3} \, ,
\end{align}
up to subleading corrections. As we have seen, these do not affect the dominant  behavior discussed in the main text. Furthermore, it follows that this form is consistent with the analytic continuation of the self-energy obtained at the QCP.

\section{Supplement for the calculation of $\myim \Sigma^R_{F,Q}$}\label{sec:negSFQ}
In the course of the calculations for the quantum component~\eqref{eq:defSFQ} of the fermionic self-energy we restricted the momentum integral to the region $v_F p \geq| E(\omega_k- \omega_p)|$ (see Eq.~\eqref{eq:SFQint1}). Here, we show that the contribution from the neglected regime of small momenta is indeed negligible.  Furthermore, we present how the estimates for the asymptotic $M^{-1}$ tails can be obtained. 

The omitted term in the self-energy reads after simplifying the angular variables like in the main text
\begin{widetext}
\begin{align}\label{eq:SFQrest}
\begin{split}
\myim \Sigma^{R , <}_{F,Q}(\mathbf k_F, \omega_k) & \simeq  -2g^2 \dkFq \int_0^{2\pi} \frac{d \varphi_{\mathbf p}}{2\pi} \int \frac{d\omega_p }{2\pi} \int_0^{|E(\omega_k-\omega_p,\varphi_{\mathbf k_F})|/v_F} \frac{dp}{2\pi} p\,  \myim \frac{1}{E(\omega_k-\omega_p,\varphi_{\mathbf k_F}) - v_F p \cos(\varphi_{\mathbf p}- \varphi_{\mathbf k_F})} \\
&\qquad\qquad\qquad\qquad\qquad\times \left[n_B(\omega_p)- \frac{T}{\omega_p} + n_F(\omega_p - \omega_k)\right]
 \frac{ \Gamma_B(p)}{(\omega_p^2 -v_F^2 A^2 p^2 -M^2)^2+ \Gamma^2_B(p)}\, .
\end{split}
\end{align}
\end{widetext}
The result on the bosonic damping rate~\eqref{eq:resGammaBsum} shows that $\Gamma_B$ is independent of the magnitude $|\mathbf p|$ in the given momentum regime. 
Next, we rescale as follows: $\bar{\omega} = \omega/T$ and $ \tilde{p} = v_F A p/[\Gamma_B(T \bar \omega_p, \varphi_{\mathbf p})]^{1/2}$.
In addition, we omit all real terms in the denominator of the fermionic Green's function since we are only interested in an upper bound:
\begin{widetext}
\begin{align}\label{eq:SFQrest1}
\begin{split}
&\myim \Sigma^{R , <}_{F,Q}(\mathbf k_F,\omega_k) \simeq- \pi \alpha \dkFq \frac{T}{\eF A^2} \int_0^{2\pi} \frac{d \varphi_{\mathbf p}}{2\pi}\int \frac{d\bar \omega_p }{2\pi} \int_0^{\frac{A|E(T(\bar \omega_k- \bar\omega_p),\varphi_{\mathbf k_F})|}{[\Gamma_B(T \bar\omega_p,\varphi_{\mathbf p})]^{1/2}}} \frac{d \tilde{p}}{2\pi} \tilde{p} \frac{1}{\myim E(T(\bar \omega _k- \bar{\omega}_p),\varphi_{\mathbf k_F})} \\
&\qquad\quad\times \left[\bar n_B(\bar \omega_p)- \frac{1}{\bar\omega_p} + \bar n_F(\bar \omega_p - \bar \omega_k)\right] 
 \frac{1}{
\left(\dfrac{T^2}{ \Gamma_B(T \bar \omega_p,\varphi_{\mathbf p})} \bar\omega_p^2 -\tilde{p}^2 -\dfrac{M^2}{\Gamma_B(T \bar \omega_p,\varphi_{\mathbf p} )}\right)^2+ 1}\, .
\end{split}
\end{align}
\end{widetext}
The asymptotic forms of the bosonic damping~\cref{eq:numSBimT0,eq:numSBimw0} indicate that $\Gamma_B(T \bar\omega_p,\varphi_\mathbf{p}) \sim T^{1/3}$, as long as $T \lesssim T_{\text{scal}}$ such that the fermionic damping $\Gamma_F(T) \sim T^{2/3}$. Since $M^2(T) \sim T^{2/3}$, only the $\tilde{p}^2$ term has to be kept in the denominator of the last term at small temperatures. In fact, this statement holds also at higher temperatures when $\Gamma_F(T) \sim T^{1/2}$ up to logarithmic corrections. With the integral 
\begin{align}
\int_0^a du \frac{u}{u^4 + 1}= \frac{1}{2} \arctan\left(
a^2\right)\, ,
\end{align} 
for $a>0$, we find then
\begin{widetext}
\begin{align}\label{eq:SFQrest2}
\begin{split}
\myim \Sigma^{R , <}_{F,Q}(\mathbf k_F,\omega_k) \simeq -\alpha \dkFq \frac{T}{4\eF A^2} \int_0^{2\pi} \frac{d \varphi_{\mathbf p}}{2\pi}\int \frac{d\bar \omega_p }{2\pi}   \frac{1}{\myim E(T(\bar \omega_k-\bar{\omega}_p),\varphi_{\mathbf k_F})}
\left[\bar n_B(\bar \omega_p)- \frac{1}{\bar\omega_p} + \bar n_F(\bar \omega_p - \bar \omega_k)\right] \\
\times \arctan\left(\frac{A^2|E(T(\bar\omega_k- \bar \omega_p),\varphi_{\mathbf k_F})|^2}{\Gamma_B(T \bar{\omega}_p,\varphi_{\mathbf p})} \right)\, .
\end{split}
\end{align}
\end{widetext}
With the scaling function~\eqref{eq:keyResSFtot2} of the fermionic self-energy we find that $E(T(\bar\omega_k- \bar \omega_p),\varphi_{\mathbf k_F}) \sim T^{2/3}$ in the low-energy limit.  Combining this with the scaling of $\Gamma_B$, we can expand the arctan around small arguments:
\begin{widetext}
\begin{align}\label{eq:SFQrest3}
\begin{split}
&\myim \Sigma^{R , <}_{F,Q}(\mathbf k_F,\omega_k) \simeq - \alpha \dkFq \frac{T}{4\eF} \int_0^{2\pi} \frac{d \varphi_{\mathbf p}}{2\pi}\int_{|\bar \omega_p| \geq \bar \omega_{\text{IR}}} \frac{d\bar \omega_p }{2\pi}  \frac{\left[\bar n_B(\bar \omega_p)- \frac{1}{\bar\omega_p} + \bar n_F(\bar \omega_p - \bar \omega_k)\right]}{\myim E(T(\bar \omega_k- \bar{\omega}_p),\varphi_{\mathbf k_F})}   
\frac{|E(T(\bar\omega_k - \bar \omega_p))|^2}{\Gamma_B(T \bar{\omega}_p,\varphi_{\mathbf p})}\, .
\end{split}
\end{align}
\end{widetext}
Here, we have introduced an infrared cut-off $\bar \omega_{\text{IR}}$ to avoid a spurious pole at $\bar\omega_p =0$ introduced by the expansion. The latter arises from the asymptotics $\Gamma_B(\omega_p \to 0, \varphi_\mathbf p) \to \alpha \omega_p/\Gamma_F(T)$ (cf. Eq.~\eqref{eq:resGammaBsum}) which is innocuous for the full expression since the arctan has a finite limit. We can estimate $\bar \omega_{\text{IR}}$ via $$|E|^2 \sim \Gamma_B \sim \alpha \omega_p/\Gamma_F(T)$$ which implies $\bar \omega_{\text{IR}} \sim T$.
With $E \sim T^{2/3}$ and $\Gamma_B \sim T^{1/3}$, the integral scales like $T^{4/3}$ up to logarithmic corrections. Furthermore, in the frequency regime $|\bar\omega_p| \lesssim \bar\omega_{\text{IR}}$ we set $\arctan((A E/\Gamma_B )^2) \to 1$ and obtain a contribution of order $\mathcal O (T^{4/3})$, too.
Consequently, the omitted terms are indeed subleading compared to the results~\eqref{eq:resSFQw0} and~\eqref{eq:resSFQdelta} when the scaling solution~\eqref{eq:resSFtot2} for the fermionic self-energy is inserted.

Now we consider the large-$\bar M$ behavior of the scaling function $\tilde{\Sigma}^R_{F,Q}$ defined in Eq.~\eqref{eq:SQFdefScal}. Based on the asymptotics given in Eq.~\eqref{eq:SFQexp} for the zeros $u_j$ in both the regimes $\bar M \gg |\bar \omega_p|$ and $\bar M \ll |\bar \omega_p|$, we obtain the simple approximation
\begin{widetext}
\begin{align}\label{eq:SQFw0}
\begin{split}
\myim \tilde{\Sigma}^R_{F,Q}(0 , \bar M \to \infty)& \simeq \!\int \frac{d\bar\omega_p }{2\pi} \!\left[\bar n_B(\bar \omega_p)- \frac{1}{\bar\omega_p} + \bar n_F(\bar \omega_p) \right]\!\!\!\bigg[
\frac{2\bar \omega_p}{\bar M^4}\log\left(\dfrac{\sqrt{e}|\bar{\omega}_p|}{\bar M^3}\right) \theta\left(\!\frac{\bar M^3}{\sqrt{e}} - |\bar{\omega}_p|\!\right)  - \frac{2\pi}{3^{3/2} \bar \omega_p^{1/3}} \theta\!\left(\!|\bar{\omega}_p|-\frac{\bar M^3}{\sqrt{e}}\!\right)\!\! \bigg] .
\end{split}
\end{align}
\end{widetext}
It yields a reasonable estimate provided that the large-$\bar M$ behavior is not dominated by the crossover regime. This is not the case as can be seen in Fig.~\ref{fig:MdepComb}. Nevertheless, an uncertainty in the $\mathcal O(1)$ prefactor arises from the value assigned for the intermediate integration boundary. Here, we have chosen $\bar M^3 /e^{1/2}$ to obtain nonpositve integrands in order to be unbiased by cancellations between both parts. Next, we introduce the new variable $v =\bar \omega_p/\bar{M}^3$. Upon sending $\bar{M} \to \infty$, $n_{F,B}$ converge to $\mp \theta(-v)$ and cancel each other while the pole at the origin is regularized by the linear frequency factor of the first term in the second bracket. As a result, we find 
\begin{align}\label{eq:TSFQM0}
\begin{split}
\myim \tilde{\Sigma}^R_F(0, \bar M \to \infty) & \simeq -\frac{4}{\bar{M}} \int_{-1/\sqrt{e}}^{1/\sqrt{e}} \frac{d v}{2\pi}\log(\sqrt{e} |v|) \\
& \quad+ \frac{2}{\bar M}\int_0^{1/\sqrt{e}} \frac{dv}{2\pi} \frac{2\pi}{3^{3/2} v^{4/3}} \\
& = \frac{2}{\bar M} \left(\frac{1}{\sqrt{e} \pi} + \frac{e^{1/6}}{\sqrt{3}}\right) = 1.7502... \frac{1}{\bar M}\, ,
\end{split}
\end{align}       
which indeed decays algebraically with $\bar M^{-1}$ in agreement with the numerical evaluation. Even the numerical prefactor is reproduced with tolerable accuracy.

Finally, we determine the $\bar M \to \infty $ asymptotics of the frequency-independent thermal correction $\myim \delta \tilde{\Sigma}^R_{F,Q}(|\bar \omega_k| \gg 1, \bar M)$ defined in Eq.~\eqref{eq:resSFQHF}. The latter definition is formally almost identical to Eq.~\eqref{eq:SQFw0} except that $\bar n_F(\bar\omega)$ is replaced by $\theta(-\bar \omega)$. Transforming the integral again to the variable $v = \bar \omega_k/M^3$ and sending $\bar M \to \infty$, leads exactly to the result of Eq.~\eqref{eq:TSFQM0} with the same uncertainty in the $\mathcal{O}(1)$ prefactor. As a consequence, we indeed obtain the dimensionfull self-energy given in Eq.~\eqref{eq:resSFQdelta} in the main text.

\section{Scaling solution for the minimal cut-off scheme}\label{sec:minCut}
We repeat the scaling analysis of Sec.~\ref{sec:bosMass}, now  with the smallest possible cut-offs $|\omega| \leq \win$ and $v_F \Lambda = \alpha /\eF$, or equivalently $h_{\min} =\alpha/\eF^2 $. 
Since $\win \ll v_F \Lambda$ anyway we can set $h_\omega=1$. 
We begin again with the bosonic mass generated from the quantum corrections $M^2_{Q}(T)$ whose most important contribution arises again from the integral~\eqref{eq:MQdom}, yet with appropriately changed cut-offs. As a result, this returns
\begin{align}
M^2_{-,Q}(T) \sim \frac{\alpha}{(h_{\min} \eF)^3} \frac{\alpha^{2/3} T^{2/3}}{\eF} \win^2 \sim \frac{\alpha^{8/3} T^{2/3}}{\eF^4} \, ,
\end{align}
which is again of the form of Eq.~\eqref{eq:MTQCR}.
In the following, we omit the index $-,Q$ since this turns out to be the dominant contribution to the inverse order parameter susceptibility in analogy to the maximal cut-off scheme. From
\begin{align}
M(T) \sim \frac{\alpha^{4/3} T^{1/3}}{\eF^2}
\end{align}
we obtain via Eq.~\eqref{eq:resGammaFQCR} 
\begin{align}
\Gamma_F(T) \sim \frac{\eF}{\alpha^{1/3}} T^{2/3}\, .
\end{align}
As stated already above we observe the identical dependence on temperature as in Sec.~\ref{sec:bosMass}. 
At sufficiently low temperatures, which we specify below, we find the scaling function
\begin{align}
\begin{split}
&\myim \Sigma^R_{F} (\mathbf{k}_F, \omega,T)   =  - \frac{\eF}{\alpha^{1/3}}T^{2/3} \tilde{\Sigma}^R_F\left(\frac{\omega_k}{T},\varphi_{\mathbf{k}}\right) \\
& \tilde{\Sigma}^R_F = \left(b_2 \dkFq + \frac{ \alpha |\dkF|^{4/3}}{8 \cdot 3^{3/2}\eF^2}\left(\frac{|\omega|}{T}\right)^{2/3}\right) \, ,
\end{split}
\end{align}
where $b_2$ is a number of order one. The nonuniversal character is implicit in this case because of the choice $h_\omega=1$. From comparing the low- and high-frequency regimes we determine the crossover scale
\begin{align}
\omega_{\lessgtr} \sim \frac{\eF^3}{\alpha^{3/2}} T \, ,
\end{align}
which is again linear in $T$. Next, we repeat the analysis that lead to Eq.~\eqref{eq:defTscal} for estimate up to which temperature scaling applies
\begin{align}
T_{\text{scal}} = 
\begin{cases}
\left(\dfrac{\alpha^{1/2}}{\eF}\right)^{17/2} \eF \, ,&\text{if } \dfrac{\alpha}{\eF^2} \gg B^4 \\ \dfrac{ \alpha^5}{B^3\eF^9}\, ,&\text{if } \dfrac{\alpha}{\eF^2} \ll B^4 
\end{cases} \, .
\end{align}
However, in the weak-coupling limit only the second case corresponding to violating $M(T) \gg \Gamma_F(T)$ is relevant due to the strong enhancement of the thermal scattering rate at simultaneously suppressed $M^2(T)$, as compared to the case of the maximal cut-off. 

Finally, we have to check that $\Sigma^R_{F,T}$ does not change the dominant scaling behavior obtained so far. To this end, we have to recalculate $M_{-,T,1,2,3}$ defined in Eqs.~\cref{eq:Mm1T,eq:Mm2T,eq:Mm3T} but with changed cut-offs. First, we note that $\omega_{M(T_{\text{scal}})} \ll \win$ in the weak-coupling limit such that $\Gamma_F(T)$ does not extend over the entire non-FL region in $(\omega_k, \mathbf k)$ space. Then $M_{-,T,1} = 0$ by symmetry reasons, whereas 
\begin{align}
M^2_{-,T,2}(T) \sim \frac{\alpha^2 T }{\eF} \frac{\win^{1/3} \omega^{5/3}_{M(T)} }{(h_{\min} \eF)^4} \sim \frac{\alpha^{1/3} T^{11/6}}{\eF^{1/2}} \, ,
\end{align}
and 
\begin{align}
M^2_{-,T,3}(T) \sim \frac{\alpha^2 T }{\eF} \frac{\win^2}{(v_F \Lambda)^4} \sim \frac{\alpha^2  T}{\eF^3} \, .
\end{align}
For temperatures below $T_{\text{scal}}$ these are negligible as compared to $M^2_{-Q}(T)$. Furthermore, $M_+^2(T)$ as well as the tails of $\myim \Sigma^R_{F,T}$ that extend beyond the range of non-FL correlations are irrelevant for the same reasons as in Sec.~\ref{sec:feedback}. As a result, the bosonic mass is again exclusively given by the contribution from the quantum component. However, the crossover from $M^2(T)\sim T^{2/3}$ to $M^2 \sim T$ with logarithmic corrections takes place at the UV frequency cut-off $T \sim \win$ in agreement with Eq.~\eqref{eq:Tcrossgen}.
\section{Irrelevant vertex corrections}\label{sec:irrVertexCorr}
In the discussion of the vertex corrections in Sec.~\ref{sec:vertexCorr} several terms were omitted. Here, we show that this is indeed justified.
First of all we have the second line of Eq.~\eqref{eq:GammaqStart}. In the limit $\mathbf p \to 0$, the angular integral becomes
\begin{align}\label{eq:angInt2}
\begin{split}
\int_0^{2\pi} \frac{d\varphi_{\mathbf q}}{(2\pi)} \frac{1}{(v_F q\cos(\varphi_{\mathbf{q}}- \varphi_{\mathbf k_F}) \pm i \myim \Sigma^R_F(\omega_q,\varphi_{\mathbf k_F}))^2} =  \\
\frac{\myim \Sigma^R_F(\omega_q,\varphi_{\mathbf k_F})}{
(v_F^2 q^2 + \myim \Sigma^R_F(\omega_q,\varphi_{\mathbf k_F})^2)}\,  ,
\end{split}
\end{align}
where we retain only the dominant frequency terms in the infrared and have set $\omega_k =0 $ directly since no terms are cancelled erroneously. Let us consider $T=0$ first. Up to numerical prefactors 
the second line of Eq.~\eqref{eq:GammaqStart} reads after rescaling
$ \omega_q = \win \hat \omega_q$ and $v_F q = \alpha^{1/3} \win^{1/3} \hat q$ 
\begin{align}\label{eq:GammaQirr}
\begin{split}
g \dkF \frac{\win}{\eF} \! \int_{-\hat{\omega}_{\max}}^{\hat{\omega}_{\max}} \!\!\!\! d\hat \omega_q \! \int_0^\infty \!\!\! d\hat q \, \frac{\hat q^2 |\hat \omega_q|}{\hat q^6+ \hat \omega_q^2}
\frac{|\hat \omega_q|^{2/3} }{\left(\hat q^2 + \dfrac{\win^{4/3}}{\alpha^{2/3}} |\hat \omega_q|^{2/3}\right)^{3/2} }  ,
\end{split}
\end{align}
where $M=0$ at QCP. Here, we have made the frequency cut-off explicit to establish UV convergence, which is necessary as we will see in the next step.
Using again the Ising nematic scaling relations $\hat{q} \sim \hat \omega_q^{1/3}$ which justifies the Landau damping form we find (again up to numerical prefactors:
\begin{align}\label{eq:GammaQirr2}
\begin{split}
g \dkF \frac{\win}{\eF} \! \int_{-\hat{\omega}_{\max}}^{\hat{\omega}_{\max}} \!\!\!\! d\hat \omega_q |\hat \omega_q|^{-1/3}\, .
\end{split}
\end{align}
Here, we have employed the condition $\hat q \gg \win^{4/3} \alpha^{-2/3} |\hat \omega_q|^{2/3}$ adapted from Eq.~\eqref{eq:ETcond1}. Thereby, we miss a logarithmic correction arising from the integration boundary $\hat q \to 0$ which, however, cannot make the term become relevant anyway. With $\wmax \sim \alpha^{1/2}$ we find that the vertex correction is of order
\begin{align}\label{eq:GammaQirr3}
g \dkF \frac{\alpha}{\eF^2} \ll g \dkF
\end{align}
and thus is irrelevant. Since no additional poles appear in the corresponding expression at finite temperatures, it is justified to neglect the second line of Eq.~\eqref{eq:GammaqStart}. Furthermore, we can conclude on the basis of this calculation that also the first and third line of Eq.~\eqref{eq:GammaQ2} are merely perturbative in the coupling constant.

Finally, the we come to the thermal vertex corrections $\Gamma^{RR(AA)}$ in the fermionic self-energy~\eqref{eq:SF2loop} that are of the form $\text{ct}(\beta \omega_q/2) \myim D^R(\mathbf q ,\omega_q) G^{R(A)} G^{R(A)}$. Repeating the analysis below Eq.~\eqref{eq:GammaTint1} leads to Eq.~\eqref{eq:GammaTint2} with the combination of fermionic Green's functions $G^R G^R$ or $G^A G^A$, respectively. The angular integral is again of the form~\eqref{eq:angInt2}. 
Then, performing the $p$ integration yields in the quantum critical regime 
\begin{align}\label{eq:GammaTirr}
\Gamma_T^{RR(AA)}(\mathbf 0,0;\mathbf k_F,0) \sim \frac{g \alpha T}{\eF M^2(T)} \dkF^3 \sim g  \frac{T^{1/3}}{\alpha^{1/6}} \, ,
\end{align} 
which is much smaller than one for temperatures $T \ll T_{\max}$.
\bibliography{Bibliography/SM-QCR}
\end{document}